\documentclass[a4paper,11pt]{article}

\usepackage[dvips]{graphicx}

\usepackage{amssymb,amsfonts,amsmath,color,multirow}

\usepackage{caption}

\usepackage{float}
\floatstyle{plaintop} 
\restylefloat{table}

\usepackage[applemac]{inputenc}
\usepackage[english]{babel} 

\usepackage{pifont} 
\newcommand{\checkm}{\ding{51}}

\usepackage{amsmath,amssymb,amsfonts,times, MnSymbol,cclicenses,bm,subfig}
\usepackage{enumerate}
\usepackage[sans]{dsfont}
\usepackage{textcomp}
\usepackage{graphicx}
\usepackage{lscape}
\usepackage{array,color,lineno,nicefrac,multirow}
\usepackage{rotating}
\usepackage{array}

\usepackage[pdftex,hmargin={2.2cm,1.8cm},vmargin={2cm,2cm}]{geometry}
\usepackage[square]{natbib}

\usepackage[pdftex,bookmarks=false,colorlinks=true,citecolor=red]{hyperref}
\usepackage[all]{hypcap}

\newcommand{\PhiFJ}{\Phi_{\mathrm{FJ}}}
\newcommand{\TFJ}{T_{\mathrm{FJ}}}
\newcommand{\kDMP}{\kappa_{\mathrm{DMP}}}

\begin{document}


\title{Epidemics in markets with trade friction and imperfect transactions}

\author{Mathieu Moslonka-Lefebvre$^{1,2,3,\#}$, Herv\'{e} Monod$^{1}$, Christopher A. Gilligan$^{2}$, \\
Elisabeta Vergu$^{1,*}$ and Jo\~{a}o A. N. Filipe$^{2,*}$}

\date{\small $^1$ INRA, UR 0341 Math\'{e}matiques et Informatique Appliqu\'{e}es, 78350 Jouy-en-Josas, France \\
$^2$ Department of Plant Sciences, University of Cambridge, Downing Street, Cambridge CB2 3EA, United Kingdom \\
$^3$ AgroParisTech, F-75005 Paris, France \\
$^\#$ Corresponding author, email: mmoslonka@jouy.inra.fr \\
$^*$ These authors contributed equally to this work \\[0.5cm]
\large 23 October 2013 \\
}

\maketitle


\begin{abstract}
Market trade-routes can support infectious-disease transmission, impacting biological populations and even disrupting causal trade. Epidemiological models increasingly account for reductions in infectious contact, such as risk-aversion behaviour in response to pathogen outbreaks. However, market dynamics clearly differ from simple risk-aversion, as are driven by different motivation and conditioned by trade constraints, known in economics as friction, that arise because exchanges are costly. Here we develop a novel economic-market model where transient and long-term market dynamics are determined by trade friction and agent adaptation, and can influence disease transmission. We specify the participants, frequency, volume, and price in trade transactions, and investigate, using analytical insights and simulation, how trade friction affects joint market and epidemiological dynamics. The friction values explored encompass estimates from French cattle and pig markets. We show that, when trade is the dominant route of transmission, market friction can be a significantly stronger determinant of epidemics than risk-aversion behaviour. In particular, there is a critical friction level above which epidemics do not occur. For a given level of friction, open unregulated markets can boost epidemics compared with closed or tightly regulated markets. Our results are robust to model specificities and can hold in the presence of non-trade disease-transmission routes. In particular, we try to explain why outbreaks in French livestock markets appear more frequently in cattle than swine despite swine trade-flow being larger. To minimize contagion in markets, safety policies could generate incentives for larger-volume, less-frequent transactions, increasing trade friction without necessarily affecting overall trade flow.
\end{abstract}


\textbf{Abbreviations:}~FTM, frictional-trade market; GSA, global sensitivity analysis; LSD, law of supply and demand; ME, market-epidemiological; RA, adaptive risk aversion; SI, Supporting Information \\

\section{Introduction}

A long-standing challenge in identifying appropriate control strategies in infectious disease epidemiology is to establish which characteristics of host contact structures drive disease spread (e.g. \citep{FilipeEtal2012PCB,KampEtal2013PLoSCB}). Adequate characterisation of population contact structure is generally difficult because individual contacts can change over time or differ in relative epidemiological relevance. When epidemics occur on a comparatively shorter time scale, epidemiological models may neglect such time variation and adequately represent changes in contact structure as responses to infection, such as recovery or elimination. Otherwise, tractable models account for inherent changes such as host ageing and demography \citep{KeelingRohani2008Book}, or change in host interactions such as dynamical rewiring of their links as typically observed for social relationships \citep{VolzMeyers2007PRSLB}. However, in these models epidemics do no affect host behaviour.

More recent studies have explored the impact of adaptive risk-aversion behaviour on disease dynamics \citep{FunkEtal2009AdaptivePNAS,FunkEtal2010AdaptiveInterface}. Risk aversion behaviour is a form of disease prevention where asymptomatic hosts reduce exposure to infection by reducing their contact rate (e.g. by staying home) and/or their probability of infection per contact (e.g. by wearing protective masks); it implies that hosts have some information about a given disease outbreak and act on their own initiative rather than relying on community measures by regulatory bodies. If this behaviour is determined by the perception of a variable risk, then it is said to be ÔadaptiveÕ risk-aversion (RA).
In the literature, RA has been expressed as a simple function of disease prevalence or outbreak awareness \citep{FunkEtal2010AdaptiveInterface}. RA has also been evaluated via complex economic optimization, where hosts arbitrate a trade-off between the benefits of interaction and the costs of infection acquired through the resulting contacts \citep{Fenichel2011AdaptivePNAS,MorinEtal2013AdaptiveNRM}.
Naturally, epidemiological models that neglect RA behaviour tend to overestimate the probability of occurrence and severity (e.g. infectious peak and cumulative cases) of epidemics \citep{FunkEtal2009AdaptivePNAS,FunkEtal2010AdaptiveInterface,Fenichel2011AdaptivePNAS,MorinEtal2013AdaptiveNRM}. To the best of our knowledge, epidemiological modelling studies have focused on adaptive human behaviour which is solely altered in response to detected outbreaks. The epidemiological significance of other behaviour, possibly with less intuitive effects, remains to be investigated.

In this paper, we investigate the development of disease epidemics and their control in systems with more general adaptive human behaviour. We focus on markets of goods, where the dynamics of potentially-infectious contacts are driven, primarily, by economic decisions very different from those underlying disease-risk aversion. Economic markets can propagate diseases through the exchange of contaminated products among market agents (e.g. farms, people, banks). Epidemics can also alter agent behaviour as a result of regulation or self awareness. Markets contributing to disease epidemics include cattle \citep{RautureauEtal2011TED}, swine \citep{LentzEtal2011PVM}, and sheep \citep{KissEtal2006Interface} trade, prostitution \citep{RochaEtal2011PCB}, and airline transportation \citep{ColizzaEtal2006PNAS}. Other types of epidemics occur through exchange of information on the Internet \citep{LloydMay2001InternetVirusScience} and exchange of debt in financial markets \citep{HaldaneMay2011Nature}. When an epidemic shock occurs in a market, differing actions and behaviour may help either to restore or disturb the balance between supply and demand. Sanatory regulation and RA aimed at reducing infectious contacts can reduce supply and demand. Market agents may try to establish new, but potentially-infectious trade relationships that could outweigh the effect of regulation and RA efforts. In other words, adjustment in supply and demand among agents could worsen the outbreak. In contrast, the establishment of trade relationships is conditioned by physical impediments, such as the time and effort needed for searching business partners, cutting deals, and delivering goods, known in labour economics as 'friction' (see e.g. \citep{Pissarides2011AER} and the model of Diamond, Mortensen and Pissarides in the supporting information (SI)). Therefore, by limiting the development of potentially-infectious trade contacts, friction can suppress epidemics. Phenomena like adjustment in supply and demand, and friction, show that human behavior in response to epidemics in markets does not simplify to regulation and RA.

In order to represent dynamics of markets and epidemics at appropriate and consistent time scales, we developed a novel economic-market model, the {\em frictional-trade market (FTM) model}, with transient and long-term dynamics determined by trade friction and agents' decisions to supply or demand. We integrate market and epidemic processes into a {\em market-epidemiological (ME)} modelling framework where trade influences disease transmission and disease control actions affect trade. We first study the behaviour of the FTM model in the absence of epidemics. Then, we investigate how market dynamics affect epidemic development, and how epidemics disrupt market dynamics in the short and long terms. We also consider two forms of response to disease outbreaks taken from the literature: the removal (inactivation) of market agents found to be infected by regulators and their later re-introduction or replacement, and an adaptive risk-aversion behaviour (RA) of market agents. Therefore, we highlight differences in concept and impact on epidemic development, between RA, which relies on individual decision-making, and market dynamics, which depends on regulatory actions, often by governmental bodies, and complex collective behaviour, as exhibited by price changes in response to shifts in supply or demand.
Finally, we extend our study beyond an isolated (e.g. national) market, by contrasting scenarios where infectious diseases are propagated differently through trade pathways with contrasting degree of openness to international trade and non-trade pathways. We expect our central results to apply to different types of markets, and illustrate applications to cattle and swine markets in France where there is detailed registration of livestock movement.

\section{Market-epidemiological modelling framework}

\subsection{Overview}

\begin{figure}[H]
\begin{center}
\includegraphics[width=\textwidth]{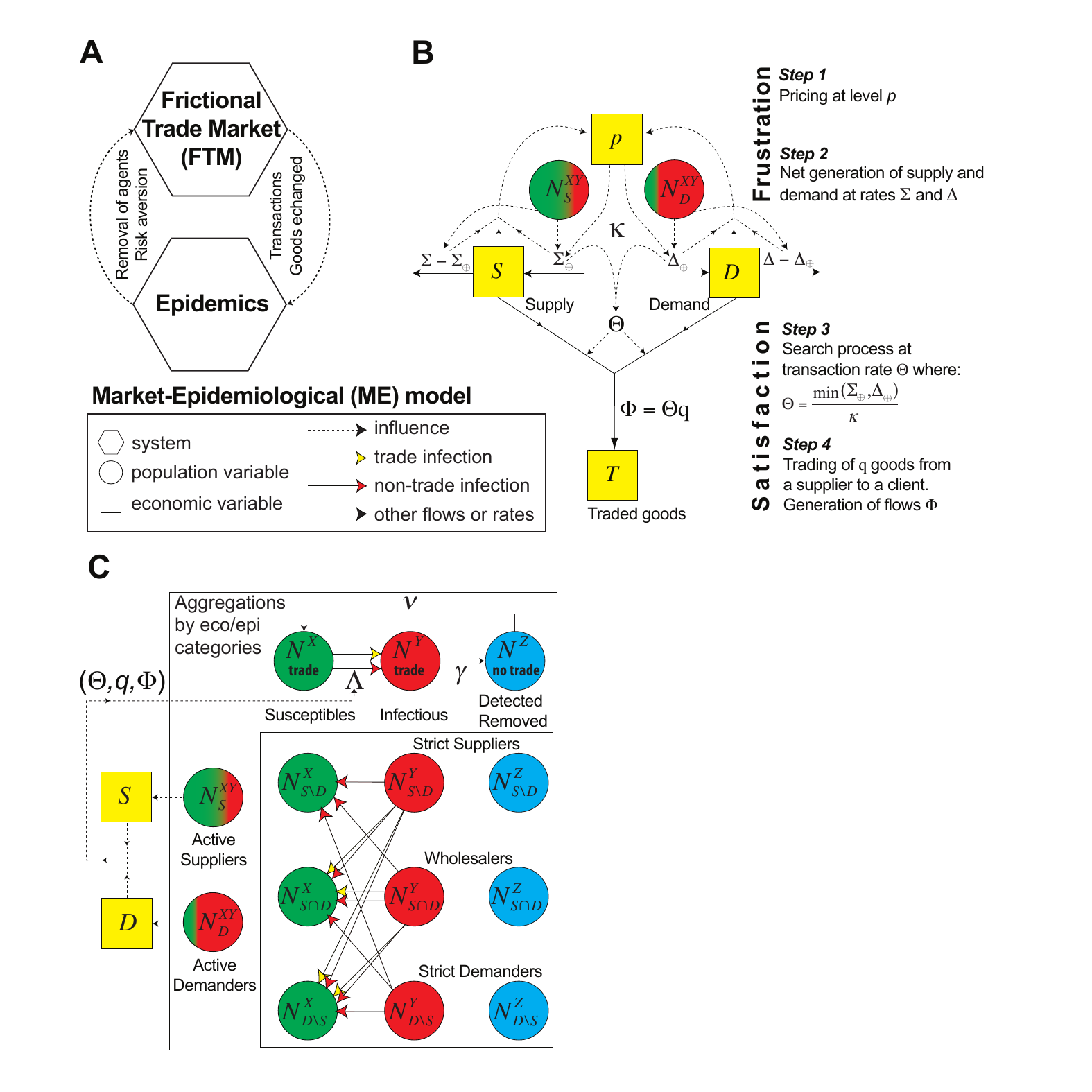}
\end{center}
\caption{\small {\bf Joint market-epidemiological modelling framework.}
\textbf{A}) General structure and key links between the market and epidemiological subsystems. \textbf{B}) Components of the frictional-trade market (FTM) model. \textbf{C}) Components of the market-epidemiological (ME) model. Yellow (red) arrows represent direction of disease transmission through trade (non-trade) routes. 
The market agents host the disease-causing pathogen. $N_A$ represents the number of agents of arbitrary type $A$ that can vary over time. $N_{S\backslash{}D}$, $N_{S\cap{}D}$ and $N_{D\backslash{}S}$ denote the numbers of {\em strict suppliers},  {\em wholesalers} (supplier and demander) and {\em strict demanders} respectively. From an epidemiological point-of-view, each agent can be in the {\em Susceptible} ($X$), {\em Infectious} ($Y$), or {\em Removed} ($Z$) state. Susceptible agents become infectious at rate $\Lambda(t)$; infectious agents are removed at rate $\gamma$; and removed agents re-enter the market (recover) at rate $\nu$. Here, ÔremovalÕ means that an infectious agent is detected and removed from the market by a regulator and becomes {\em inactive}. 
Each agent, whether a strict supplier, wholesaler, or strict demander, can be in each of the epidemiological states. For example, $N_{S}^X$ and $N_{S}^Y$ denote the number of susceptible and infectious suppliers, respectively. Therefore, there are $N_{S}^{XY} = N_{S\cap{}D}^{XY} + N_{S\backslash{}D}^{XY}$ active suppliers.
Supply and demand stocks $S$ and $D$ quantify the {\em willingness to trade} or {\em frustration} of suppliers and demanders respectively. Supply stocks consist of goods held by sellers, while demand stocks represent goods wanted by buyers and are virtual.  The {\em trade flow} $\Phi$ out of the supply and demand stocks, aggregates the transactions between supply and demand agents that agree to exchange some of their stock at price $p$. Transactions occur at {\em transaction rate} $\Theta$ and in each transaction an {\em average transaction stock} $q$ is exchanged. At {\em the reference market} dynamic equilibrium, $q$ is equal to $\kappa$, {\em the coefficient of friction}. For equilibrium trade flow $\Phi^*$ kept constant, increasing $\kappa$ leads to $\Theta^* = \Phi^* / \kappa$. Hence same goods per time unit are exchanged less frequently but in greater amounts per transaction (see ÔResultsÕ). 
\label{fig:coupledmodel}}
\end{figure}

We develop a novel theoretical framework for the propagation of infectious diseases in economic markets where the exchanged goods can transmit an infectious organism between market agents (Fig.~\ref{fig:coupledmodel}A). In order to represent this process, we link a model of an economic market system and a model of an epidemiological system. Each model dynamics can exist per se, i.e. epidemics can occur in host populations unaffected by markets, and markets often operate without disease outbreaks through trade routes. However, by building a system that links the dynamics of these subsystems we can study their interdependencies. 
As the epidemiological model we use is a simple adaptation of a standard compartmental epidemiological model, it is introduced later with brief explanation. The dynamic economic-market model, however, is novel, and is derived in detail.
A key property of this model is its coefficient of friction, which characterizes a market's
inherent dynamics and response to disturbance caused, for example, by disease outbreaks (see the ÔResultsÕ section).

\subsection{A frictional-trade market (FTM) model}

Here we develop a FTM model for the dynamics of markets (without epidemics) where goods of a single type are exchanged for money in {\em transactions} between suppliers (sellers) and demanders (buyers), and all transactions at given time are based on a single price per good. The model tracks the dynamics of {\em extensive} state variables (i.e.\ with a ÔsizeÕ or ÔscaleÕ), such as stocks, number of agents, and trade flow, and {\em intensive} state variables (i.e.\ global indices), such as price of goods. Usually, price, trade flow, and number of agents are observable market quantities, while overall stocks are not observable.

Our model market is defined, at each time $t$, by an overall {\em price} per good $p(t)$, and overall stocks of {\em supply} $S(t)$ and {\em demand} $D(t)$. Supply stocks consist of goods held by sellers, while demand stocks represent goods wanted by buyers and are virtual. The stocks quantify the {\em willingness to trade} of suppliers and demanders.
Our FTM model integrates many complex mechanisms and dynamical feedbacks that have seemingly not been explored concurrently in the literature (see SI).
To keep our model parsimonious, we hence model market dynamics at a whole-market level, where agent-level stocks and transactions are approached by average values per agent. 
Inspired by well-known population dynamics models (e.g. \citep{May1977ThresholdsEcosystemsNature,DurrettLevin1994MEcologicalModelsComparisonsTPB}), our whole-market-scale model is a population-level description of agents and stocks with mass-action interactions (transactions).
At this simplified level of description, we define the market model through temporal change in overall stocks. Each {\em stock} ($S$ and $D$) is created at a specific {\em net creation rate} and depleted through a {\em trade flow}, represented by the rate equation:
\begin{equation}
\frac{d \ \text{[stock]}}{dt} = \text{[net creation rate]} - \text{[trade flow]} \ .
\label{mmodel1}
\end{equation}

The {\em net creation rate}, of supply stock $\Sigma$ or demand stock $\Delta$, is composed of: 1) a {\em production rate}, $\Sigma_\oplus(p,N_S)$ or $\Delta_\oplus(p,N_D)$, respectively, that depends on current price $p$ and numbers of supply agents $N_S$ and demand agents $N_D$, 2) a net {\em loss rate} $L$, e.g.\ spoilage of supply goods and loss of demanders interest (positive loss) or multiplication of livestock (negative loss), and 3) an {\em external flow} of stock $E$, e.g., import or export of raw materials or goods:
\begin{eqnarray}
\text{[net creation rate]}  & = &  \text{[production rate]} \ - \  \text{[loss rate]} \ + \  \text{[external flow]} ~,     \nonumber \\
\Sigma & = &  \Sigma_{\oplus}(p, N_S) \ - \  L_S \ + \  E_S  \ \ \ \ \ (\text{supply}) ~,      \nonumber \\
\Delta & = &  \Delta_{\oplus}(p, N_D) \ - \ L_D \ +\  E_D    \ \ \ \  (\text{demand}) \ .
\label{mmodel1b}
\end{eqnarray}

Following the economic literature (see SI for details), production rates are defined as:
\begin{equation}
\begin{gathered}
\Sigma_{\oplus}(p, N_S) = N_S~\sigma_0~p^{\varepsilon_S}~, \hfill \\
\Delta_{\oplus}(p, N_D) = N_D~\delta_0~p^{-\varepsilon_D}~, \hfill
\label{prodrates1}
\end{gathered}
\end{equation}
where $\sigma_0$ and $ \delta_0$ are the reference \textit{per-agent} production rates in supply and demand at the reference price $p = p_0 = 1$, and $\varepsilon_S \ge 0$ and $\varepsilon_D \ge 0$ are the price elasticities of supply and demand respectively. Notice that $\Sigma_{\oplus}(p, N_S)$ increases while $\Delta_{\oplus}(p, N_D)$ decreases with increasing price.
Furthermore, we assume that the loss rates are directly proportional to stocks, i.e.\ $L_S=r_S S(t)$ and $L_D=r_D D(t)$ with $r_S, r_D$ constants, and that external flows ($E_S, E_D$) are constant.

The {\em trade flow} $\Phi(t)$ out of the supply and demand stocks, aggregates the transactions between supply and demand agents that agree to exchange some of their stock (i.e., exchange of supply stock for demander's money). Transactions occur at {\em transaction rate} $\Theta(t)$ and in each transaction an {\em average transaction stock} $q(t)$ is exchanged; therefore, the trade flow is:
\begin{eqnarray}
\text{[trade flow]} &=& \text{[transaction rate]} ~ \text{[average transaction stock]} ~, \nonumber \\
\Phi(t)  &=& \Theta(t) ~ q(t) \ .
\label{mmodel1c}
\end{eqnarray}
First we define $q$. The average per-agent supply and demand stocks are $S(t)/N_S$ and $D(t)/N_D$.
Here we take the number of market agents to be constant in time, but later, when considering epidemics in markets, we allow for removal of infectious agents and their subsequent re-introduction after sanitation measures.
In this model we assume that, once a pair of supply and demand agents has been identified and agreed to transact, they exchange the maximum possible per-agent stock (a best-possible match): $S(t)/N_S$ if there is excess demand ($D(t)/N_D \ge S(t)/N_S$), and $D(t)/N_D$ if there is excess supply ($S(t)/N_S \ge D(t)/N_D$). Hence, the average transaction stock, conditional on best-possible matching, is:
\begin{equation}
q(t) = \min \{S(t)/N_S ; D(t)/N_D\} \ . \label{mmodel1d}
\end{equation}
The transaction rate $\Theta(t)$ in (\ref{mmodel1c}) is determined by a driving factor, the {\em urge-to-exchange rate}, and a {\em limiting factor} that compounds multiple constraints, such as the {\em search} for a trading partner and the logistics of stock {\em delivery}. Hence:
\begin{eqnarray}
\text{[transaction rate]} & = &  \frac{\text{[urge-to-exchange rate]}}{\text{[limiting factor]}} ~, \nonumber \\
\Theta(t) & = & \frac{\min \{\Sigma_{\oplus}(p, N_S) ; \Delta_{\oplus}(p, N_D)\}}{\kappa} \ .
\label{mmodel1e}
\end{eqnarray}
We assume that the {\em urge-to-exchange rate} is determined predominantly by present decisions to increase stocks and, thus, by the production rates in (\ref{mmodel1b})-(\ref{prodrates1}); more specifically, it is determined by the current maximum
possible rate of exchange of indivisible goods between the two sides of the market, $\min \{\Sigma_{\oplus}(p, N_S), \Delta_{\oplus}(p, N_D)\}$. Here, the urge-to-exchange rate is not determined by the net creation rates in (\ref{mmodel1b}) because these include loss rates $L$ and external flows $E$ that are likely to depend on the overall stocks $S$ and $D$, which usually are imperfectly know or unquantifiable; however, the current price $p(t)$, which determines the production rates, is known to market agents.
In addition, we represent the limiting factor of the transaction rate by a dimensionless coefficient $\kappa$ that, through physical analogy, we call coefficient of {\em friction} or {\em inverse fluidity} of the market (see more below).

To finalize the specification of our market model, we need to specify price dynamics. The net {\em net willingness to trade} or {\em frustration} at a given time is the excess in demand, $D(t)-S(t)$. While different relationships between price and other state variables can be specified, we assume for definiteness, and in agreement with the literature (see SI), that changes in log price are directly related, via a dimensionless coefficient $\mu$, to changes in net willingness to trade,
\begin{eqnarray}
\frac{d \text{[price]}}{dt} & = &  \mu \frac{d\text{[net willingness to trade]}}{dt}\text{[price]} ~, \nonumber \\
\frac{d p}{dt} & = & \mu \frac{d(D-S)}{dt} ~ p \ ,
\label{mmodel.dprice}
\end{eqnarray}
which can be solved explicitly as an exponential relationship (see~(\ref{sys:eco})).

With these specific assumptions, and in the absence of disturbances such as epidemics, our FTM dynamics are defined by the equations (represented diagrammatically in Fig.~\ref{fig:coupledmodel}B):
\begin{equation}
\begin{gathered}
\frac{dS}{dt} = \overbrace { \underbrace{N_S \sigma_0 p^{\varepsilon_S}}_{\Sigma_\oplus} - r_S S + E_S  }^\Sigma  - \overbrace {  \underbrace{\tfrac{\min \{ \Sigma_\oplus ; \Delta_\oplus \}}{\kappa}}_{\Theta} \underbrace{\min \{ \tfrac{S}{N_S} ; \tfrac{D}{N_D} \} }_q}^\Phi ~, \hfill \\
\frac{dD}{dt}  = \overbrace { \underbrace{N_D \delta_0 p^{- \varepsilon_D}}_{\Delta_\oplus} - r_D D + E_D }^\Delta  - \overbrace { \Theta q}^\Phi ~, \hfill \\
p(t) = p(0) \exp{\left\{\mu\left[D(t)-S(t) - \left(D(0)-S(0)\right)\right]\right\}} \ . \hfill
\label{sys:eco}
\end{gathered}
\end{equation}
The special case where $r_S = r_D = 0$ and $E_S = E_D = 0$ is referred to as the {\em reference market}. 
Hereafter, we use a star in superscript to denote market variables at equilibrium in the reference market in the absence of disturbances such as epidemics (e.g. $\Phi^*$ is the equilibrium trade flow). 

At a market (macroscopic) level, our model has two determinants of trade flow (see (\ref{mmodel1c})-(\ref{mmodel1e})): the coefficient of friction $\kappa$, and the average transaction stock $q$. 

We explain first the interpretation and significance of friction. An increase in $\kappa$ reduces the transaction rate (see (\ref{mmodel1e})); likewise, in physical systems friction is a macroscopic manifestation of resistance to movement. The constraints that underlie frictional trade at the microscopic (transaction) level include partner {\em search} and stock {\em delivery}; therefore, we may think that $\kappa$ has at least two components:
\begin{equation}
\kappa = \kappa_{\text{search}} + \kappa_{\text{delivery}} \ .
\label{mmodel1g}
\end{equation}
As an example, suppose $\kappa_{\text{delivery}}$ is the dominant component and 1000 goods are produced, consumed and traded per agent per year. If a minimum of $\kappa=10$ goods are delivered in a single shipment, the transaction rate is at most $\Theta=1000/10=100$ per agent per year. If, however, the nature of the goods and transportation mean it is more viable to ship a minimum of 100 goods,
then the transaction rate would be at most $\Theta=1000/100=10$ per agent per year. This illustration involves simplifications; in practice the macroscopic coefficient $\kappa$ is unlikely to associate so directly with a microscopic quantity like minimum shipment size. In our reference 
market,
$\kappa$ turns out to be the average stock exchanged per transaction when the market is at equilibrium $q^*$ (Results section \textit{Trade without stock loss: the reference market}); this suggests $\kappa$ must exceed, but can still be arbitrarily larger than the minimum shipment size.
In real markets, of money and financial products, for example, goods can be subdivided almost indefinitely, so we expect $\kappa$ to be close to 0, which translates into an almost frictionless market, in line with the high liquidity of monetary and financial markets.
In contrast, for indivisible goods such as livestock, the minimum shipment size is at least 1, so we expect $\kappa \ge 1$.
Our analyses of empirical data on livestock markets (See Material and Methods and SI) suggests that $\kappa \approx 3$ for cattle and $\kappa \approx 72$ for pigs, which, if these markets were at equilibrium and matched the assumptions of our reference market, would be the mean number of animals exchanged per transaction.

The second determinant of trade flow is the average transaction stock $q$ (see (\ref{mmodel1c}) and (\ref{mmodel1d})).
A characteristic of this market model is that the match between supply and demand is generally imperfect, as there are {\em residuals} in supply or demand stocks after each transaction (imposed by the min function in (\ref{mmodel1d}) when there is excess stock per agent). These residuals lead to a degree of transient Ôexcess frustrationÕ in market agents whose duration depends on the ÔfluidityÕ (or conversely, the ÔfrictionÕ) of the market.
As trade flow depletes both stocks $S$ and $D$ and is an observable quantity, the cumulative trade flow over a period, $T([t_0,t_f]) = \int _{t_0}^{t_f} \Phi(t) dt$, is a measurable indicator of the evolution of satisfaction (or frustration) of market agents.
Our explicit representation of market transactions driven by imperfect and frictional individual-level supply and demand, from which potentially long-lasting non-equilibrium market dynamics can emerge depending on the coefficient of friction (Results section {\em Market dynamics without shocks - effect of trade friction}), seems fundamentally different from current economic models (see SI for a comparative review of existing market models).

\subsection{Market-epidemiological (ME) model with risk aversion}

To investigate how disease epidemics and economic markets can influence each other,
we model the spread and control of infectious diseases in markets by incorporating a standard epidemiological
(E) model into our FTM model.
We call this aggregate {\em market-epidemiological} (ME) model. 
In order to compare and integrate our framework with the literature, we include in the model adaptive risk aversion (RA) behaviour by the market agents.

The market agents are the population hosting the disease-causing pathogen. 
We use notation $\mathcal{N}_A$ to represent a set of agents of arbitrary type $A$, and $N_A$ to represent their number, which can vary over time.
An agent can be a {\em strict supplier}, a {\em wholesaler} (supplier and demander), or a {\em strict demander} \citep{PautassoEtal2010JAE}. The corresponding sets of agents are $\mathcal{N}_{S\backslash{}D}$, $\mathcal{N}_{S\cap{}D}$, and $\mathcal{N}_{D\backslash{}S}$; and, the total number of agents is $N=N_{S\backslash{}D} + N_{S\cap{}D} + N_{D\backslash{}S}$.
The markets is hence composed of $N_S=N_{S\backslash{}D}+N_{S\cap{}D}$ suppliers and $N_D=N_{S\cap{}D}+N_{D\backslash{}S}$ demanders.
We use a standard 'SIRS' epidemiological model \citep{AndersonMay1991Book} where each agent (host) can be in the {\em Susceptible} ($X$), {\em Infectious} ($Y$), or {\rm Removed} ($Z$) state (the notation $XYZ$ is preferred to $SIR$ to avoid confusion with the market model notation).
Susceptible agents become infectious at rate $\Lambda(t)$ (the force of infection); infectious agents are removed at rate $\gamma$; and removed agents re-enter the market (recover) at rate $\nu$ (Fig.~\ref{fig:coupledmodel}C).
Here, ÔremovalÕ means that an infectious agent is detected and removed from the market by a regulator and becomes {\em inactive}. The infectious period ($1/\gamma$) is, therefore, the average time during which an infectious agent remains {\em active}, which is determined by the swiftness of the regulators (we assume that the biological infectious period is significantly longer than this anthropomorphic infectious period). The recovery period ($1/\nu$) is the quarantine and sanitation time during which an infectious agent remains {\em inactive}, and is generally determined by regulators and agents.
Each agent, whether a strict supplier, wholesaler, or strict demander, can be in each of the epidemiological states (Fig.~\ref{fig:coupledmodel}C).
For example, $N_{S\cap{}D}^X$ and $N_{S\cap{}D}^Y$
denote the number of susceptible and infectious wholesalers, respectively.
Therefore, there are $N_{S\cap{}D}^{XY} $ = $N_{S\cap{}D}^X + N_{S\cap{}D}^Y$ {\em active}
wholesalers and 
$N_{S}^{XY} = N_{S\cap{}D}^{XY} + N_{S\backslash{}D}^{XY}$ active suppliers.

In specifying how disease spreads in the market model, we consider the general case where the 
pathogen 
can be transmitted both through trade routes ($tr$) and non-trade routes (${\bar tr}$).
We assume that transmission through trade occurs in the direction of transactions, i.e.\ through the shipment of contaminated stock from active infected suppliers ($\mathcal{N}_{S}^{Y}$) to active non-infected demanders ($\mathcal{N}_{D}^{X}$), while transmission through non-trade routes occurs from active infected agents ($\mathcal{N}^{Y}$) to active non-infected agents ($\mathcal{N}^{X}$). We also allow for import of contaminated stock though external flow $E$ (see (\ref{mmodel1b})).
In this case, the force of infection on demanders has terms associated with transmission through trade and non-trade routes, and a risk-aversion factor:
\begin{equation}
\Lambda(t) = \left[ \Lambda_{tr}(t) + \Lambda_{\overline{tr}}(t) \right] P_{RA}(t) \ .
\label{FOI1}
\end{equation}
The term for trade routes is $\Lambda_{tr}(t)=\beta_{tr} \cdot N_S^Y/N_S^{XY}$, i.e.\ a rate of transmission via trade $\beta_{tr}$, times the probability of transacting with an active supplier that is infected, $N_S^Y/N_S^{XY}$. The rate $\beta_{tr} = P_{tr}(q) \cdot \Theta(t)/N_D^{XY}$, i.e.\ the probability $P_{tr}(q)$ of acquiring infection from the average transaction stock $q(t)$ shipped from an infected supplier during a single transaction, times the transaction rate per active demander $\Theta(t)/N_D^{XY}$. If each of the $q$ units of stock has similar and independent probability of infection $\phi$, the probability of no infection from the stock is $(1-\phi)^q$, and the probability that the demander is infected during a single transaction by at least one unit of stock is $P_{tr}(q) = 1 - (1-\phi)^q$.
Similarly, the force of infection via non-trade routes, is $\Lambda_{\overline{tr}}(t)=\beta_{\overline{tr}} \cdot N^Y/N^{XY}$, the rate of transmission $\beta_{\overline{tr}}$ per active agents via yet unspecified transmission routes, times the probability of contacting an infected active agent, $N^Y/N^{XY}$. Finally:
\begin{equation}
\begin{gathered}
\Lambda_{tr} =
\overbrace {\underbrace{ [ 1 - (1 - \phi)^q ] }_{P_{tr}(q)} \frac{\Theta }{N_D^{XY}} }^{\beta_{tr}}  \frac{N_S^Y}{N_S^{XY}} ~, \hfill \\
\Lambda_{\overline{tr}} = \beta_{\overline{tr}} ~ \frac{N^Y}{N^{XY}} \ . \hfill
\label{FOI2}
\end{gathered}
\end{equation}

Following the literature \citep{FunkEtal2010AdaptiveInterface}, we include RA in our model by allowing agents to reduce their probability of infection per transaction, or per non-trade contact, according to the level of disease detection. We assume that RA reduces the probability of infection per contact by a factor $0 \le P_{RA} \le 1$ given by:
\begin{equation}
P_{RA} = \left(1 - \frac{N^Z}{N}\right)^{\alpha} ,
\label{PRA}
\end{equation}
where $\alpha \ge 0$, and $N^Z$ is the number of inactive (detected) agents.
When $\alpha = 0$ there is no RA ($P_{RA} = 1$), while in the limit $\alpha \to \infty$ RA is maximal ($P_{RA} = 0$).

In addition to how epidemics affect the active agent population, we consider how epidemics affect their stocks.
When an infected supplier is removed (at rate $\gamma$) its share of the stock, $S/N_S^{XY}$ is also removed; hence the rate of removal of infected supplier stock is $\rho_S = \gamma N_S^{Y} (S/N_S^{XY})$.
As the generic stock-loss rates in the FTM model without epidemics (equations (\ref{mmodel1b}) and (\ref{sys:eco})) are analogous to $\rho_S$, in this paper we neglect losses other than $\rho_S$ by setting $r_S = r_D = 0$.

Finally, the ME model is defined by the dynamics of market stocks (which generalizes (\ref{sys:eco})) and agents under an epidemic (we show wholesaler equations, see the SI for other agents):
\begin{equation}
\begin{gathered}
\frac{dS}{dt} = \overbrace { \underbrace{N_S^{XY} \sigma_0 p^{\varepsilon_S}}_{\Sigma_\oplus} - \gamma \tfrac{N_S^Y}{N_S^{XY}}S + E_S  }^\Sigma  - \overbrace {  \underbrace{\tfrac{\min \{ \Sigma_\oplus ; \Delta_\oplus \}}{\kappa}}_{\Theta} \underbrace{\min \{ \tfrac{S}{N_S^{XY}} ; \tfrac{D}{N_D^{XY}} \} }_q}^\Phi ~, \hfill \\
\frac{dD}{dt} = \overbrace { \underbrace{N_D^{XY} \delta_0 p^{- \varepsilon_D}}_{\Delta_\oplus} - \gamma \tfrac{N_D^Y}{N_D^{XY}}D + E_D }^\Delta  - \overbrace { \Theta q}^\Phi ~, \hfill \\
\frac{dN_{S \cap D}^X}{dt} = \nu N_{S \cap D}^Z -  \Lambda N_{S \cap D}^X ~,  \hfill \\
\frac{dN_{S \cap D}^Y}{dt}= \Lambda N_{S \cap D}^X - \gamma N_{S \cap D}^Y ~, \hfill \\
\frac{dN_{S \cap D}^Z}{dt} = \gamma N_{S \cap D}^Y - \nu N_{S \cap D}^Z ~, \hfill
\label{eq:ecoepi}
\end{gathered}
\end{equation}
where the market price $p(t)$ is given by (\ref{mmodel.dprice}) and the forces of infection by (\ref{FOI1})-(\ref{PRA}).

We will compare epidemics in the ME model with those in simpler models.
In particular, we will consider the disease-free market equilibrium state ($q^*=\kappa$ and $\Theta^*=\Phi^*/\kappa)$) as a system without explicit market dynamics, whose trade transmission rate (in (\ref{FOI2})) is:
\begin{equation}
\beta_{tr}^*
= P_{tr}(q^*) \frac{\Theta^* }{N_D^{XY}}        
= [ 1 - (1 - \phi)^{\kappa} ] \frac{\Phi^* }{\kappa N_D^{XY}} ~.
\label{eq:beta01}
\end{equation}
In addition, when comparing with the literature on epidemiological models with RA, we will take the limit of (\ref{eq:beta01}) in a frictionless (immediate-equilibration) market ($\kappa \to 0$)
\begin{equation}
\beta_{tr}^0 = \ln \left(\frac{1}{1-\phi}\right) \frac{\Phi^* }{N_D^{XY}} ~,
\label{eq:beta02}
\end{equation}
which is identical to known functional forms of the transmission rate (see e.g. \citep{KeelingRohani2008Book}).
As in previous epidemiological models incorporating host adaptive behaviour 
driven by
health 
economics and other factors \citep{FunkEtal2009AdaptivePNAS,Fenichel2011AdaptivePNAS,MorinEtal2013AdaptiveNRM}, we allow for RA-driven reduction in transmission rate ((\ref{eq:beta02}) or, more generally, that in (\ref{FOI2})) through the RA factor (\ref{PRA}), as in (\ref{FOI1}).
However, by incorporating frictional-market dynamics, our model differs from those in the epidemiological literature, which, to our knowledge are comparable to frictionless markets.

\section{Results}
To help understanding the implications of our new theoretical framework, we study the FTM and ME models using a bottom-up approach (Table~\ref{tab:models}).
We first analyse our FTM model in the absence of epidemics (Fig.~\ref{fig:coupledmodel}B), and then explore the integrated ME model where epidemics and trade influence each other (Fig.~\ref{fig:coupledmodel}A-C). 

\begin{table}[h]
{\scriptsize
\caption{{\bf Overview of the models investigated in our study.} \hfill \label{tab:models}}
\begin{center}
\begin{tabular}{lcccccc}
 & \textbf{frictional trade} & \textbf{price dynamics} & \textbf{stock loss} & \textbf{external flows} & \textbf{epidemics} & \textbf{risk aversion} \\ 
\hline
\textbf{Market dynamics without shocks (FTM)} &  &  &  &  &  &  \\ 
\hline
Reference market & \checkm & \checkm & $\cdot$ & $\cdot$ & $\cdot$ & $\cdot$ \\ 
Trade with stock loss (special case) &  \checkm & $\cdot$ & \checkm & $\cdot$ & $\cdot$ & $\cdot$ \\ 
Trade with stock loss (general case) & \checkm & \checkm & \checkm & $\cdot$ & $\cdot$ & $\cdot$ \\ 
Trade with external flows & \checkm & \checkm & $\cdot$ & \checkm & $\cdot$ & $\cdot$ \\ 
Numerical illustrations & \checkm & \checkm & \checkm & $\cdot$ & $\cdot$ & $\cdot$ \\ 
\hline
\textbf{Market dynamics with epidemics} &  &  &  &  &  &  \\ 
\hline
Frictionless epidemiological model & $\cdot$ & $\cdot$ & $\cdot$ & $\cdot$ & \checkm & \checkm \\ 
Market-epidemiological (ME) model &  \checkm & \checkm  & \checkm  & \checkm  & \checkm  & \checkm  \\ 
\hline
\multicolumn{7}{l}{Ticks (dots respectively) represent the mechanisms included (not included respectively) in each model.}
\end{tabular}
\end{center}
}
\end{table}

\subsection{Market dynamics without shocks - effect of trade friction}
The FTM model introduces the notion of imperfect transactions with friction, bringing together differing economic models (see SI).
To assess the impacts of friction on transient and long-run trade dynamics \textit{per se}, we analyse the FTM model in the absence of epidemic shock (system~(\ref{sys:eco})).
We start by exploring the reference market where stock losses and external flows are neglected. We then analyse the cases of non-negligible stock losses ($r_S \neq 0$ or $r_D\neq 0$) and symmetric imports ($E_S=E_D=E \ge 0$). In all cases, transients and steady-states are investigated analytically and numerically.

\subsubsection{Trade without stock loss: the reference market}
In the {\em reference market} we neglect stock loss ($L_S = L_D = 0$) and external flows ($E_S = E_D =0$), and denote the equilibrium value of state variable with a \textit{star} in superscript. 
This market has an infinite number of equilibria (see SI for proof):
\begin{equation}
\begin{gathered}
(S^* = S_{min}^*; D^* \ge D_{min}^*; p=p^*) ~, \hfill \\
(S^* \ge S_{min}^*; D^* = D_{min}^*; p=p^*) ~, \hfill 
\label{eq:ref_marketeq}
\end{gathered}
\end{equation}
where $S_{min}^* =  \kappa N_S$ and 
$D_{min}^* =  \kappa N_D$ 
are the minimal stocks of supply and demand for which the market is equilibrated, and $p^*~=~(\frac{N_D \delta_0}{N_S \sigma_0})^\frac{1}{\varepsilon_S + \varepsilon_D}$ is the unique equilibrium price, obtained by solving $\Sigma_\oplus (p^*) = \Delta_\oplus (p^*)$.
The equilibria in supply and demand depend on
the initial conditions (see SI); there is hence an infinite number of unstable equilibria $(S^*,D^*)$ with a switched fixpoint: either $S^* = S_{min}^*$ or $D^* = D_{min}^*$.
Since $p^*$ is unique, trade flow at equilibrium 
is unique and given by:
\begin{equation}
\Phi^* = \Sigma_\oplus (p^*) = \Delta_\oplus (p^*) = {[{N_S}{\sigma _0}]^{\frac{{{\varepsilon _D}}}{{{\varepsilon _S} + {\varepsilon _D}}}}}{[{N_D}{\delta _0}]^{\frac{{{\varepsilon _S}}}{{{\varepsilon _S} + {\varepsilon _D}}}}} ~.
\label{eq:flows_ref_marketeq}
\end{equation}
The market converges asymptotically to reference flow ($\Phi(t) \to \Phi^*$) and price ($p(t) \to p^*$) for any initial conditions or external perturbations (see SI for proof). The famous law of supply and demand (LSD) is a particular case (see SI for proof). The LSD implies that supply should equal demand when price is equilibrated; which is a very special case of our model with unique equilibrium
$S^* = D^* = \max \{S_{min}^*, D_{min}^*\}$ (Fig. S2D).
We study this case analytically, and then return to the general model formulation (system~(\ref{sys:eco})) as there is little empirical support for the LSD \citep{McCauley2009Book}.

\subsubsection{Trade with stock loss: detailed analysis of a special case}

To study market transient behaviour, we consider initial conditions and parameter values that enable us to solve system (\ref{sys:eco}) analytically. We set $[S(t_0)=D(t_0) \ge 0;p(t_0)=p^*]$ at initial time $t_0$ and track trade flow $\Phi$ until equilibration. This set of initial conditions is compatible with the LSD since $S(t_0)=D(t_0)$ and $p(t_0)=p^*$. Once accumulated over time at rates $\Sigma_\oplus$ and $\Delta_\oplus$, supply and demand stocks are converted through trade ($\Phi$) and losses (at rates $r_S S$ and $r_D D$). For simplicity, we consider symmetrical losses ($r_S = r_D = r$). In this case, equations (\ref{sys:eco}) are symmetrical, which, with the above initial conditions ensures stocks remain symmetrical ($S(t)=D(t)$ for $t\ge 0$) and price remains constant, $p(t)=p^*$ (see SI). Therefore, system~(\ref{sys:eco}) reduces to:
\begin{equation}
\dot S = \Phi^* - ( r+ \frac{a \Phi^*}{\kappa} ) S ~,
\label{sys:eco2}
\end{equation}
where $a = \min \{ \tfrac{1}{N_S}, \tfrac{1}{N_D} \}$ is a dimensionless constant, and can be solved analytically to give (see SI):
\begin{equation}
\Phi(t)= \Phi^* \frac{\Phi^*}{ \Phi^* + \frac{r \kappa}{a}  } [1-e^{ -t (r+ \frac{a \Phi^*}{\kappa})  } ].
\label{eq:analytic_flows}
\end{equation}
In the long term ($t\to\infty$), flow diverges ($\Phi \to \infty$) if $r \le - \frac{a \Phi^*}{\kappa}$, while, if $r > - \frac{a \Phi^*}{\kappa}$ flow converges to:
\begin{equation}
\Phi_{eq} = \Phi^* \frac{\Phi^*}{ \Phi^* + \frac{r \kappa}{a}} ~,
\label{eq:analytic_eqflows}
\end{equation}
where {\em eq} in subscript denotes equilibrium values in the general case (in contrast with the special case of the reference market where equilibrium is denoted by a star in superscript).
In the more realistic case where losses are strictly positive ($r > 0$), trade flow is sub-optimal ($\Phi_{eq} < \Phi^*$), i.e., friction $\kappa$ and $r$ have an overall negative impact on equilibrium flow. If, however, the losses were negative with $r \in ] - \frac{a \Phi^*}{\kappa}, 0 [$, trade flow would be over-optimal ($\Phi_{eq} > \Phi^*$). When losses are negligible ($r = 0$), we recover the reference market flow ($\Phi_{eq} = \Phi^*$).

\subsubsection{Trade with stock loss: the general case}
We now consider the more general case of a market with asymmetric positive losses ($r_S \ge 0$ and $r_D\ge 0$), and no external flows ($E_S=E_D=0$), and consider arbitrary initial conditions $[S(0); D(0) ;p(0)]$ (see SI for derivations).
From~(\ref{sys:eco}), we deduce that equilibrium flow is always suboptimal:
\begin{equation}
\Phi_{eq} \le \min \{ \Sigma_\oplus^{eq} ,\Delta_\oplus^{eq} \} \le \Phi^* = \min \{ \Sigma_\oplus^* ,\Delta_\oplus^* \} ~.  \hfill \\
\label{eq:suboptimalflows1}
\end{equation}
Then, two cases arise. When the equilibrium limit of the average transaction stock $q_{eq}  \equiv \min \{ \tfrac{S_{eq}}{N_S}, \tfrac{D_{eq}}{N_S} \} $ is bounded by the per agent supply ($ q_{eq} = \tfrac{S_{eq}}{N_S}$), we find:
\begin{equation}
\Phi_{eq} = \Sigma_\oplus^{eq} \frac{\min \{ \Sigma_\oplus^{eq} ,\Delta_\oplus^{eq} \}}{\min \{ \Sigma_\oplus^{eq} ,\Delta_\oplus^{eq} \} + r_S \kappa N_S} ~.
\label{eq:suboptimalflows3}
\end{equation}
Conversely, when $q_{eq}$ is bounded by the per agent demand ($ q_{eq} = \tfrac{D_{eq}}{N_D}$), we have, by symmetry, that $ q_{eq}$ is given by (\ref{eq:suboptimalflows3}) but with $N_S$ and $r_S$ replaced by $N_D$ and $r_D$ respectively.
Importantly, equation (\ref{eq:suboptimalflows3}) generalises the special case of equation (\ref{eq:analytic_eqflows}) and, likewise, imply that $r_S$, $r_D$ and $\kappa$ have a negative impact on equilibrium flow. Since the dynamics with stock loss are not fully analytically tractable in the general case, we resort to extensive numerical simulations to confirm the key influence of $r_S$, $r_D$ and $\kappa$ on trade dynamics (see Global Sensitivity Analysis (GSA) of the FTM model in the SI).

\subsubsection{Trade with external flows}

To examine the impact of external flows on market dynamics, we consider, for simplicity, positive and symmetric external inflows ($E_S = E_D = E > 0$) and neglect losses ($r_S = r_D = 0$).
Symmetry ensures equilibrium price and trade rate are the same as in the reference market, $p_{eq} = p^*$ and $\Theta_{eq} = \Theta^*$, while trade flow and average stock exchanged per transaction increase to $\Phi_{eq} = \Phi^* + E$, the solution to $\frac{dS}{dt} = \frac{dD}{dt} = 0$, and $q_{eq} = \kappa \frac{\Phi^* + E}{\Phi^*} > \kappa $, as $q=\Phi/\Theta$.
Similarly to (\ref{eq:ref_marketeq}), equilibrium supply and demand have two infinite sets of possible values: either $S_{eq} = \kappa N_S \frac{\Phi^* + E}{\Phi^*}$ and $D_{eq} \ge  \kappa N_D \frac{\Phi^* + E}{\Phi^*}$, or $S_{eq} \ge \kappa N_S \frac{\Phi^* + E}{\Phi^*}$ and $D_{eq} =  \kappa N_D \frac{\Phi^* + E}{\Phi^*}$.
Hence, external flow increases trade flow through the average stock exchanged, but not the transaction rate, which is determined by number of agents, trade friction, and price.

\subsubsection{Numerical illustrations}

\begin{figure}[H]
\begin{center}
\includegraphics[width=\textwidth]{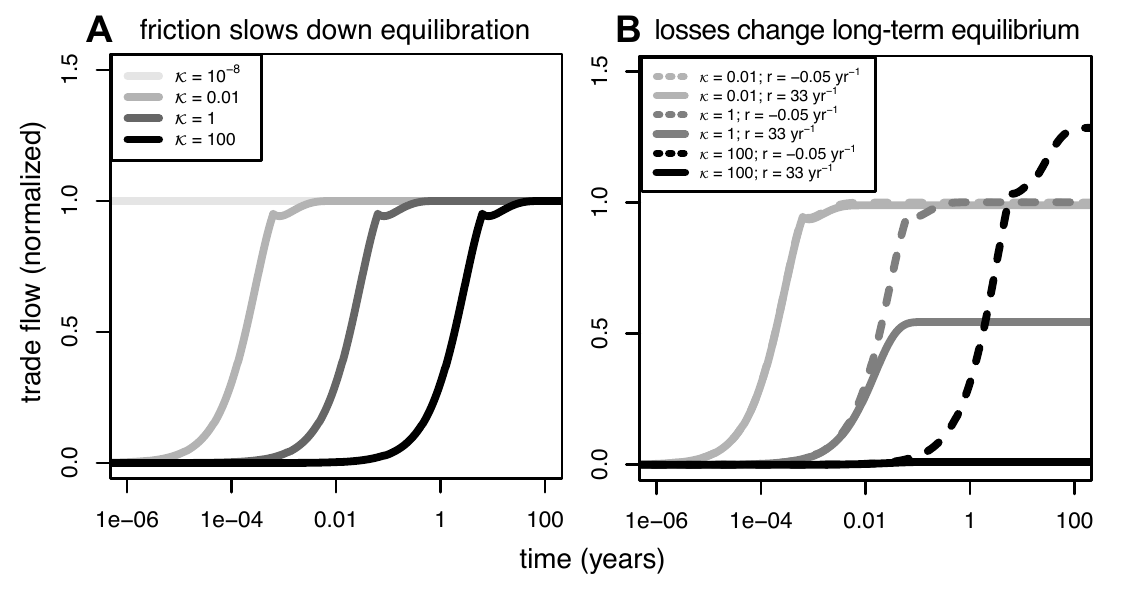}
\end{center}
\caption{{\bf The influence of frictional-trade on transient and long-term market dynamics.}
Evolution of normalized trade flow ($\frac{\Phi(t)}{\Phi^*}$) for variable levels of friction (from $\kappa = 10^{-8}$Êin light gray to $\kappa = 100$ in dark) when losses are negligible ($r=0~yr^{-1}$, \textbf{A}) or non-negligible ($r \in \{-0.5, 33 \}~yr^{-1}$ in dashed and plain lines respectively, \textbf{B}). Initial conditions are set to $[S(0)=0,D(0)=0,p(0)=1.2p^*]$. Other parameters are set from the French cattle estimates (see SI).
\label{fig:transientdynamics}}
\end{figure}

To confirm and extend the analytical insights on the impact of trade friction and losses on trade dynamics, we now explore the market model numerically. We use $[S(0)=0; D(0)=0; p(0)= 1.2~p^*]$ as initial condition, and consider symmetric losses ($r_S=r_D=r$) and no external flows ($E_S = E_D = 0)$. As the initial price is not equilibrated, trade flow equilibrates over time in a way that depends on market characteristics such as trade friction $\kappa$ and stock loss rate $r$ (Fig.~\ref{fig:transientdynamics}). Increasing $\kappa$, drastically slows down market equilibration (Fig.~\ref{fig:transientdynamics}A-B). In addition, the equilibrium flow depends on $r$. Without stock losses ($r=0$), flow converges to the reference market level $\Phi^*$ (Fig.~\ref{fig:transientdynamics}A). However, equilibrium flow is sub-optimal, $\Phi_{eq} \le \Phi^*$, when there are positive losses ($r > 0$) and friction is large enough; conversely, flow is over-optimal when losses are negative $r < 0$ and friction is large (Fig.~\ref{fig:transientdynamics}B). Scenarios with a wide range of loss rates $r_S$ and $r_D$, external flows $E_S$ and $E_D$, and initial conditions, explored via GSA, confirm these findings (see SI), which also agree with our previous analytical findings ((\ref{eq:analytic_eqflows}) and (\ref{eq:suboptimalflows3})). Note that a negative loss rate corresponds to exponential inflows of supply and demand stocks, a scenario that may appear to contrast real markets; we include it to show the general scope of the model.
Overall, we find that, in the model, friction can increase market equilibration time by several orders of magnitude, while stock loss andÊexternal stock flow can alter the long-term state of the market. We expect these parameters to play a central role in understanding trade dynamics in markets and, therefore, in the epidemiology of trade-driven diseases. 

\subsection{Market dynamics with epidemic shocks}
We now explore the dynamics of the ME model, where epidemics and trade dynamics can influence each other (Fig.~\ref{fig:coupledmodel}A-C and system (\ref{eq:ecoepi})).

\subsubsection{Relative impacts of trade friction and adaptive risk aversion}
\begin{figure}[H]
\begin{center}
\includegraphics[width=\textwidth]{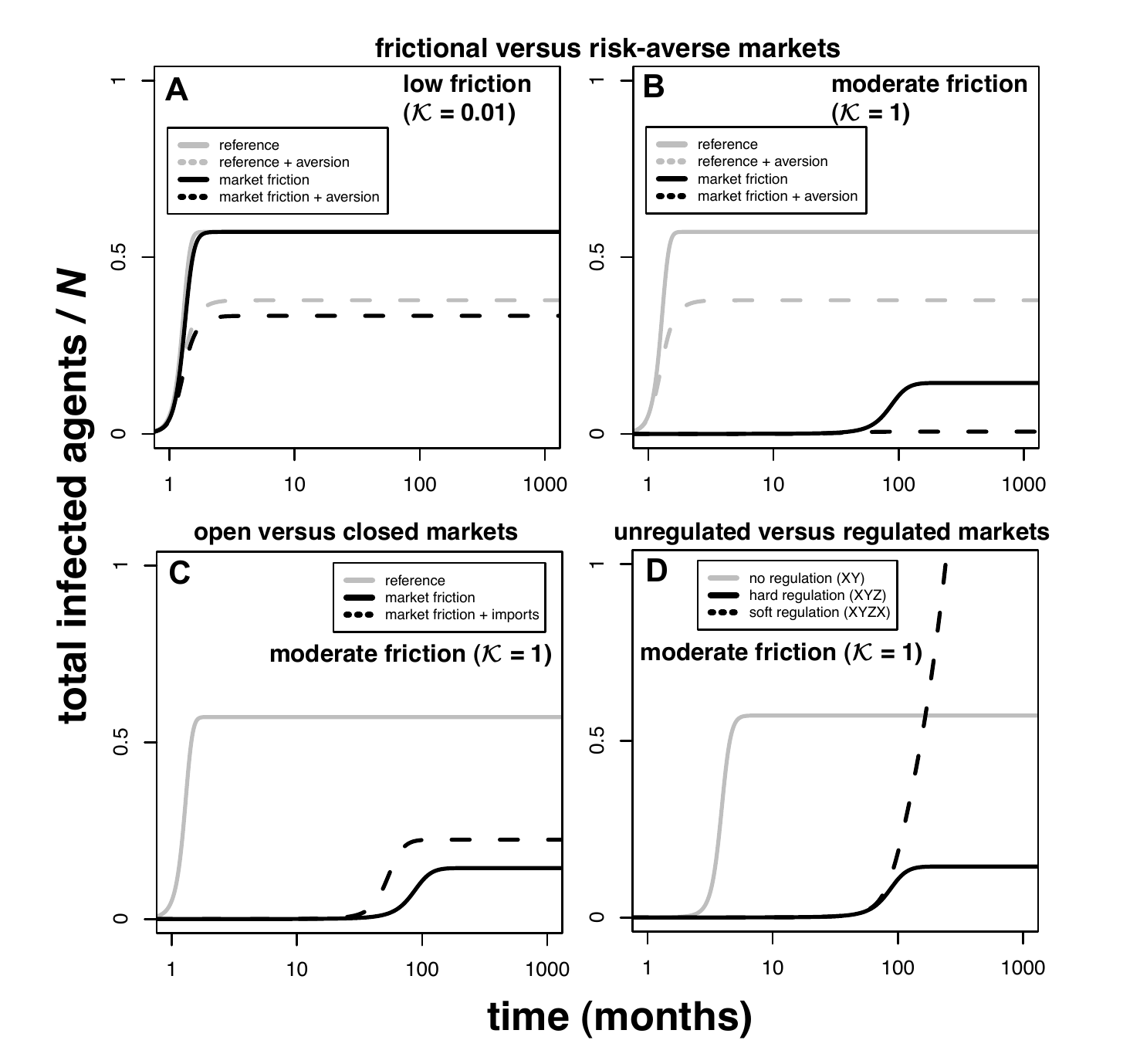}
\end{center}
\caption[Impacts of frictional-trade dynamics with risk aversion, imports and regulation on disease dynamics]
{{\bf Impacts of frictional-trade dynamics with risk aversion, imports and regulation on disease dynamics.}
Evolution of total infected agents normalized by the number of agents as function of time for various types of frictional-trade dynamics with risk aversion, imports and regulation. We parameterize our model to replicate the 2001 outbreak of Foot-and-Mouth Disease in the UK ($\phi = 0.9780$, $\gamma = 33~yr^{-1}$, $\nu = 0~yr^{-1}$; leading to an XYZ model \textbf{A-C}) assuming trade is the only path of transmission ($R_0^{tr} = R_0$). \textit{Impacts of friction versus adaptive risk aversion (\textbf{A,B})}. Friction is either low ($\kappa = 0.01$, \textbf{A}) or moderate ($\kappa = 1$, \textbf{B}). The trade transmission rate is either set to its frictionless reference value $\beta_{tr}^0$ (in grey; see also (\ref{eq:beta02})) or to its frictional-market value $\beta_{tr}$ (in black; see also (\ref{FOI2})). RA is either negligible ($\alpha = 0$, plain lines) or non-negligible ($\alpha = 8$, dashed line).  \textit{Impacts of open versus closed markets (\textbf{C})}. The level of friction is set to $\kappa = 1$ and the dashed line represents the impact of doubling trade flow by setting $E = \Phi^*$. Other lines are as in \textbf{B}. \textit{Impacts of unregulated versus regulated markets (\textbf{C})}. Regulation is inexistent ($\gamma = 0~\text{yr}^{-1}$;Êin grey), soft ($\gamma = 33~\text{yr}^{-1}$, $\nu = 1~\text{yr}^{-1}$; in dashed black) or hard ($\gamma = 33~\text{yr}^{-1}$, $\nu = 0~\text{yr}^{-1}$; in plain black). The level of friction is set to $\kappa = 1$ and the transmission rate to its frictional-market value. Initial conditions are set to start from an equilibrated market $[S(0)= \kappa N_S \tfrac{\Phi^* + E}{\Phi^*},D(0)= \kappa N_S \tfrac{\Phi^* + E}{\Phi^*},p(0)=p^*]$ and one agent per market category is initially infected (one strict supplier, one wholesalers, one strict demander). Other parameters are set from the French cattle estimates (see SI). When friction is neglected ($\kappa \to 0$), we have $R_0 = 4$, the reference FMD value. When friction is included, we have $R_0 <4$.
\label{fig:FrictionsAdaptation_FMD}}
\end{figure}

We parameterize our model to mimic the 2001 outbreak of Foot-and-Mouth Disease in the UK, and explore the impacts of frictional-trade and RA behaviour on epidemic dynamics (Fig.\ref{fig:FrictionsAdaptation_FMD}A-B).
We assume that the market is equilibrated before epidemic onset and trade is the only path of pathogen transmission.
The trade-transmission rate has either a frictionless-market value $\beta_{tr}^0$ (equation (\ref{eq:beta02}); grey in Fig.\ref{fig:FrictionsAdaptation_FMD}A-B) or its corresponding frictional-market value $\beta_{tr}$ (equation (\ref{FOI2}); black in Fig.\ref{fig:FrictionsAdaptation_FMD}A-B).
When market friction is very low (Fig.\ref{fig:FrictionsAdaptation_FMD}A), infection reaches the same endemic level with or without friction.
The inclusion of RA (dashed lines) reduces the number of infected agents, and does so similarly with or without friction (Fig.\ref{fig:FrictionsAdaptation_FMD}A). This reduction in infections in the frictionless market is in agreement with the literature (e.g. \citep{FunkEtal2010AdaptiveInterface}), and is expected as RA decreases the force of infection in response to an outbreak (\ref{FOI1}).
When market friction has a significant level (Fig.\ref{fig:FrictionsAdaptation_FMD}B), the endemic level is considerably lower in the frictional than in the frictionless market.
Again, the inclusion of RA has a similar effect on the endemic with and without friction, but this is comparatively less important than the effect of a significant increase in friction ($\kappa$ from 0.01 to 1, Fig.\ref{fig:FrictionsAdaptation_FMD}A-B).
Overall, our results suggest trade friction can suppress trade-driven disease transmission significantly, possibly more than in Fig.\ref{fig:FrictionsAdaptation_FMD}B, and to a significantly greater extent than RA behaviour, as our analyses of cattle market data suggest $\kappa = 3.4 > 1$. In addition, the combined effects of market friction and RA can lead to epidemic elimination (Fig.\ref{fig:FrictionsAdaptation_FMD}B) when trade is the only pathway of transmission.

\subsubsection{Impact of market friction level on epidemics}
\begin{figure}[H]
\begin{center}
\includegraphics[width=\textwidth]{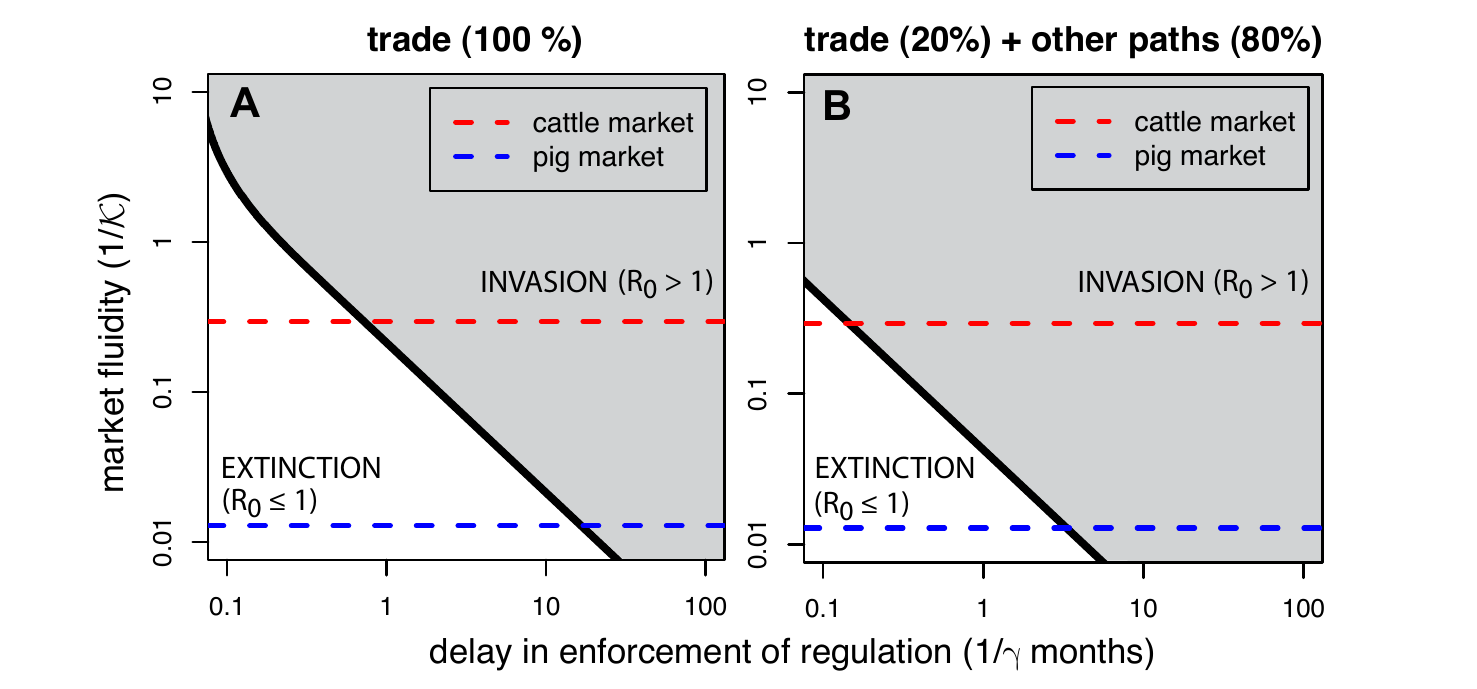}
\end{center}
\caption[Maximal delay to enforce regulation allowed to prevent invading epidemics depending on market fluidity and other paths of transmission.]
{{\bf Maximal delay to enforce regulation allowed to prevent invading epidemics depending on market fluidity and other paths of transmission.}
Probability of invasion $P_I$ for various levels of market fluidity $\frac{1}{\kappa}$, delays in enforcement of regulation $\frac{1}{\gamma}$ and intensity of other paths of transmission (\textbf{A-B}). For each panel, the black curve represent the equation $R_0=1$ that separates the $(\frac{1}{\kappa},\frac{1}{\gamma})$ space into two subspaces: the area under (above) the curve leads to an extinction (invasion) of the disease, i.e. $P_I = 0$ ($P_I = 1$). The epidemic is either only caused by trade (the black curve is given by: $R_0^{tr} (\kappa_c, \gamma_c) = 1~R_0 = 1; \textbf{A}$) or by trade and other paths of transmission (the black curve is given by: $R_0^{tr} (\kappa_c, \gamma_c) = 0.2~R_0 = 0.2 ; \textbf{B}$), where $\frac{1}{\gamma_c}$ is the maximal delay to enforce regulation corresponding to the critical amount of friction $\frac{1}{\kappa_c}$, the only value of $\frac{1}{\kappa}$ for which $R_0=1$. The dashed red (blue) line represent the estimated level of market fluidity of the French cattle market (French swine market). Other parameters are as in Fig. \ref{fig:FrictionsAdaptation_FMD}.
\label{fig:Control_R0}}
\end{figure}

Increasing trade friction reduces the severity (Fig.\ref{fig:FrictionsAdaptation_FMD}A-B and GSA of the ME model in the SI) and magnitude of the peak (Fig.S4 and GSA in the SI) of epidemics when trade routes are the only pathway of transmission.
Trade friction is also a key determinant of the epidemic threshold as assessed by the basic reproduction number $R_0$, a fundamental epidemiological summary. $R_0$ is the average number of susceptible agents infected by a single infectious agent propagated in an initially disease-free agent population \citep{AndersonMay1991Book}. In a deterministic framework, the pathogen eventually dies out if $R_0 \le 1$; while if $R_0 > 1$, the pathogen eventually invades the population.
In the general case with both trade and non-trade pathogen transmission (\ref{FOI1}), $R_0$ is given by (see SI):
\begin{equation}
\begin{gathered}
R_0 = \frac {R_0^{tr} + R_0^{\overline{tr}} + \sqrt{ (R_0^{tr} - R_0^{\overline{tr}} )^2 + 4 R_0^{tr} R_0^{\overline{tr}} \frac{N_S N_D}{N N_{S \cap D}} }}{2} ~, \hfill \\
R_0^{tr} = \frac{\beta_{tr}}{\gamma} \frac{N_{S \cap D}}{N_S} ~, \hfill \\
R_0^{\overline{tr}} = \frac{\beta_{\overline{tr}}}{\gamma}  ~, \hfill 
\label{eq:R0}
\end{gathered}
\end{equation}
where $R_0^{tr}$ ($R_0^{\overline{tr}}$ respectively) is the value of $R_0$ when trade (non-trade) provides the only pathway of pathogen transmission.
When trade is the only transmission route ($R_0^{\overline{tr}}=0$), and inserting expression (\ref{FOI2}) for $\beta_{tr}$, yields (noting that $N_D^{XY}= N_D$ in this context):
\begin{equation}
R_0 = \frac{P_{tr}(q_{eq})}{\gamma} \frac{N_{S \cap D} \Theta_{eq} }{N_D N_S} \le \frac{1}{\gamma} \frac{N_{S \cap D} \Phi^*}{\kappa N_D N_S} ~,
\label{eq:R0_tradeonly}
\end{equation}
since $P_{tr}(q_{eq}) \le 1$ and $\Theta_{eq} = \frac{\min \{ \Sigma_\oplus^{eq} ,\Delta_\oplus^{eq} \}}{\kappa} \le \frac{\Phi^*}{\kappa}$. 
Therefore, $R_0$ vanishes in the limit when the market friction is large:
\begin{equation}
\mathop {\lim }\limits_{\kappa \to \infty} R_0  = 0
\label{eq:R0_tradeonly_limit}
\end{equation}
Result (\ref{eq:R0_tradeonly_limit}) stands for \textit{any} modelling choice for $P_{tr}(q)$, including our current $P_{tr}(q) = [1-(1- \phi)^q]$. In addition to its mechanistic interpretation, this choice has the advantage of yielding a finite value for $R_0$ (equation (\ref{eq:R0_tradeonly})) in the limit of negligible friction ($\kappa\to 0$, when $\beta_{tr}\to\beta_{tr}^0$ given by (\ref{eq:beta02})):
\begin{equation}
R_0 = \frac{\ln (\frac{1}{1-\phi})}{\gamma} \frac{N_{S \cap D} \Phi^* }{N_D N_S}~,
\label{eq:R0_tradeonly_ref}
\end{equation}
which allows comparison with existing  epidemiological models that implicitly assume $\kappa=0$.
Increasing trade friction can cause decrease in $R_0$ up to the critical point where $R_0 < 1$ (Fig.~\ref{fig:Control_R0}A).
Provided that the delay in enforcing regulations is small enough, this result also stands when trade is not the main transmission pathway (Fig.~\ref{fig:Control_R0}B).
Therefore, accounting for trade friction is central to the estimation of epidemic thresholds in markets. This finding is confirmed by a GSA of $R_0$ in response to variation of its composing parameters (see GSA in SI).
We can also use our expression of $R_0$ to rank the relative risk of sustaining an epidemic for various markets. As an example, French swine markets are characterized by a larger coefficient of friction ($\kappa = 71.7$; see SI) than the French cattle market ($\kappa = 3.4$). Since $R_0^{tr}(\text{swine}) / R_0^{tr}(\text{cattle}) \approx 0.8$ for $\phi$ and $\gamma$ kept constant, trade of swine is less likely to sustain epidemics than trade of cattle. This result would appear counter-intuitive for typical epidemiological models, because trade flow is larger in swine than cattle (see SI).

\subsubsection{Open international market versus closed national market}
When we consider an open market by including imports in the model, we find that international trade can boost epidemics moderately in comparison with closed national markets (Fig. \ref{fig:FrictionsAdaptation_FMD}C).
Imports increase stocks, and thus the average stock exchanged $q$ (\ref{mmodel1d})
and probability of infection $P_{tr}(q) = 1 - (1-\phi)^q$ per transaction. However, imports do not affect the transaction rate $\Theta$. Therefore, as the force of infection involves the product $P_{tr}(q) \Theta$ and $P_{tr} \le 1$, the effect of imports on the force of infection, and thus on epidemics, is limited. Another limiting factor is the current level of control measures, i.e. removal of infected agents after a given period of infectiousness.

\subsubsection{Markets with differing level of disease regulation}
In our model, regulation to control or prevent disease spread in markets can be implemented in two ways: by removing infected agents at rate $\gamma$ or by limiting their re-introduction, after sanitation, at rate $\nu$.
We show that epidemics can be mitigated by increasing the exit rate of infected agents and/or decreasing their reentry rate 
(Fig. \ref{fig:FrictionsAdaptation_FMD}D).
The GSA suggests that increasing the removal rate $\gamma$ may be more efficient at mitigating disease than decreasing the rate of agent re-introduction $\nu$ (see SI).
In particular, epidemics can be eradicated by increasing $\gamma$ (Fig.~\ref{fig:Control_R0}A-B). Our analytical summaries of $R_0$ (equations (\ref{eq:R0})) provide estimates of the maximal delay in the enforcement of regulation $1/\gamma_{max}$ that still allows prevention of epidemics for various types of markets and combinations of transmission pathways.

\section{Discussion}
Market trade routes propagate epidemics differently from other transmission pathways due to the unique contact structure that emerges from the willingness of agents to sell and buy goods (Fig.~\ref{fig:coupledmodel}).
This willingness arises from the business motivation of economic agents and the inherent features of trade dynamics, which may differ from other epidemic-conducive human behaviour.
Trade markets involve recurrent interaction events (transactions) between suppliers and demanders (and ultimately, consumers), and in each transaction there is a variable volume of goods exchanged ($q$ in our model) that may be contaminated and lead to the spread of infection. This notion of transaction contrasts with the economic literature, particularly with the notion of match in labour economics (see SI). Transactions are extremely important from an epidemiological point of view because they form a dynamic contact structure among agents that can support disease transmission. In our model, the frequency of trade, i.e., of transactions, is limited by a coefficient $\kappa$ compounding multiple constraints referred to as trade \textit{friction}. Causes of such friction include the search for a trading partner and the logistics of stock delivery. $\kappa$ can be easily derived from available trade flow datasets such as trade of livestock (see SI for an application to exchanges of swine and cattle in France). We believe that the notion of frictional dynamical transactions is an improvement over existing market models (see SI for details) that may help to better understand the interactions between trade and disease epidemiology for many infectious diseases.

Taken together, our findings show that frictional markets are associated with a specific response to infectious diseases that contrasts with the response of other complex systems that sustain epidemics and are often assumed to be frictionless. In particular, the coefficient of friction $\kappa$ is a central parameter governing trade and disease dynamics: $\kappa$ can increase market equilibration time by several orders of magnitude (Fig. \ref{fig:transientdynamics}) and suppress trade-driven disease transmission to a significantly greater extent than RA behaviour (Fig. \ref{fig:FrictionsAdaptation_FMD}A-B). The outcomes of our model suggest that to minimize contagion in markets, $\kappa$ could be increased to allow for larger-volume, less-frequent transactions, without necessarilyÊaffecting overall trade flow, and therefore, business activity. However, increasing friction may be difficult to achieve in practice due to practical constraints underlying $\kappa$ (e.g. trucks have a limited size). The model also suggests that in markets with a given level of friction $\kappa$, international trade (Fig. \ref{fig:FrictionsAdaptation_FMD}C) and regulatory measures (Figs. \ref{fig:FrictionsAdaptation_FMD}D and \ref{fig:Control_R0}A-B) have strong, but contrasting, influence on trade-driven epidemics. In today's globalised world, a key question is how to mitigate epidemics efficiently in open international markets. The GSA of the ME model suggests that increasing the intervention rate $\gamma$ is more efficient at mitigating disease than decreasing the business re-establishment rate $\nu$, irrespective of imports (see SI).

From a cross-disciplinary perspective, our FTM model can be used to approximate market dynamics in the absence of empirical data. Our model can also be used to predict the impacts of disturbances such as epidemics on trade dynamics (Fig. S5). A {\em continuous-time ex-ante} dynamical assessment of market resiliency to the disturbance of disease dynamics is an improvement over the current health economics literature, which tends to focus on \textit{static ex-post} assessments (e.g. \citep{SolimanEtal2010CP}) or \textit{discrete-time ex-ante equilibrium-based} approaches (e.g. \citep{ZhaoEtal2006EcoEpiFMD_ARER}), which often will be at odds with the multiple and contrasting time scales encompassed by economic and epidemic processes. The use of $R_0$ could also be of importance in economics. While the dynamics of isolated-market models such as the FTM model tend to return to the original equilibrium after a disturbance, the ME model shows that contagion shocks can lead either to full recovery or long-lasting recession of the economy depending on whether $R_0$ is less or greater than 1. From a purely financial perspective, a contagion process may be e.g. caused by propagation of rumours or imitation of bad financial practices.

Our model can be extended to account for heterogeneous contact structure such as individual-based networks of trading agents (e.g. \citep{AtalayEtal2011PNAS}). An explicit trade-agent model would account for transactions involving identified pairs of agents who jointly decide to exchange goods against money. For simplicity, this pairing process is assumed here to be governed by a homogenous mixing process. However, where trade flow and the number of business partners are positively correlated at agent level, which is likely in many cases, we expect heterogeneous contact structure to boost epidemic development in comparison with homogeneous settings \citep{KampEtal2013PLoSCB}. A key open question we are currently investigating concerns the conditions under which realistic levels of friction can also mitigate epidemics propagated on heterogenous markets.

\section{Materials and Methods}
\subsection{Data analyses}
To parameterize our FTM model and estimate the range of friction encompassing real markets, we analyse trading datasets of two livestock markets: cattle (BDNI dataset) and swine (BDPorc dataset). The BDNI and BDPorc datasets are respectively managed by the French ministry in charge of Agriculture and the French professional union BDPorc. Each dataset details movements of living animals occurring in France among all economic agents involved in the supply chain, from strictly breeding farms to slaughterhouses with various categories of wholesalers in between (e.g. breeding-fattening farms, strictly fattening farms, markets, dealers). To estimate our model parameters, we extract and reconstruct from a subset of each dataset a table of individual transactions detailing: the pair of supplier-demander involved during each transaction, the date and the associated volume of goods exchanged (see SI for details on data subset included in our study). Traceability is imposed by the regulator at the scale of individual animals for cattle and batches of animals for swine. It follows transactions can be directly extracted from the BDNI dataset but not from the BDPorc dataset. Inferring detailed transactions from the BDPorc dataset requires some reconstruction hypotheses that are detailed in the SI.

\subsection{Global sensitivity analyses (GSA)}
In addition to initial conditions of state variables, our FTM and ME models include many parameters. To assess the robustness of our key analytical and numerical findings to uncertainty and variability in parameter values and initial conditions, we carry out two GSA on key economic and/or epidemiological outputs: one GSA for the FTM model; one GSA for the ME model. We rank the relative importance of all parameters and initial conditions of potential importance with an improved version of the Morris method, a GSA technique used to screen the importance of factors in high-dimensional models \citep{CampolongoEtal2007MorrisOptimalDesignEMS}. In a nutshell, the improved Morris method can discriminate the sign and overall influence of factors at a low computational cost and minor risk of error (see SI).

\section*{Acknowledgments}

Many thanks to Samuel Alizon, Hugues Beyler, Caroline Bidot, Pauline Ezanno, Yann Kervinio, Julien Fosse, Etienne Geoffroy, Natacha Go, Bhagat Lal Dutta, Franois Moutou, Emilie Moyne, Marco Pautasso and St\'{e}phane Robin for helpful comments and insights. We are grateful to the French Ministry in charge of Agriculture and to the professional union BDPorc for granting us access to the cattle and swine datasets respectively. EV, HM and MML would like to thank the French Ministries in charge of Agriculture and Environment and the INRA MIA Department for financial and operational support. JANF and CAG were funded by DEFRA and USDA.

\clearpage


\appendix

\section*{Supporting Information}

\setcounter{table}{0}
\setcounter{figure}{0}
\setcounter{equation}{0}

\renewcommand{\thetable}{S\arabic{table}}
\renewcommand{\thefigure}{S\arabic{figure}}
\renewcommand{\theequation}{S\arabic{equation}}


\section{A comparative introduction to markets and market models}

Understanding how markets emerge and operate remains a key open and controversial question in economics \citep[e.g.][]{Hahn1982StabilityHME,Katzner2010EqRED}. As trade-driven epidemics are impacted by market dynamics, we believe it is important to clarify the central concepts associated with markets. We also present market models widely used in the economic literature as a point of comparison to understand and justify the introduction of our frictional-trade market (FTM) model. 

\subsection{A tentative definition of markets}

A market is an institution where voluntary exchanges of goods and services occur between economic agents. A market can hence be described as a network composed of agents in interaction \citep{Rosenbaum2000RSE,Goyal2009}. An agent is an entity which pursues its own interests through some kind of economic optimization. Agents have generally divergent interests resolved through exchanges and price definition \citep{Guesnerie1996Book,Callon1998SR}. Examples of agents include individuals, businesses, countries or even sets of such entities \citep{Goyal2009}. Note however that the concept of market is still debated among economists and remains largely ambiguous \citep{Rosenbaum2000RSE}. Besides the literature in economics, sociologists such as \citet{Callon1998SR} argue that complex institutions such as markets cannot be reduced to networks. Operational fields such as marketing \citep{Sissors1966JM} even define markets as peculiar group of people who do not necessarily form social networks (e.g. people who consume the same brand or people of similar age). 

Two types of markets are distinguished in our study: \textit{trading markets} where goods are exchanged against money and \textit{labour markets} where unemployed workers look for vacant jobs and companies seek to hire new workers. 

\subsection{A comparative review of existing market models}

We review and compare existing market models that inspired our FTM model. As we focus on trade-driven epidemics, trading markets constitute the core of this review. We also briefly sketch search and matching labour models since we transpose their key concept of friction in our own model. 

\subsubsection{Trading market models}

Mathematically, a trading market can be formalized explicitly at the agent-level through network models \citep{Goyal2009,AtalayEtal2011PNAS} or implicitly through mass action compartmental models where agents are aggregated by categories \citep{MasColellEtal1995,SolimanEtal2010CP}. Here, markets are described as compartmental models corresponding to aggregated trade networks. We restrict ourselves to market models composed of two categories of agents with complementary interests: suppliers who wish to produce and sell goods if they receive money in compensation, and demanders who wish to buy and consume goods by providing money in exchange. We further assume that only two categories of units are exchanged against one another: goods against money. All goods are assumed to be identical. The price of one unit of money, known as the \textit{num\'{e}raire} and denoted $p_0$, is assumed to remain constant and equal to 1. $p_0$ is the standard against which the relative value $p(t)/p_0$ of one unit of good is computed. A good hence represents here a typical product exchanged on the market against $p(t)/p_0$ units of money. We write throughout $p(t)$ for $p(t)/p_0$ that we simply refer to as the price. Apparently generic concepts such as supply and demand highly depend on the market model and are hence only defined on a case-by-case basis.

\paragraph{The partial equilibrium model} ~\\

Economists essentially conceive market models working close to equilibrium rather than accounting for full transient dynamics and tipping points \citep{MayEtal2008Nature,McCauley2009Book}. Quoting the reference textbook in microeconomics \citep{MasColellEtal1995}: \textit{A characteristic feature that distinguishes economics from other scientific fields is that, for us, the equations of equilibrium constitute the center of our discipline. Other sciences, such as physics or even ecology, put comparatively more emphasis on the determination of dynamic laws of change. The reason, informally speaking, is that economists are good (or so we hope) at recognizing a state of equilibrium but are poor at predicting precisely how an economy in disequilibrium will evolve [...]. One of the difficulties in this area is the plethora of plausible disequilibrium models. Although there is a single way to be in equilibrium, there are many different ways to be in disequilibrium.}

Here, we introduce such an equilibrium-focused market model, called the partial equilibrium (PE) model, that still constitutes the basis of modern economic theory. In contrast with general equilibrium models, the term partial means that the PE model is restricted to a market where only one type of goods is produced. The PE model neglects population structure and transients to yield a simple relationship between the equilibrium values of supply, demand, and price \citep{MasColellEtal1995}. The market is said to be equilibrated at a price equalizing supply and demand. To find such an equilibrium, the model specifies how supply and demand for a given good, denoted $Q_S$ and $Q_D$ respectively, change as a function of price $p$. $Q_S(p)$ means that suppliers will supply $Q_S(p)$ goods and demanders will demand $Q_D(p)$ goods at price $p$. Supply increases in price while demand decreases in price. The slopes associated with $Q_S(p)$ and $Q_D(p)$ are governed by quantities known as price elasticity of supply and demand and denoted respectively $\varepsilon_S(Q_S,p)$ and $\varepsilon_D(Q_D,p)$. Elasticities are defined as the relative change of supply or demand in response to the relative change in price:  
\begin{equation}
\begin{gathered}
\varepsilon_S(S,p) \equiv \frac{dQ_S}{dp}\frac{p}{Q_S} ~, \hfill \\
\varepsilon_D(D,p) \equiv - \frac{dQ_D}{dp}\frac{p}{Q_D} ~. \hfill
\label{eqS:PE_elasticities}
\end{gathered}
\end{equation}
Note the negative sign in the definition of $\varepsilon_D$ so that both elasticities can only take positive values. 

If we assume that elasticities are constants, it follows by integration that $Q_S(p)$ and $Q_D(p)$ follow simple power-laws with respect to price: 
\begin{equation}
\begin{gathered}
Q_S = Q_{S_0} \left(\tfrac{p}{p_0}\right)^{\varepsilon_S} ~, \hfill \\
Q_D = Q_{D_0} \left(\tfrac{p}{p_0}\right)^{- \varepsilon_D} ~. \hfill
\label{eqS:PE_supply_demand}
\end{gathered}
\end{equation}
where $Q_{S_0}$ and $Q_{D_0}$ are the reference quantities supplied and demanded at the reference price $p = p_0 = 1$. 

From a microeconomic perspective that falls outside the scope of this review, supply and demand curves described by equations \ref{eqS:PE_supply_demand} correspond respectively to the sum of production functions when suppliers maximize their profits and the sum of consumption functions when demanders maximize their utilities \citep[see Chapter 10.C in][]{MasColellEtal1995}. 

The value of state variables at equilibrium, denoted by the sign \textit{eq}Êin subscript, is found by solving the equation $Q_S(p) = Q_D(p)$, yielding:
\begin{equation}
\begin{gathered}
Q_{S_{eq}} =  Q_{S_0} {p_{eq}}^{\varepsilon_S} ~, \hfill \\
Q_{D_{eq}}  = Q_{D_0} {p_{eq}}^{- \varepsilon_D} ~, \hfill \\
p_{eq} = \left(\tfrac{Q_{D_0}}{Q_{S_0}}\right)^{\frac{1}{\varepsilon_S+\varepsilon_D}} ~. \\
\label{eqS:PE_eq}
\end{gathered}
\end{equation}
Equations (\ref{eqS:PE_eq}) are in agreement with the so-called law of supply and demand (LSD). The latter stipulates that the equilibrium price $p_{eq}$ should increase if the reference demand $Q_{D_0}$ is permanently increased (e.g. through a growing population). Conversely, $p_{eq}$ should fall if the reference supply $Q_{S_0}$ is increased (e.g. when new businesses enter the market). Notice however that relationships (\ref{eqS:PE_eq}) tells us nothing about \textit{how quickly} and \textit{by which trading mechanisms} such an equilibrium emerge and shift when disrupted. While inapplicable to investigate the impact of market dynamics on epidemics, the PE model can still prove useful in practice. For instance, we can use a PE approach to quantify \textit{a posteriori} the economic losses induced by epidemics. The PE method consists in comparing the initial and final state of the market with respect to a past outbreak \citep[see e.g.][for examples]{SolimanEtal2010CP}.

\paragraph{Tat\^{o}nnement model of price dynamics} ~\\

The partial equilibrium (PE) model neglects temporal dynamics. Building upon the PE model, Samuelson introduced an additional relationship between state variables, referred to in the literature as the Walras-Samuelson tat\^{o}nnement (WST) model \citep[see Chapter 17.H in][]{MasColellEtal1995}. The WST model postulates that price is updated based on the difference between demand and supply, known in the literature as the Ôexcess demandÕ and denoted $Q_{E}(t) \equiv Q_{D}(t) - Q_{S}(t)$. The full WST model is given by: 
\begin{equation}
\begin{gathered}
Q_S(t) = Q_{S_0} \left(\tfrac{p(t)}{p_0}\right)^{\varepsilon_S}  ~,  \hfill \\
Q_D(t) = Q_{D_0} \left(\tfrac{p(t)}{p_0}\right)^{- \varepsilon_D}  ~,  \hfill \\
Q_E(t) = Q_D(t) - Q_S(t)  ~, \hfill \\
\frac{dp}{dt} = f(Q_{E}(t)) ~~~\text{with:}~ \frac{\partial f}{\partial Q_E} \ge 0 ~. \hfill
\label{eqS:WS_tato1}
\end{gathered}
\end{equation}

In practice, the relationship $ \frac{dp}{dt} = f(Q_{E}(t))$ is usually assumed to scale either linearly or logarithmically with $Q_{E}(t)$ \citep{MasColellEtal1995, AndersonEtal2004PriceJET}. Expressing price as function of supply and demand only, the price component of the WST model simplifies to: 
\begin{equation}
\begin{gathered}
\textit{linear assumption:}~~~ \frac{dp}{dt} = \lambda (Q_{D} - Q_{S}) ~,  \hfill \\
\textit{logarithmic assumption:}~~~ \frac{dp}{dt} = \lambda~p~(Q_{D} - Q_{S}) ~, \hfill
\label{eqS:WS_tato2}
\end{gathered}
\end{equation}
where $\lambda > 0$ is a rate parameter controlling the speed of adjustment. While the linear model is extremely simple, the logarithmic model has the advantage of always yielding positive prices when $p(t_0) > 0$.

The WST model (equations (\ref{eqS:WS_tato1})-(\ref{eqS:WS_tato2})) assumes that price will increase with respect to its current value when demand exceeds supply, and decrease when supply exceeds demands. This price adjustment process corresponds to the LSD. The LSD can hence refer both to price dynamics in a given market with a unique equilibrium (equations (\ref{eqS:WS_tato1})-(\ref{eqS:WS_tato2}) of the WST model) or to the evolution to new equilibrium prices in response to external processes affecting the market (equations (\ref{eqS:PE_eq}) of the PE model).

While price is now a dynamical state variable, the total number of goods $Q(p)$ actually traded out of $Q_{S}(p)$ and $Q_{D}(p)$ is not specified by the WST model. Instead, the WST model assumes that no goods can actually be exchanged on the market prior equilibration, i.e. prior $Q_{S} = Q_{D}$. The weakness of this assumption is clearly stated and criticized in the economic literature and the conditions needed for the economic equilibrium $[Q_{S_{eq}},Q_{D_{eq}},p_{eq}]$ to be well-behaved seem very complicated \citep{Hahn1982StabilityHME,MasColellEtal1995,AndersonEtal2004PriceJET,McCauley2009Book,Kitti2010PriceJEDC}. We can conclude that even apparently simple concepts such as supply and demand have no clear real-world interpretation in simple trading market models such as the PE and the WST models.

\paragraph{Disequilibrium market models} ~\\

In practice, trade also occurs when the market is not equilibrated \citep{Hahn1982StabilityHME}. Although unsatisfactory from a conceptual perspective \citep{MasColellEtal1995,Katzner2010EqRED} and lacking empirical support \citep[e.g.][]{Hahn1982StabilityHME,AndersonEtal2004PriceJET}, disequilibrium (DE) market models were introduced to approximate the functioning of markets out of equilibrium \citep[see the seminal contributions of][]{HahnNegishi1962Econometrica,FairJaffee1972Econometrica}. As for the WST model, a common assumption in DE models is that price dynamics are governed by the excess demand (see e.g. equations (\ref{eqS:WS_tato2})). However, in contrast with the WST model, goods can be exchanged even when the market is not equilibrated. 

Two types of contributions to DE dynamics should be distinguished due to their contrasted objectives and assumptions: theoretically-motivated and empirically-motivated studies. One the one hand, theoretical economists developed complex DE models to demonstrate the stability of general equilibrium (GE) market models. A GE model is a generalization of the PE model to account for multiple markets in interaction. While insightful to investigate the convergence conditions in GE models where exchanges of goods are only based on initial endowments \citep{HahnNegishi1962Econometrica}, theoretical approaches proved poorly applicable to GE models with production of new goods, the ones that matter for epidemics \citep[see][for detailed explanations by a key author in the field]{Hahn1982StabilityHME}. One the other hand, applied economists developed simpler DE market models to estimate key economic parameters from empirical time series based on less restrictive assumptions. We hence only present in details the core model underlying empirically-motivated DE models: the DE model by Fair and Jaffee (FJ) \citep{FairJaffee1972Econometrica,Quandt1988Book,LeeEtal2011QF}. The most parsimonious FJ model is given by \citep[Chapter 2 of][]{Quandt1988Book}:
\begin{equation}
\begin{gathered}
Q_{S}(\tau _k) = w_{0,S}~W_S (\tau _k) + w_{1,S}~p(\tau _k) + \chi_S(\tau _k) ~, \hfill \\
Q_{D}(\tau _k) = w_{0,D}W_D (\tau _k) - w_{1,D}p(\tau _k) + \chi_D(\tau _k) ~, \hfill \\
Q({\tau_k}) = \min \{ {Q_S}({\tau _k}),{Q_D}({\tau _k}) \} ~, \hfill \\
p({\tau_k}) = p({\tau_{k-1}})  + \lambda \left(Q_{D}(\tau _k) - Q_{S}(\tau _k)\right) + \chi_p(\tau _k) ~, \hfill
\label{eqS:FJ1}
\end{gathered}
\end{equation} 
where $\tau_k$ is the $k^{th}$ period of time and $Q({\tau_k})$,  $Q_S({\tau_k})$ and $Q_D({\tau_k})$ are the total quantities traded, supplied, demanded at price $p(\tau _k)$ during period $\tau_k$ respectively. $W_S(\tau _k)$ and $W_D(\tau _k)$ are vectors of  variables other than price that influence $Q_S(\tau _k)$ and $Q_D(\tau _k)$ respectively. $w_{0,S}$, $w_{1,S}$, $w_{0,D}$, $w_{1,D}$, and $\lambda$ are parameters to estimate and $\chi_S(\tau _k)$, $\chi_D(\tau _k)$ and $\chi_p(\tau _k)$ are error terms.

Interestingly, the FJ model emphasises that while supply and demand are non-observable, quantities traded and prices can be measured. Though some state variables are non-observable, parameter estimation is still possible because of a key dependency introduced between the state variables: at each time period, the total quantity traded (observed variable) is given by the {\em minimum} of supply and demand (non-observable variables). Thanks to the $\min$ function, supply and demand can now be interpreted as \textit{willingnesses} to supply and demand goods. While far more realistic than the WST model to understand trade dynamics, the FJ model suffers from a major drawback: the residuals in willingnesses to trade, i.e. the leftovers implied by ${Q_S} \ne {Q_D}$, are not re-injected in the evolution of supply and demand. The functional shapes of supply and demand should have been modified to account for disequilibrium in trade. In addition, the contact process underlying disequilibrium trade is not specified, which makes the model inapplicable to epidemiological settings. 

\subsubsection{Labour market models: key insights from search and matching theory}

\paragraph{The importance of friction: the labour market model of Diamond, Mortensen and Pissarides} ~\\

While inapplicable as such to model disease dynamics on trading markets, labour market models provide inspiring concepts to grasp the contact process underlying market dynamics. Following \citet{Nobel2010SearchFrictions}, we focus on the reference labour market model of Diamond, Mortensen and Pissarides (DMP). The DMP model assumes a labour market operating on continuous time with a fixed number of labour force participants $L$. Let $u$ denote the fraction of unemployed workers. It follows $uL$ workers are unemployed and $(1-u)L$ workers are employed. Let $v$ denote the fraction of $L$ that corresponds to vacant positions, so that $vL$ is the number of vacant positions while $(1-v)L$ is the number of non-vacant positions. In the general case, $u \ne v$. Here, we restrict our analysis of the DMP model to unemployment dynamics, i.e. we only specify a model for $\frac{du}{dt}(u,v)$. This restriction is sufficient to highlight the key contributions of the DMP model that we transpose in the FTM model. 

In the DMP model, jobs are destroyed at per capita rate $\gamma$. Unemployed workers find a job at per capita rate $\nu$. In contrast with $\gamma$ that is assumed constant, $\nu$ depends on $uL$ and $vL$. To fully specify $\nu$, the DMP model introduces a matching function $\Psi=m(uL,vL,\kappa)$ where $\Psi$ represents the rate of successful job matches at the population level that result from the joint-search efforts of the $uL$ unemployed workers to find jobs and of companies to fill their $vL$ vacancies. $\kappa$ is a parameter controlling the intensity of search and matching friction in the market. As expected intuitively, $m$ is chosen so that $\Psi$ increases with $uL$ and $vL$ and decreases with $\kappa$. By construction, $\nu$ is  given by $\nu = \frac{\Psi}{uL}$. The dynamics of $u$ are hence described by the differential equation: 
\begin{equation}
\frac{du}{dt} =  \frac{m(uL,vL,\kappa)}{uL} uL - \gamma (1-u) L
\end{equation}

We now assume that unemployment reaches a steady-state. Setting $\frac{du}{dt} = 0$ yields a key relationship between $u_{eq}$ and $v_{eq}$, known in the labour economics as the Beveridge curve:
\begin{equation}
u_{eq} = \frac{\gamma}{\gamma + \frac{m\left(u_{eq}L,v_{eq}L,\kappa\right)}{u_{eq}L}}
\end{equation}

Based on the properties of the matching function $m$, the Beveridge curve implies that equilibrium unemployment $u_{eq}$ and vacancies $v_{eq}$ are negatively related. If we set the ratio $v_{eq}/u_{eq}$ constant, increasing the level of friction $\kappa$ will increase both $u_{eq}$ and $v_{eq}$. In other words, by decreasing the rate of successful matches $\Psi=m(uL,vL,\kappa)$ between agents, friction makes the labour market worse. 

\paragraph{Understanding the causes of friction: the mechanistic determinants of the matching function} ~\\

While providing us with intuitions on the effects of imperfect matches on labour markets, the DMP model neither specifies the origins of $\kappa$ nor the shape of the matching function $m$. $m$ and $\kappa$ are essentially black boxes \citep{Pissarides2001IESBS}. Recent studies managed to uncover the mechanisms underlying $m$ and $\kappa$ \citep{Lagos2000JPE,BurdettEtal2001JPE,Stevens2007IER}, i.e. the local contact structure underlying friction in labour markets. Such studies assume time is discrete so that $M = m(uL,vL,\kappa)$ now corresponds to a number of successful matches and is not a rate any more. Here, we only present the taxicab model of \citet{Lagos2000JPE} as its matching function $M = m(uL,vL,\kappa)$ closely resembles our transaction rate $\Theta$. 

The taxicab model explicitly describes the labour market with an agent-based model based on taxicabs transporting passengers to their desired location on a spatial grid. In this model, $uL$ represents the number of people who want to exit their current location to move to another location. $vL$ represents the number of taxicabs. A passenger can only move on the grid if a free taxicab is present on its current location. Each taxicab can only transport one passenger at a time. When inside a taxicab, a passenger decides it destination. Free taxicabs decide where they want to pick up passengers. $M$ denotes the total number of successful cab-passenger meetings occurring over the grid. \citet{Lagos2000JPE} shows that at the population-level, $M$ can be simply expressed as: 
\begin{equation}
M = \min \{ uL, \frac{1}{\kappa} vL \} \le \min \{ uL, vL \}. 
\end{equation}
where $\kappa \ge 1$ increases with the heterogeneity in preferences expressed by taxicabs and passengers for specific spatial locations.

The case $\kappa = 1$ corresponds to the minimal amount of friction in the taxicab model where $M = \min \{ uL, vL \}$. In contrast, the case $\kappa \to \infty$ corresponds to the maximal amount of friction where $M = 0$. The case where $\kappa \le 1$ is not allowed in the taxicab model because taxicabs and people cannot be divided in smaller units. In contrast with the transaction rate $\Theta$ of the FTM model, notice that friction only affects $vL$ and not $uL$. This stems from the asymmetric properties of taxicabs (that can move freely and decide where they want to pick new passengers) and passengers (who can decide where they want to go but cannot move by themselves). In the FTM model, both suppliers and demanders can move freely and search for each other, so we assume friction applies both to suppliers and demanders, in a symmetric way.

\subsection{Correspondence between the FTM model and existing market models}

While developing new concepts such as imperfect transactions with friction, the FTM model 
builds upon the four existing market models reviewed above (PE, WST, FJ and DMP). 
We highlight here the key differences and commonalities between the five models. In the following, supply, demand and the total number of goods traded are denoted respectively $T$, $S$ and $D$ when they are generated based on continuous processes, while they are denoted $Q$, $Q_S$ and $Q_D$ in the discrete case. 

\subsubsection{Correspondence with the partial equilibrium (PE) model}

The FTM model assumes that supply and demand are respectively created at rates $\Sigma_{\oplus}(p, N_S)$ and $\Delta_{\oplus}(p, N_S)$ given by:

\begin{equation}
\begin{gathered}
\Sigma_{\oplus}(p, N_S) = N_S~\sigma_0~p^{\varepsilon_S} ~, \hfill \\
\Delta_{\oplus}(p, N_D) = N_D~\delta_0~p^{-\varepsilon_D} ~. \hfill
\end{gathered}
\end{equation}
where elasticities are defined the same way as the PE model by replacing stocks with rates (see equations (\ref{eqS:PE_elasticities}) - (\ref{eqS:PE_eq})). As it defines creation rates rather than stocks of supply and demand, the FTM model is a generalization of the PE model to dynamical settings.

\subsubsection{Correspondence with the Walras-Samuelson tat\^{o}nnement model (WST) of price dynamics}

The FTM model assumes that variations in price are directly related, via a dimensionless coefficient $\mu$, to changes in net willingness to transact:
\begin{equation}
\frac{d p}{dt} =  \mu \frac{d(D-S)}{dt} p =   \mu (\Delta - \Sigma) p ~.
\label{Smmodel.dprice}
\end{equation}

As the WST model requires that trade only occurs at equilibrium, assuming that price updating is based on the excess demand stock ($Q_D - Q_S$; see equations (\ref{eqS:WS_tato2})) is identical to assuming that price updating is based on the excess demand creation rate ($\Delta_{\oplus} - \Sigma_{\oplus}$). In other words, the WST model implicitly assumes that stocks of supply ($Q_S$) and demand ($Q_D$) are identical to our creation rates of supply ($\Sigma_{\oplus}$) and demand ($\Delta_{\oplus}$) as far as price dynamics are concerned. The two models are hence fully equivalent when losses and external flows are neglected in the FTM model, i.e. when $\Sigma = \Sigma_{\oplus}$ and $\Delta = \Delta_{\oplus}$. 

\subsubsection{Correspondence with the disequilibrium market model by Fair and Jaffee (FJ)}

Our definition of the average transaction stock exchanged from a supplier to a demander ($q$ in the main text) is directly inspired by the FJ model. Here we show that the two market approaches are not equivalent: the FTM model involves less restrictive assumptions on how trade is carried out compared with the FJ model. Our market model is more general than the FJ model as it accounts for various levels of friction. 

\paragraph{Implications of the hypotheses of Fair and Jaffee} ~\\

The key contribution of the discrete-time FJ model (see equations (\ref{eqS:FJ1})) essentially boils down to: 
\begin{equation}
Q({\tau_k}) = \min \{ {Q_S}({\tau _k}),{Q_D}({\tau _k}) \} ~.
\label{eqS:FJ_essential}
\end{equation}

We extend the FJ model to continuous time. To give a fair comparison of the models, we assume that supply and demand are created at net rates $\Sigma(t)$ and $\Delta(t)$, the same as in the FTM model. We denote by $\PhiFJ(t)$ the trade flow generated by the FJ model. Our objective is to characterize the value of $\PhiFJ(t)$ or its integral $\TFJ(t)$ as function of parameters and functions common to both models. 

Generalized to continuous time, equation (\ref{eqS:FJ_essential}) implies:
\begin{equation}
\begin{gathered}
\underbrace {\int\limits_{t - dt}^t {\PhiFJ (u)du} }_{\scriptstyle{\text{Total number of goods traded from $t - dt$ to $t$}}} = \min \Bigg[{\underbrace {S({t_0}) + \int\limits_{{t_0}}^{t - dt} {\Sigma (u)du}  - \int\limits_{{t_0}}^{t - dt} {\Phi_{FJ} (u)du} }_{\text{Residual supply at $t - dt$}}} + \underbrace {\int\limits_{t - dt}^t {\Sigma (u)du} }_{\scriptstyle{\text{Creation of supply from $t - dt$ to $t$}}}; \\
{\underbrace {D({t_0}) + \int\limits_{{t_0}}^{t - dt} {\Delta (u)du}  - \int\limits_{{t_0}}^{t - dt} {\Phi_{FJ} (u)du} }_{{\text{Residual demand at $t - dt$}}}} + {\underbrace {\int\limits_{t - dt}^t {\Delta (u)du} }_{\scriptstyle{\text{Creation of demand from $t - dt$ to $t$}}}}\Bigg] ~.
\label{eqS:FJ2}
\end{gathered}
\end{equation}

After simplifications, equation (\ref{eqS:FJ2}) leads to:
\begin{equation}
\TFJ(t) = T({t_0}) + \min \left[S({t_0}) + \int\limits_{{t_0}}^t {\Sigma (u)du}~;~D({t_0}) + \int\limits_{{t_0}}^t {\Delta (u)du} \right] ~,
\label{eqS:FJ3}
\end{equation}
where $\TFJ(t)$ is the total number of goods traded from time $t_0$ to time $t$ under the FJ model. 

\paragraph{Implications of the hypotheses of the FTM model} ~\\

The generic version of the FTM model reads:
\begin{equation}
\begin{gathered}
\frac{dS}{dt} =  \Sigma(t) - \Phi(t), \hfill \\
\frac{dD}{dt} =  \Delta(t) - \Phi(t)~, \hfill \\
\frac{dT}{dt} =  \Phi(t)~. \hfill
\label{sys:generic_market}
\end{gathered}
\end{equation}

This directly implies:
\begin{equation}
\begin{gathered}
\int\limits_{{t_0}}^t {\Phi (u)du}  = S({t_0}) + \int\limits_{{t_0}}^t {\Sigma (u)du}  - S(t) ~,\\
\int\limits_{{t_0}}^t {\Phi (u)du}  = D({t_0}) + \int\limits_{{t_0}}^t {\Delta (u)du}  - D(t) ~.
\label{eqS:GenericEcoDynamics2}
\end{gathered}
\end{equation}

In particular, equalities (\ref{eqS:GenericEcoDynamics2}) lead to: 
\begin{equation}
\begin{gathered}
T(t) = T({t_0}) + \min \left\{ S({t_0}) + \int\limits_{{t_0}}^t {\Sigma (u)du}  - S(t)~;~D({t_0}) + \int\limits_{{t_0}}^t {\Delta (u)du}  - D(t) \right\} ~,
\label{eqS:SimplestEcoDynamics3}
\end{gathered}
\end{equation}
where $T(t)$ is the total number of goods traded from time $t_0$ to time $t$ under the FTM model. 

\paragraph{Comparison of the two models} ~\\

We immediately notice:
\begin{equation}
T(t) \le \TFJ(t)~~\forall t ~.
\label{eqS:FM_v_FJ1}
\end{equation}

The two quantities are equal at market equilibrium when the residual supply and demand stocks are equal and vanish. The FJ model is hence a special case of our reference market model where: 1) we assume that the LSD is respected (so that $S^* = D^* = \max \{ \kappa N_S, \kappa N_D \}$; see the analysis of the integral price model with $h=0$ in the third section); 2) we neglect friction (so that $\kappa \to 0$). Taken together, special cases 1) and 2) imply $S^* = D^* = 0$, which leads to $T \to \TFJ$. 

Intuitively, the two approaches are equivalent only when the market fully clears, i.e. when each good produced is instantly shipped and consumed. Our approach allows for the more general case where the LSD is not necessarily respected and where the time needed for a transaction is not negligible any more. In other words, accounting for imperfect transactions with friction implies that supply and demand stocks can never vanish, even at market equilibrium.

\subsubsection{Correspondence with the labour market model of Diamond, Mortensen and Pissarides (DMP)}

The DMP model was designed to approximate labour markets while the FTM model was designed to approximate trading markets. It follows the two models differ in nature, notably as regard to the type of exchanges and friction involved. We also show that their conclusions are different in scope and highlight what we believe is our key economic contribution in contrast with existing market models: the notion of imperfect trade transaction with friction. 
 
Both models involve two market sides and interaction events between them that depend on friction. The DMP model introduces a matching function $\Psi(uL,vL,\kDMP)$ that closely resembles our transaction rate $\Theta(N_S,N_D,\sigma_0,\delta_0,p,\kappa)$, where $\kDMP$ and $\kappa$ are the coefficients of friction associated with the DMP model and the FTM model respectively. We recall that $\Psi$ represents the rate of successful job matches at the population level that result from the joint-search efforts of the $uL$ unemployed workers to find jobs and of companies to fill their $vL$ vacancies. $\Theta$ represents the transaction rate between the $N_S$ suppliers and the $N_D$ demanders. 

The types of interaction between pairs of agents involved in labour and trading markets are not equivalent. Labour markets involve on-off matches between workers and employers: either an encounter is successful and a vacancy is filled by a worker or the pairing is rejected. Trading markets involve recurrent transactions between suppliers and demanders, and each transaction is associated with a variable number of goods exchanged $q$. More precisely: 
\begin{itemize}
\item A \textit{job match} is stable for a certain period of time (a new employee is likely to keep her/his job for a few days at the very least), is usually unrepeated for a given pair of agents (a person who gets fired by a company is unlikely to get hired again in the same company) and corresponds to a binary process (a person is either employed or unemployed).
\item In contrast \textit{a trade transaction} is a transient process (once a trade is realized, the physical interaction stops), is generally repeated in time for a given pair of agents (a supplier and demander who get on well are likely to interact multiple times per time period) and corresponds to a continuous process (a variable number of goods is exchanged from a given supplier to a given demander).
\end{itemize}
The types of friction hence spanned by $\kDMP$ and $\kappa$ differ: while $\kDMP$ aggregates job search and matching friction, $\kappa$ represents trade friction (= partner search friction + stock delivery friction). The ranges encompassed by $\kappa_{DMP}$ and $\kappa$ are also different: though $\kDMP$ must satisfy $\kDMP \ge 1$ because you cannot exchange less than one "unit" of labour, $\kappa$ satisfies $\kappa \ge 0$ because you can have more transaction events per time unit than production/consumption events per time unit. Note that the average transaction stock $q \equiv \min \{ \frac{S}{N_S}, \frac{D}{N_D} \}$ of the FTM model is associated with another type of dynamical friction: stock-matching friction, since the per capita supply does not necessarily match the per capita demand. This type of friction is not accounted for by the DMP model. 

The models finally differ in their conclusions. The DMP model shows that friction in job matches is essential to explain the existence of persistent unemployment in labour markets. Our model not only shows that friction in transactions explains the persistence of residual supply and demand stocks in trading markets, but friction can also modify the joint-dynamics of markets and epidemics (see results in the main text). This point stands whether epidemics are included or not, and suggests that the FTM model can also prove useful from a purely economic point of view. Generally speaking, we believe the notion of dynamical transactions with friction is an improvement over existing trading market models.

\section{Model parametrization and estimation of friction from French cattle and swine livestock exchange data}

\subsection{Presentation of the datasets}

To parameterize the FTM model and estimate the range of friction encompassing real markets, we analyze trading datasets of two livestock markets: cattle (BDNI dataset) and swine (BDPorc dataset). The BDNI and BDPorc datasets are respectively managed by the French ministry in charge of Agriculture (\url{http://agriculture.gouv.fr/identification-et-tracabilite}) and the French professional union BDPorc (\url{http://www.bdporc.com/}). Each dataset details movements of living animals occurring in France among all economic agents involved in the supply chain, from strictly breeding farms to slaughterhouses with various categories of wholesalers in between (e.g. breeding-fattening farms, strictly fattening farms, markets, dealers). Imports and exports are also included. Traceability is imposed by the regulator at the scale of individual animals for cattle and batches of animals for swine. Declaration of cattle and swine movements is compulsory in France since January 1999 and July 2009 respectively. 

\subsection{Extraction of individual transactions from the datasets}

To estimate the parameters of the FTM model, we need to extract for each dataset a table of individual transactions detailing: the pair of supplier-demander involved during each transaction, the date and the associated volume of goods exchanged. We only extract such detailed transactions for a subset of the datasets. We focus on cattle data for civil year 2009 and pig data for civil year 2010. Following the literature on networks of livestock exchanges in France \citep{RautureauEtal2011TED, RautureauEtal2012PigsAnimal}, we neglect imports, exports and movements of animals to slaughterhouses and rendering plants. Imports and exports are already accounted for in the FTM model as free parameters ($E_S$ for supply and $E_D$ for demand, with $E>0$ for imports and $E<0$ for exports). While concentrating a large part of the total number of goods exchanged (about 50\% for cattle \citep{RautureauEtal2011TED} and 80 \% for pigs \citep{RautureauEtal2012PigsAnimal}), slaughterhouses and rendering plants are positioned at the very bottom of the supply chain. They are hence probably associated with a negligible risk of infection \citep{RautureauEtal2012PigsAnimal}. Note that transactions involving topological dead-ends such as slaughterhouses could be easily included in the FTM model as we distinguish wholesalers from strict demanders. However, as the FTM model describes mass-action interactions among identical agents within their respective hierarchical categories (strict suppliers, wholesalers and strict demanders), we would largely overestimate the infectious risk associated with trade by over-estimating the fractions of transactions and average transaction stocks that contribute to infection. In other words, even if such dead-end movements are important for the market, we neglect them in favor of more epidemiological realism.

\subsubsection{Extraction of transactions from the BDNI dataset (cattle)}

Since the scale of traceability is known at the animal level for cattle, transactions can be directly extracted from the BDNI dataset. 

\subsubsection{Extraction of transactions from the BDPorc dataset (swine)}

Swine traceability is only known at the scale of batches of pigs transported by the same truck during a round of transportation. A round is a movement of swine involving the same truck that starts a journey empty, loads pigs from one or more holdings and finishes the journey empty after successful delivery to one or more holdings. Loading and delivery events may occur several times and in any order during a given round. Inferring detailed transactions from the BDPorc dataset hence requires some reconstruction hypotheses that are detailed here. 

The inference problem is the following: we know how much each agent is contributing, with whom and when it is interacting, but we do not know which pigs precisely reach a given destination. In other words, we need to infer the quantities of pigs exchanged $q_{i,j}^r$ from agent $i^r$ to agent $j^r$ during round $r$. The problem is identical to the reconstruction of sex acts on weighted sexual contact networks, where we know the partnerships, how much sex acts each individual has, but we do not know how many sex acts are associated with each existing partnership. We hence use the same methodology of reconstruction as \citet{Moslonka-LefebvreEtal2012JTB}. In the following, the letter $i$ denotes individual suppliers and the letter $j$ individual demanders. $\widehat{x}$ means that quantity $x$ is a raw data that will be subsequently modified by a reconstruction treatment. 

Data reconstruction is applied to each round $r$ associated with \textit{at least} $n^r$ suppliers (since we remove some suppliers such as importers) and \textit{at least} $m^r$ demanders (since we remove some demanders such as slaughterhouses). For each round $r$, we know that: $i^r$ is a supplier supplying $\widehat{s_i^r}$ pigs to at least $m^r$ demanders, $j$ is a demander receiving $\widehat{d_j^r}$ pigs from at least $n^r$ sellers and that $i^r$ may be connected to $j^r$ because the same truck visited these two agents during the same round. However, we do not know for sure if $i^r$ actually shipped goods to $j^r$ as more than two agents can be involved per round, i.e. we do not know the value of $q_{i,j}^r$. The reconstruction of $q_{i,j}^r$ operates in four steps:

\begin{itemize}

\item \textbf{Step 1: Calculation of the maximal number of goods exchanged.} To account for cut-offs e.g. induced by neglected slaughterhouses and imports, we calculate the total number of goods $q^r$ that is exchanged during round $r$ among the $A^r$ \textit{non-neglected} agents. $A^r$ is made of $n^r$ suppliers and $m^r$ demanders. The $n^r$ suppliers supply $\widehat{s^r} = \sum_{i=1}^{n^r}  \widehat{s_i^r}$ goods. The $m^r$ demanders receive $\widehat{d^r} = \sum_{j=1}^{m^r}  \widehat{d_j^r}$ goods. We assume $q^r = \min \{ \widehat{s^r} ; \widehat{d^r} \} $. 

\item \textbf{Step 2: Normalization of the number of goods exchanged at the round level.} The quantities $\widehat{s_i^r}$ and $\widehat{d_j^r}$ are normalized so that the total quantity of pigs supplied matches the total quantity of pigs demanded within a round: 
\begin{equation*}
\begin{gathered}
s_i^r \leftarrow \widehat{s_i^r} \frac{q^r}{\widehat{s^r}}~,  \hfill \\
d_j^r \leftarrow \widehat{d_j^r} \frac{q^r}{\widehat{d^r}}~,  \hfill
\end{gathered}
\end{equation*}
where the notation $x \leftarrow y$ means that we replace the value $x$ by the value $y$. After normalization, we have by construction: $\sum_i  s_i^r = \sum_j  d_j^r = q^r$. 

\item \textbf{Step 3: Partitioning of potential goods exchanged.} Since traceability is only known at the round-level, we do not know the precise destination of each pig. We hence need to specify a preference function describing for each agent how much goods she/he wishes to trade with each of its potential partners. Let the asymmetric preferences functions $[i^r \to j^r]^\text{sell}$ and $[j^r \to i^r]^\text{buy}$ represent the total amount of goods that $i^r$ wishes to sell to $j^r$ and $j^r$ buy from $i^r$ during round $r$ respectively. Here we assume that agents favour preferentially their biggest potential partners with a proportional model:
\begin{equation*}
\begin{gathered}
~ [i^r \to j^r]^\text{sell}  = s_i^r \frac{d_j^r}{\sum_j  d_j^r} ~, \hfill \\
~ [j^r \to i^r]^\text{buy} = d_j^r \frac{s_i^r}{\sum_i  s_i^r} ~. \hfill
\end{gathered}
\end{equation*}

\item \textbf{Step 4: Reconstruction of actual goods exchanged.} For each pair of trading agents ($i^r$,$j^r$), the actual number of goods exchanged $q_{i,j}^r$ is assumed to be the minimum of their respective preferences: 
\begin{equation*}
q_{i,j}^r = \min \{ [i^r \to j^r]^\text{sell} ; [j^r \to i^r]^\text{buy} \} = \frac{s_i^r d_j^r}{q^r}  ~.
\end{equation*}


\end{itemize}

All $q_{i,j}^r$ are reconstructed by iterating Steps 1-4 for all pairs of agents involved in round $r$. Straightforward calculations show that $\sum_{i,j} q_{i,j}^r = q^r$. 

\subsection{Estimation of model parameters}

\subsubsection{The assumptions on the FTM model}

To enable quick estimations, we assume that the two livestock markets analyzed are equilibrated and are not submitted to any kind of major shock. We also neglect losses ($L_S = L_D = 0$) and external flows ($E_S = E_D = 0$). We hence recover the reference market model (see main text). We finally assume that the reference price is equal $p^* = 1$. We deduce the following relationships between model parameters: 
\begin{equation*}
\begin{gathered}
\Phi^* = \Theta^* \kappa ~, \hfill \\
\Sigma^*  = \Sigma_\oplus^* = N_S \sigma_0 = \Phi^* ~, \hfill \\
\Delta^*  = \Delta_\oplus^* = N_D \delta_0 = \Phi^* ~, \hfill \\
\end{gathered}
\end{equation*}
with $N_S = N_{S\backslash{}D} + N_{S\cap{}D}$ and $N_D = N_{D\backslash{}S} + N_{S\cap{}D}$. 

We only need to estimate $\Phi^*$, $\Theta^*$, $N_{S\backslash{}D}$, $N_{S\cap{}D}$ and $N_{D\backslash{}S}$ to deduce all the other parameters. 

\subsubsection{Determination of key parameters from data}

Let $t_1$ and $t_2$ be the extreme dates of the data we wish to consider for parameter estimation. The corresponding interval of time is given by: $\Delta t = t_2 - t_1$. For each market, agents are numbered from $1$ to $N$. Let $a \in [| 1; N |]$ denote the index of an agent. When referring to a pair of agents, we reserve the index $i \in [| 1; N |]$ for sellers and the index $j \in [| 1; N |]$ for demanders. For cattle data, a round $r$ corresponds to a unique transaction involving a supplier and a demander. For swine data, a round $r$ is a set of transactions involving potentially multiple suppliers and demanders. By definition, $q_{i,j}^r = 0$ means that $i$ and $j$ are not connected during round $r$, while $q_{i,j}^r > 0$ means that supplier $i$ and demander $j$ are connected by a unique transaction. 

Let $s_a([t_1; t_2]) $ and $d_a([t_1; t_2])$ denote the total number of goods supplied and received by $a$ over $[t_1; t_2]$ respectively. By definition we have: 
\begin{equation*}
\begin{gathered}
s_a([t_1; t_2]) = \sum_{r \in [t_1; t_2]} \sum_j q_{a,j}^r \hfill \\
d_a([t_1; t_2]) = \sum_{r \in [t_1; t_2]} \sum_i q_{i,a}^r \hfill
\end{gathered}
\end{equation*}
where $r \in [t_1; t_2]$ means that round $r$ occurs within period $[t_1; t_2]$. 

It follows $N_{S\backslash{}D}$, $N_{S\cap{}D}$ and $N_{D\backslash{}S}$ are given by: 
\begin{equation*}
\begin{gathered}
N_{S\backslash{}D}([t_1; t_2]) = \sum_{a} \mathds{1}_{ \{s_a([t_1; t_2]) \ne 0\} \&  \{d_a([t_1; t_2]) = 0\} } \hfill \\
N_{S\cap{}D}([t_1; t_2]) = \sum_{a} \mathds{1}_{ \{s_a([t_1; t_2]) \ne 0\} \&  \{d_a([t_1; t_2]) \ne 0\} } \hfill \\
N_{D\backslash{}S}([t_1; t_2]) = \sum_{a} \mathds{1}_{ \{s_a([t_1; t_2]) = 0\} \&  \{d_a([t_1; t_2]) \ne 0\} } \hfill
\end{gathered}
\end{equation*}
where $\mathds{1}_O$ is the indicator function of event $O$, i.e. $\mathds{1}_O = 1$ when $O$ is true and $\mathds{1}_O = 0$ when $O$ is false.

Reference trade flow $\Phi^*$ and transaction rate $\Theta^*$ are estimated as: 
\begin{equation*}
\begin{gathered}
\Phi^*([t_1; t_2]) =\frac{\sum_{r \in [t_1; t_2]} \sum_{i,j} q_{i,j}^r}{\Delta t} \hfill \\
\Theta^*([t_1; t_2]) =\frac{\sum_{r \in [t_1; t_2]} \sum_{i,j} \mathds{1}_{q_{i,j}^r \ne 0}}{\Delta t} \hfill
\end{gathered}
\end{equation*}

\subsubsection{Average values of model parameters}

Apart from the coefficient of friction (see below), we are solely interested by the average value of model parameters over one year of transactions. Table \ref{stab:parms} contains the parameter values derived from the French cattle and swine markets with $\Delta t = 1$ year.

\begin{table}[H]
\caption{\bf{Parameter values for cattle and swine markets.} \hfill \label{stab:parms}}
\begin{center}
\begin{tabular}
{p{2cm}p{6cm}rrp{2cm}}
parameter & meaning & cattle & swine & unit \\
\hline
$\kappa$ & coefficient of friction & 3.4 & 71.7 & none \\
$\Phi^*$ & trade flow & 7,578,476~~ & 8,075,973~~ & per year \\
$\Theta^*$ & transaction rate & 2,224,182~~ & 112,683~~ & per year \\
$\sigma_0$ & per-agent production rate in supply & 39~~ & 1,474~~ & per year \\ 
$\delta_0$ & per-agent production rate in demand & 64~~ & 761~~ & per year \\ 
$N_S$ & number of suppliers & 193,354~~ & 5,480~~ & none \\
$N_D$ & number of demanders & 118,503~~ & 10,619~~ & none \\
$N_{S\backslash{}D}$ & number of strict suppliers & 88,761~~ & 1,314~~ & none \\
$N_{S\cap{}D}$ & number of wholesalers & 104,593~~ & 4,166~~ & none \\
$N_{D\backslash{}S}$ & number of strict demanders & 13,910~~ & 6,453~~ & none \\
\hline
\end{tabular}
\end{center}
\end{table}

\subsubsection{Dynamics of the coefficients of friction}

\begin{figure}[H]
\begin{center}
\includegraphics[width=0.6\textwidth]{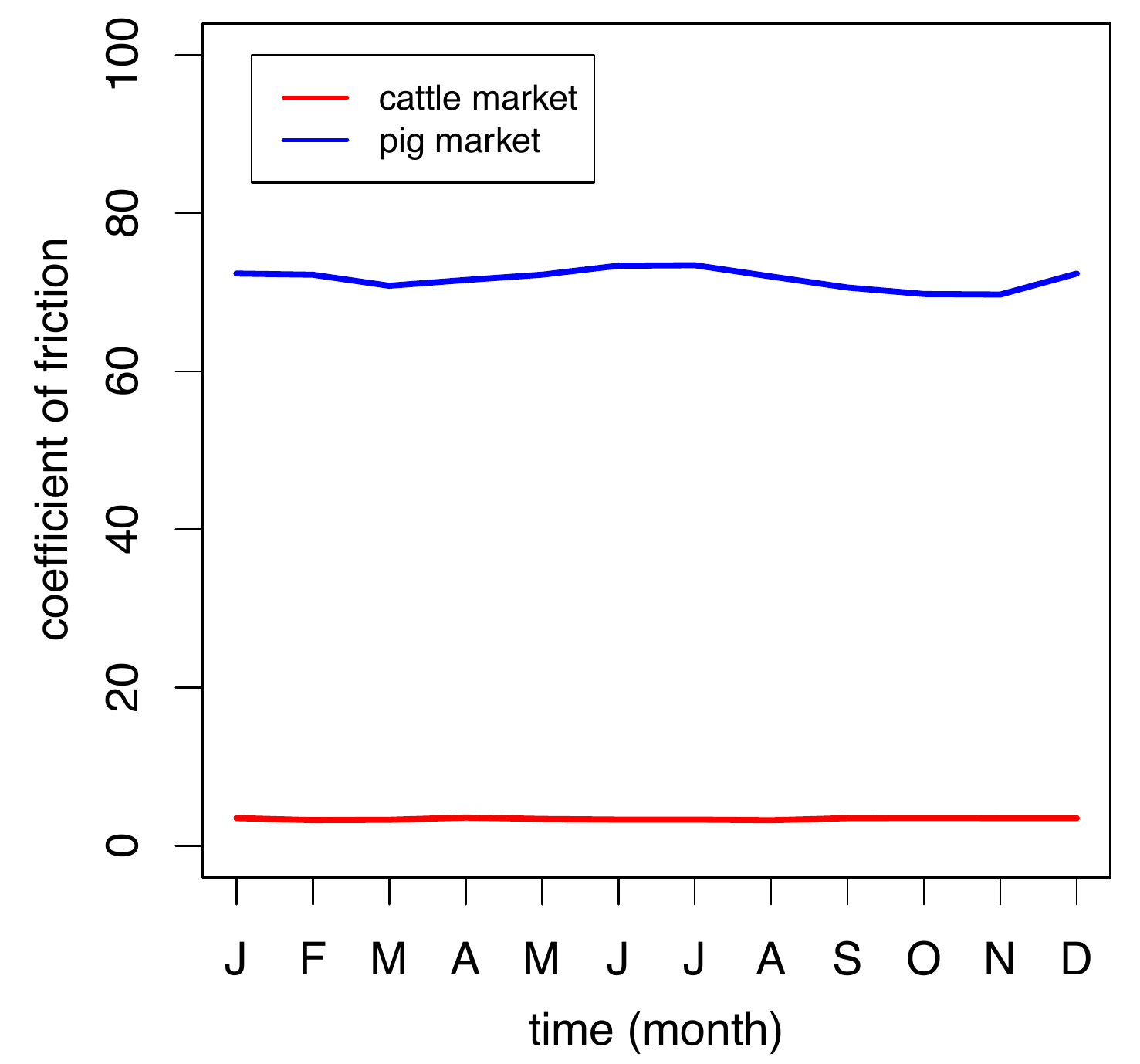}
\end{center}
\caption[Dynamics of the coefficients of friction for cattle and swine markets]
{{\bf Dynamics of the coefficients of friction for cattle and swine markets.} For each market, the coefficient of friction is calculated every month. 
\label{sfig:kappa_time}}
\end{figure}

For each market, we calculate the monthly value of the coefficient of friction $\kappa$ and set $\Delta t = 1/12$ year. We show in Fig. \ref{sfig:kappa_time} that $\kappa$ is constant for each market.

\section{Supplementary results on the frictional-trade market (FTM) model}

This section presents detailed proof and additional results derived from the FTM model. 

\subsection{Existence and unicity of solutions}

For any $a \ge 0$ and $b \ge 0$, we consider the function $f$ defined as: 
\begin{align*}
\mathbb{R}^{+2} &\to \mathbb{R}^+ \\
f: (x,y) &\mapsto f(x,y)=\min(ax;by) ~. \\
\end{align*}

Property: $f$ is Lipschitz continuous with Lipschitz constant $M = \max(a;b)$.

\null

Proof is as follows:  

\null

We first rewrite $f$ as: 
\begin{equation*}
f(x,y) = \frac{ax + by - |ax -by| }{2} ~.
\end{equation*}

Then we have: 
\begin{align*}
f(x_2,y_2) - f(x_1,y_1) &=  \frac{a(x_2 - x_1) + b(y_2-y_1) - ( | ax_2 - by_2 | - | ax_1-by_1 | )  }{2} \\
&\le  \frac{a |x_2 - x_1| + b |y_2-y_1| + |~| ax_2 - by_2 | - | ax_1-by_1 |~|  }{2} ~. \\
\end{align*}

By the reverse triangle inequality, we have: 
\begin{equation*}
f(x_2,y_2) - f(x_1,y_1)  \le \frac{a |x_2 - x_1| + b |y_2-y_1| + | a(x_2 - x_1) - b(y_2 -y_1) |  }{2} ~.
\end{equation*}

The triangle inequality finally yields: 
\begin{equation*}
f(x_2,y_2) - f(x_1,y_1)  \le a |x_2 - x_1| + b |y_2-y_1| \le \max(a;b) (|x_2 - x_1| + |y_2-y_1|) ~.
\end{equation*}

Apart from the min function, the other functions used in the FTM and ME models are canonical and known to be Lipschitz on sets of epidemiological and economic relevance. We deduce directly, by usual operations (composition, multiplication, addition), that all functions implemented in the FTM and ME models (the right-hand sides of the ordinary differential equations in the main text) are Lipschitz. From the existence and uniqueness theorem \citep[Chapters 7 and 17 in][]{HirschEtal2004DynamicalSystemsBookElsevier}, it follows that solutions to the associated ordinary differential equations are unique and exist for all initial conditions. 

\subsection{Equilibria and stability analyses}

We analyse the FTM model analytically with a bottom-up approach. First, we define and analyse a \textit{reference market} without stock loss nor external flows. Second we show that the \textit{law of supply and demand} (LSD) is spanned by the reference market. Third we explore the impacts of \textit{stock loss} on trade dynamics. Fourth we analyse the influence of \textit{external flows} on trade dynamics.

\begin{figure}[H]
\begin{center}
\includegraphics[width=\textwidth]{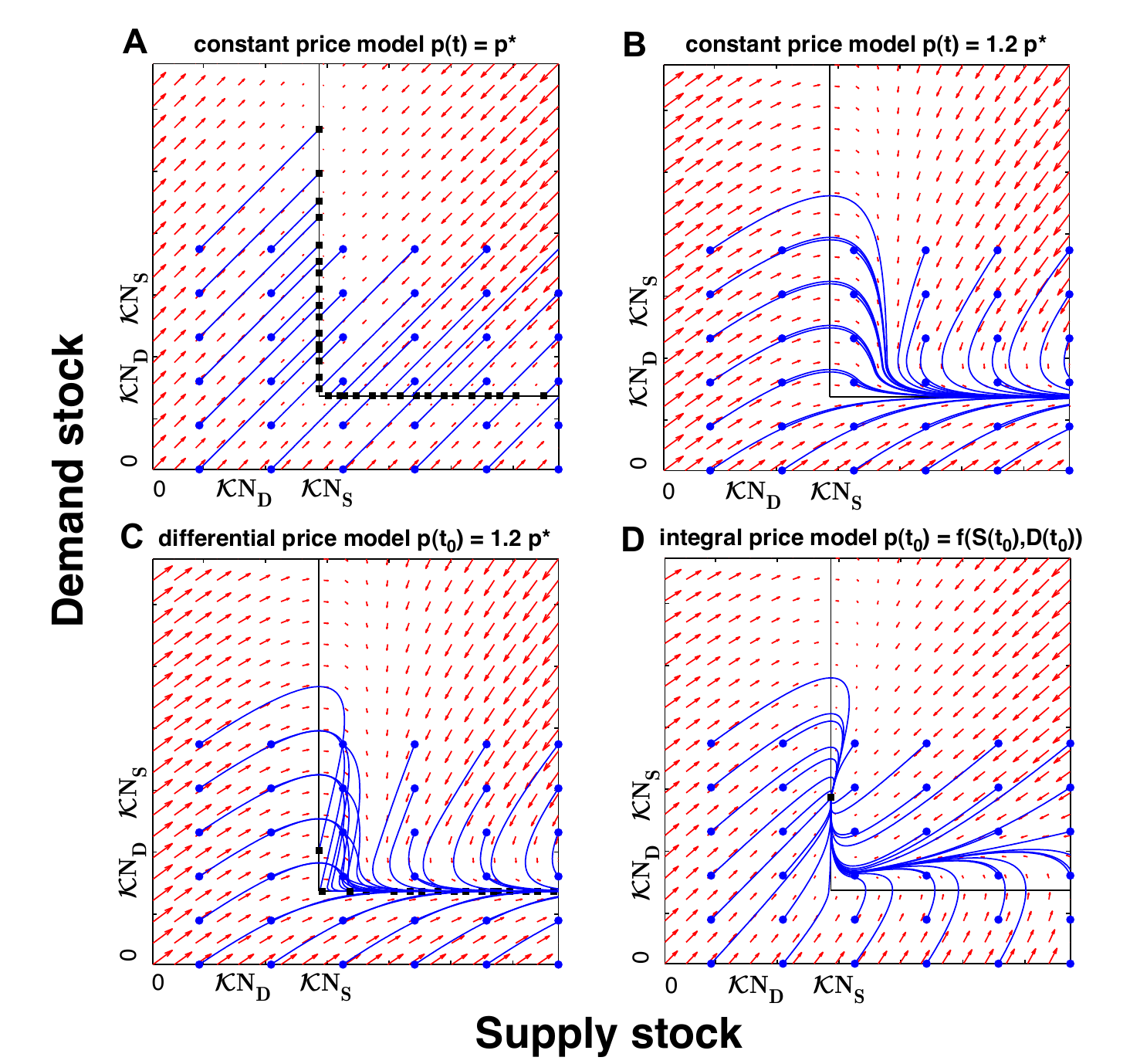}
\end{center}
\caption[Impacts of various price models on reference market dynamics]
{{\bf Impacts of various price models on reference market dynamics.}
\small Joint-evolution of the demand stock $D(t)$ (y-axis) as function of the supply stock $S(t)$ (x-axis) for various price dynamics: constant price model ($\mu = 0$) with $p= p^*$ and $p= 1.2~p^*$ (Panels \textbf{A-B}), differential price model with $p= 1.2~p^*$ (Panel \textbf{C}) and integral price model corresponding to the LSD (Panel \textbf{D}). For the four cases, market dynamics correspond to the reference market with $\kappa = 1$. We set $\mu = 0$ for the constant price model and $\mu = 10^{-6}$ for the differential and integral price models. Other parameters are from the cattle dataset. Red arrows represent the vector field governing the dynamics of $S$ and $D$. For any initial condition $(S(t_0), D(t_0))$, each arrow hence represents the vector $\textbf{v}: (\frac{dS}{dt}, \frac{dD}{dt})_{S=S(t_0),D=D(t_0),p=p(t_0)}$. The blue dots represent a subset of the latter initial conditions for which we explored the full trajectories of the model numerically; the blue lines and the black squares represent the associated dynamics and steady-states  respectively. The vertical ($S = S_{min}^* = \kappa N_S$) and horizontal ($D = D_{min}^* = \kappa N_D$) black lines represent the theoretical equilibria $(S^*,D^*)$ of the differential price model. The point $S^* = D^* = \max \{ \kappa N_S, \kappa N_D \} $ is the only equilibrium of the integral price model that does not violate the LSD, as predicted theoretically and observed numerically. 
\label{sfig:phasediag}}
\end{figure}

\subsubsection{Trade without stock loss nor external flows: the reference market}

We neglect stock loss ($L_S = L_D = 0$) and external flows ($E_S = E_D =0$). This case is referred to as the \textit{reference market} and the value of its variables at equilibrium are denoted throughout with a \textit{star} in superscript. Market dynamics reduce to: 
\begin{equation}
\begin{gathered}
\frac{dS}{dt}  =  \overbrace {N_S~\sigma_0~p^{\varepsilon_S}}^{\Sigma_\oplus}  - \overbrace{ \frac{ \min \{ \Sigma_\oplus ; \Delta_\oplus \} }{\kappa} }^\Theta ~ \overbrace{ \min \left\{ \frac{S}{N_S} ; \frac{D}{N_D} \right\}}^q ~, \hfill \\
\frac{dD}{dt}  = \overbrace {N_D~\delta_0~p^{- \varepsilon_D}}^{\Delta_\oplus} - \overbrace{ \Theta q }^\Phi ~, \hfill \\
\frac{dp}{dt} = \mu \frac{d(D-S)}{dt} ~ p = \mu (\Delta_\oplus - \Sigma_\oplus)p ~. \hfill
\label{sys:ref_market}
\end{gathered}
\end{equation}

The case $\mu = 0$ corresponds to a model where price is assumed to remain constant and equal to its initial value, i.e. $p(t) = p(t_0)$. Such a model is known in the economic literature as a Ôfixprice modelÕ and corresponds to an example of Ôprice stickinessÕ \citep{Silvestre2008FixpriceNPDE}. The subcase $p(t_0) = p^*$ is treated below (see also Fig. \ref{sfig:phasediag}\textbf{A}). Due to the functional shapes of $\Sigma_\oplus$ and $\Delta_\oplus$, the subcase $p(t_0) \ne p^*$ implies that either $\frac{dS}{dt}$ or $\frac{dD}{dt}$ is never equal to 0. The market is hence diverging in either supply or demand and $p(t_0)$ is a bifurcation parameter. For instance, the constant price model with $p= 1.2p^*$ diverges in supply (Fig. \ref{sfig:phasediag}\textbf{B}). The more interesting case $\mu > 0$ is analyzed below. 

\paragraph{Equilibria.}~

We assume from now that $\mu > 0$, i.e. we allow for price dynamics. 

By construction, $p=p^* = \left(\frac{N_D \delta_0}{N_S \sigma_0}\right)^\frac{1}{\varepsilon_S + \varepsilon_D}$ equalizes $\Sigma_\oplus$ and $\Delta_\oplus$. Given trade flow $\Phi = \Theta q$ acts symmetrically on supply and demand, it follows $p=p^*$ is the only value of $p$ for which the market is equilibrated. As $p=p^*$ is unique, reference trade flow $\Phi^*$ is equal to $\Sigma_\oplus(p^*) = \Delta_\oplus(p^*)$ and is unique. $\Theta(p^*) = \Theta^*$ is also unique. 

$\Phi^* $ is the only value of $\Phi$ that satisfies $\frac{dS}{dt} = \frac{dD}{dt} = 0$. To determine the equilibrium values of supply and demand, we hence solve $q(S^*,D^*) = \min \{ \frac{S^*}{N_S}; \frac{D^*}{N_D} \} = \frac{\Phi^*}{\Theta^*} = \kappa$. Due to the $\min$ function, the system is characterised by two sets of infinite equilibria: either  $\{ S^* = \kappa N_S$ and $D^* \ge \kappa N_D \}$ or $\{ S^* \ge \kappa N_S $ and $D^* = \kappa N_D \}$. $S_{min}^* = \kappa N_S$ and $D_{min}^* = \kappa N_D$ are the minimal stocks of supply and demand for which the market is equilibrated. Trade flow at equilibrium is hence given by:
\begin{equation}
{\Phi^*} = {[{N_S}{\sigma _0}]^{\frac{{{\varepsilon _D}}}{{{\varepsilon _S} + {\varepsilon _D}}}}} ~ {[{N_D}{\delta _0}]^{\frac{{{\varepsilon _S}}}{{{\varepsilon _S} + {\varepsilon _D}}}}} ~.
\end{equation}

While trade flow at the reference market equilibrium is unique and does not depend on $\kappa$, $S^*$ and $D^*$, i.e. persistent frustrations, are positively associated with $\kappa$. 

\paragraph{Stability.}~

From system (\ref{sys:ref_market}), we notice that $\frac{dp}{dt}$ is a function of variables $p$, $\Sigma_\oplus(p,N_S)$ and $\Delta_\oplus(p,N_D)$. Since $N_S$ and $N_D$ are constant in the FTM model, $\frac{dp}{dt}$ only depends on $p$. 

We first analyse the global stability of $p$. We define the function $V(p) = (p - p^*)^2$. $V$ has the following properties:
\begin{equation}
\begin{gathered}
V(p^*) = 0 ~\text{and}~ V(p)>0 ~\text{for}~ p \ne p^* ~, \hfill \\
\frac{dV}{dt} = 2 \frac{dp}{dt} (p - p^*) = 2 \mu p \left(\Delta_\oplus (p,N_D) - \Sigma_\oplus (p,N_S)\right) (p - p^*) < 0 ~\text{for}~ p \ne p^* ~,\hfill
\end{gathered}
\end{equation}
since by construction $\Delta_\oplus (p,N_D) < \Sigma_\oplus (p,N_S) \Leftrightarrow p > p^*$.  

We conclude $V$ is a strict Liapunov function and the point $p=p^*$ is asymptotically stable \citep[Section 9.2 of][]{HirschEtal2004DynamicalSystemsBookElsevier}. 

Given that the point $p=p^*$ is eventually reached by the system, we only need to explore the joint-dynamics of ($S,D$) from the point of time $t_x$ where $p(t_x)=p^*$. We do not know the precise values of $t_x$ and $[S(t_x),D(t_x)]$. We can still conclude on the stability of system (\ref{sys:ref_market}) if we can analyse the dynamics of the system ($S,D$) with $p=p^*$ for \textit{any} initial conditions $[S(t_0), D(t_0)]$ because $[S(t_0), D(t_0)] \supset [S(t_x),D(t_x)]$. 

From now on, we analyse such a system ($S,D$) with initial conditions $[S(t_0), D(t_0)]$ and where $p=p^*$. Since $p=p^*$, we know that: $\Sigma_\oplus = \Delta_\oplus = \Phi^*$. It follows $\frac{dS}{dt} = \frac{dD}{dt}~, \forall t$. The equality of the derivatives means that there is a linear vector field associated with any initial values of $(S,D)$. In other words, the dynamics of $(S,D)$ are hence fully described by a line $D(t) = S(t) + a$ where $a$ is a constant that depends on initial conditions. For $t=t_0$, $D(t_0) = S(t_0) + a$, which leads to $a = D(t_0) - S(t_0)$. Full market dynamics are hence given by the line:
\begin{equation}
 D(t)= S(t) + D(t_0) - S(t_0) ~.
\label{eqS:line_pstar}
\end{equation}

The direction of the dynamics along the line described by equation (\ref{eqS:line_pstar}) are found by solving the inequality $\frac{dS}{dt}~=~\frac{dD}{dt}~<0$:
\begin{equation*}
\begin{gathered}
\frac{dS}{dt} < 0 \Leftrightarrow \Phi^* - \Theta^* q < 0 ~,  \hfill \\
\frac{dS}{dt} < 0 \Leftrightarrow q(t) = \min \left\{ \frac{S(t)}{N_S}; \frac{D(t)}{N_D} \right\} > \kappa ~,  \hfill \\
\frac{dS}{dt} < 0 \Leftrightarrow [S(t) > \kappa N_S ~~\text{AND}~~ D(t) > \kappa N_D] ~, \hfill
\end{gathered}
\end{equation*}
which is equivalent by logical negation to: $\frac{dS}{dt} > 0 \Leftrightarrow [S(t) < \kappa N_S~~\text{OR}~~D(t) < \kappa N_D$]. Those equivalences prove that the exchange volume $q(t)$ always converges to $\kappa$. It follows the system is globally stable in trade flow, i.e. $\Phi(t) \to \Phi^*$ for any initial conditions. 

Given relation (\ref{eqS:line_pstar}) and our stability analysis, the position of $(D(t_0) - S(t_0))$ with respect to $(D_{min}^* - S_{min}^*)$ separates the state space $(S,D)$ in two subparts that determine the final state of the system:
\begin{equation*}
\begin{gathered}
\text{If $(D(t_0) - S(t_0)) \ge (D_{min}^* - S_{min}^*)$:}~~~(S^* = S_{min}^*~,~D^* = S_{min}^* + (D(t_0) - S(t_0)) ) ~. \hfill \\
\text{If $(D(t_0) - S(t_0)) \le (D_{min}^* - S_{min}^*)$:}~~~(S^* = S_{min}^* - (D(t_0) - S(t_0))~,~D^* = D_{min}^*) ~. \hfill \\
\end{gathered}
\end{equation*}
The case $(D(t_0) - S(t_0)) \ge (D_{min}^* - S_{min}^*)$ corresponds to an initial positive excess demand which implies a limiting supply at equilibrium. The case $(D(t_0) - S(t_0)) \le (D_{min}^* - S_{min}^*)$ corresponds to an initial negative excess demand which implies a limiting demand at equilibrium. 

Since the equilibria in supply and demand depend on initial conditions, there is an infinite number of unstable equilibria $(S^*,D^*)$ with a switched fixpoint: either $S^* = S_{min}^*$ or $D^* = D_{min}^*$. 

In conclusion, the system always converges to reference flows ($\Phi(t) \to \Phi^*$) and reference price ($p(t) \to p^*$) for any initial conditions or external perturbations. In other words, our model market is globally stable in (measurable) price and flows while extremely unstable in (hidden) frustrations. Our key findings are summarized graphically in Fig. \ref{sfig:phasediag}\textbf{C}.

\subsubsection{The law of supply and demand (LSD) is spanned by the reference market}

One could argue that our pricing model is unsatisfactory since $S$ and $D$ are not necessarily identical at equilibrium. The LSD indeed stipulates that price should increase in response to any excess demand $D - S > 0$. There is little empirical support for the LSD \citep{McCauley2009Book}, which is consistent with the fact that $S$ and $D$ are not observable. For comparative purposes with the existing literature, we nevertheless investigate the LSD analytically and show that it is a special case spanned by the reference market. 

By integration of system~(\ref{sys:ref_market}), we have the equivalence:
\begin{equation}
\begin{gathered}
(\mathcal{P}_d): ~~ \frac{d p}{dt} = \mu \frac{d(D-S)}{dt} ~ p ~\text{with initial conditions: $[S(t_0) = S_0,D(t_0) = D_0,p(t_0) = p_0]$}~, \hfill \\
\Leftrightarrow \\
(\mathcal{P}_i): ~~ p(t) = p_0 \exp{\left(\mu\left[D(t)-S(t) - (D_0-S_0)\right]\right)}~,\hfill
\end{gathered}
\end{equation}
where $(\mathcal{P}_d)$ and $(\mathcal{P}_i)$ represent the differential and integral formulations of our pricing model $\mathcal{P}$, and $S_0$, $D_0$ and $p_0$ are constants. 

System (\ref{sys:ref_market}) is hence equivalent to a new system where: 
\begin{equation}
\begin{gathered}
\frac{dS}{dt}  =  \overbrace {N_S~\sigma_0~p^{\varepsilon_S}}^{\Sigma_\oplus}  - \overbrace{ \frac{ \min \{ \Sigma_\oplus ; \Delta_\oplus \} }{\kappa} }^\Theta \overbrace{ \min \left\{ \frac{S}{N_S} ; \frac{D}{N_D} \right\}}^q \hfill ~, \\
\frac{dD}{dt}  = \overbrace {N_D~\delta_0~p^{- \varepsilon_D}}^{\Delta_\oplus} - \overbrace{ \Theta q }^\Phi ~, \hfill \\
p(t) = p_0 \exp{\left(\mu\left[D(t)-S(t) - (D_0-S_0)\right]\right)} ~. \hfill
\label{sys:ref_market_Pi}
\end{gathered}
\end{equation}

Importantly, (\ref{sys:ref_market_Pi}) is only identical to (\ref{sys:ref_market}) is we take the same initial conditions, i.e. if we set $[S(t_0) = S_0,D(t_0) = D_0,p(t_0) = p_0]$. Now we analyse system (\ref{sys:ref_market_Pi}) in the case where the constants $[S_0, D_0, p_0]$ are not necessarily equal to $[S(t_0),D(t_0),p(t_0)]$. From now, we only consider the case where $p_0 = p^*$. We want to find the values $[S_0, D_0]$ that are compatible with the LSD. 

Since $(\mathcal{P}_i) \Rightarrow (\mathcal{P}_d)$ (the converse proposition is not true any more), we have the same properties as already described in the analysis of system (\ref{sys:ref_market}). But we have new properties thanks to the $(\mathcal{P}_d)$ model. We know: $p(t) = p^* \exp{\left(\mu\left[D(t)-S(t) - (D_0-S_0)\right]\right)}$. Since $p = p^*$ is asymptotically stable, we solve equation $p(t) = p^*$ at equilibrium to yield a new relationship: 
\begin{equation}
D^* - S^* = D_0 - S_0 ~, \hfill
\label{eqS:constraint_LSD}
\end{equation}
that is specific to the $(\mathcal{P}_d)$ model.

\paragraph{Equilibria.}~

At equilibrium, we have now two relationships between $S^*$ and $D^*$: 
\begin{equation}
\begin{gathered}
D^* - S^* = D_0 - S_0 ~, \hfill \\
q(S^*,D^*) = \min \left\{ \frac{S^*}{N_S}; \frac{D^*}{N_D} \right\} = \kappa ~, \hfill \\
\label{sys:ref_market_Pi_eq1}
\end{gathered}
\end{equation}
which lead to a unique equilibrium whose value depends on the position of $(D_0 - S_0)$ with respect to $(D_{min}^* - S_{min}^*)$:
\begin{equation}
\begin{gathered}
\text{If $(D_0 - S_0) \ge (D_{min}^* - S_{min}^*)$:}~~~(S^* = S_{min}^*~,~D^* = S_{min}^* + (D_0 - S_0))~. \hfill \\
\text{If $(D_0 - S_0) \le (D_{min}^* - S_{min}^*)$:}~~~(S^* = D_{min}^* - (D_0 - S_0)~,~D^* = D_{min}^*)~. \hfill \\
\label{sys:ref_market_Pi_eq2}
\end{gathered}
\end{equation}

From (\ref{sys:ref_market_Pi_eq2}), we notice that the LSD corresponds to the case $(D_0 - S_0) = 0$, i.e. $S_0=D_0$. If $N_D > N_S$, the LSD is hence given by $S^* = D^* = \kappa N_S$. While if $N_D < N_S$, the LSD is given by $S^* = D^* = \kappa N_D$. For both cases, the LSD leads to $S^* = D^* = \max \{ \kappa N_D, \kappa N_S  \}$. 

Note that if agents were rational, they would try to minimize their stocks while keeping trade flows $\Phi^*$ constant: this corresponds instead to the case $(D_0 - S_0) = (D_{min}^* - S_{min}^*)$, i.e. $S_0 = D_0 - (D_{min}^* - S_{min}^*)$, for which $(S^* = S_{min}^*, D^* = D_{min}^*)$ according to (\ref{sys:ref_market_Pi_eq2}).

For any constants $(S_0,D_0)$ taken as a reference, the mismatch between supply and demand can be assessed with a generalisation of the excess demand, denoted $E(S_0,D_0)$, that accounts for imperfect transactions with friction:
\begin{equation}
E(S_0,D_0) \equiv D(t) - S(t) - (D_0 - S_0)
\label{generalexcessdemand}
\end{equation}

\paragraph{Stability.}~

We know $(\mathcal{P}_i) \Rightarrow (\mathcal{P}_d)$. It follows our former analysis stands and the LSD equilibrium $[p^*,S^*=D^*=\max \{ \kappa N_D, \kappa N_S  \}]$ is asymptotically stable. The integral version of our pricing model is hence compatible with the expected properties of the LSD. As predicted, the integral price model with $S_0=D_0$ converges to a unique equilibrium $S^* = D^* = \max \{ \kappa N_S , \kappa N_D \} = \kappa N_S$ in the case of cattle data (Fig. \ref{sfig:phasediag}\textbf{D}). 

\subsubsection{Trade with stock loss: the general case}

We now allow for stock loss in the FTM model, which leads to the system: 
\begin{equation}
\begin{gathered}
\frac{dS}{dt}  =  \overbrace {N_S~\sigma_0~p^{\varepsilon_S}}^{\Sigma_\oplus} - r_S S - \overbrace{ \frac{ \min \{ \Sigma_\oplus ; \Delta_\oplus \} }{\kappa} }^\Theta \overbrace{ \min \left\{ \frac{S}{N_S} ; \frac{D}{N_D} \right\}}^q ~, \hfill \\
\frac{dD}{dt}  = \overbrace {N_D~\delta_0~p^{- \varepsilon_D}}^{\Delta_\oplus} - r_D D - \overbrace{ \Theta q }^\Phi ~, \hfill \\
\frac{dp}{dt} =  \mu \frac{d(D-S)}{dt} ~ p = \mu \left(\Delta(D,p) - \Sigma(S,p)\right)~p ~, \hfill
\label{sys:ref_market_and_L}
\end{gathered}
\end{equation}
with $\Sigma(S,p) = \Sigma_\oplus - r_S S$ and $\Delta(D,p) = \Delta_\oplus - r_D D$. Notice that in contrast with the reference market (system~(\ref{sys:ref_market})), $\frac{dp}{dt}$ now depends on $S$, $D$ and $p$.

\paragraph{Equilibria.}~

When stock loss is included, we only know that the number of equilibria is not unique and we cannot derive the equilibrium values of state variables explicitly. Let \textit{eq} in subscript denote equilibrium values in the general case (in contrast with the special case of the reference market where equilibrium is denoted by a star in superscript). At equilibrium, $\Sigma(S_{eq},p_{eq})=\Delta(D_{eq},p_{eq})$ is indeed the only independent equation of system~(\ref{sys:ref_market_and_L}) which has three unknowns $(S_{eq},D_{eq},p_{eq})$. However we can still extract useful information on the system. If we e.g. only consider positive stock loss ($r_S \ge 0$ and $r_D\ge 0$), we know for sure from~(\ref{sys:ref_market_and_L}) that:
\begin{equation}
\begin{gathered}
\Phi_{eq} = \Sigma_\oplus^{eq} - r_S S_{eq} \le  \Sigma_\oplus^{eq} ~, \hfill \\
\Phi_{eq} = \Delta_\oplus^{eq} - r_D D_{eq} \le  \Delta_\oplus^{eq} ~. \hfill \\
\label{eqS:suboptimalflows1}
\end{gathered}
\end{equation}

From~(\ref{eqS:suboptimalflows1}) we conclude that equilibrium flows are necessarily suboptimal as:
\begin{equation}
\Phi_{eq} \le \min \{ \Sigma_\oplus^{eq} ,\Delta_\oplus^{eq} \} \le \Phi^* = \min \{ \Sigma_\oplus^* ,\Delta_\oplus^* \}   \hfill \\
\label{eqS:suboptimalflows2}
\end{equation}

A deeper analysis enables us to quantify the effects of $\kappa$, $r_S$ and $r_D$ on equilibrium trade flow. We know in the general case that: 
\begin{equation}
\begin{gathered}
\Phi_{eq} = \Sigma_{eq} = \Delta_{eq} = \Sigma_\oplus^{eq} - r_S S_{eq} = \Delta_\oplus^{eq} - r_D D_{eq} ~, \hfill \\
q_{eq} \equiv \frac{\Phi_{eq}}{\Theta_{eq}} \Rightarrow \min \{ \frac{S_{eq}}{N_S}; \frac{D_{eq}}{N_D} \} = \frac{\kappa \Phi_{eq}}{\min \{ \Sigma_\oplus^{eq} ,\Delta_\oplus^{eq} \}} ~. \hfill 
\label{eqS:suboptimalflows3}
\end{gathered}
\end{equation}

Then two cases need to be considered depending on the limiting factor of $q_{eq}$.

If $q_{eq}$ is limited by the per capita supply, i.e. if $ q_{eq} = \tfrac{S_{eq}}{N_S}$, we find from (\ref{eqS:suboptimalflows3}) that $S_{eq} = \frac{\kappa N_S \Sigma_\oplus^{eq}}{\min \{ \Sigma_\oplus^{eq} ,\Delta_\oplus^{eq} \} + r_S \kappa N_S}$. We hence deduce the value of $q_{eq}$. Given that $\Phi_{eq} = \Theta_{eq} q_{eq}$, we find: 
\begin{equation}
\Phi_{eq} = \Sigma_\oplus^{eq} \frac{\min \{ \Sigma_\oplus^{eq} ,\Delta_\oplus^{eq} \}}{\min \{ \Sigma_\oplus^{eq} ,\Delta_\oplus^{eq} \} + r_S \kappa N_S}
\label{eqS:suboptimalflows4}
\end{equation}

If $q_{eq}$ is limited by the per capita demand, i.e. if $ q_{eq} = \tfrac{D_{eq}}{N_D}$. We directly find by symmetry:
\begin{equation}
\Phi_{eq} = \Delta_\oplus^{eq} \frac{\min \{ \Sigma_\oplus^{eq} ,\Delta_\oplus^{eq} \}}{\min \{ \Sigma_\oplus^{eq} ,\Delta_\oplus^{eq} \} + r_D \kappa N_D}
\label{eqS:suboptimalflows5}
\end{equation}  

Importantly, equations (\ref{eqS:suboptimalflows4})-(\ref{eqS:suboptimalflows5}) imply that $r_S$, $r_D$ and $\kappa$ have a negative impact on equilibrium flow. In particular, equations (\ref{eqS:suboptimalflows4})-(\ref{eqS:suboptimalflows5}) imply:
\begin{equation}
\mathop {\lim }\limits_{r_S \kappa \to \infty} \Phi_{eq}  = \mathop {\lim }\limits_{r_D \kappa \to \infty} \Phi_{eq}  = 0
\label{eqS:suboptimalflows6}
\end{equation}

Since the dynamics with stock loss are not fully analytically tractable in the general case, we resort to extensive numerical simulations to confirm the key influence of $r_S$, $r_D$ and $\kappa$ on trade dynamics (see Global Sensitivity Analysis (GSA) of the FTM model). In addition, the special case analysed thereafter enables us to get exhaustive analytical insights on the impact of stock loss on market dynamics.

\paragraph{Stability.}~

We cannot conclude analytically on the stability of system~(\ref{sys:ref_market_and_L}) in the general case.

\paragraph{Trade with stock loss: detailed analysis of a special case.}~

To study market transient behaviour and confirm our general findings, we consider initial conditions and parameter values that enable us to solve system (\ref{sys:ref_market_and_L}) analytically. We set $[S(t_0)=D(t_0) \ge 0;p(t_0)=p^*]$ at initial time $t_0$ and track trade flow $\Phi$ until equilibration. This set of initial conditions is compatible with the LSD since $S(t_0)=D(t_0)$ and $p(t_0)=p^*$. Once accumulated over time at rates $\Sigma_\oplus$ and $\Delta_\oplus$, supply and demand stocks are converted through trade ($\Phi$) and losses (at rates $r_S S$ and $r_D D$). For simplicity, we consider symmetrical losses ($r_S = r_D = r$). Since we start from $S(t_0)=D(t_0)$ and $p(t_0)=p^*$ as initial conditions, and as $N_S$ and $N_D$ do not change over time, we have $p(t)=p^*$. Hence, when $r$ is set, the coefficient of friction $\kappa$ is the only factor influencing market dynamics. Since $p(t)=p^*$, it follows that $\Sigma_\oplus (t) = \Delta_\oplus (t) = \Phi^*$ and  $\Theta (t)= \tfrac{\Phi^*}{\kappa}$. Using symmetry arguments in equations describing $S$ and $D$ in system~(\ref{sys:ref_market_and_L}) and given that we assumed $S(t_0)=D(t_0)$ and $r_S = r_D = r$, we have $S(t)=D(t)~,\forall t$. We hence have $q(t) = a S(t)$ where $a = \min \{ \tfrac{1}{N_S}, \tfrac{1}{N_D} \}$ is a dimensionless constant. From the expression of $\Theta$ and $q$, trade flow $\Phi$ can be written: $\Phi(t) = \frac{a \Phi^* S(t)}{\kappa}$. Hence, system~(\ref{sys:ref_market_and_L}) reduces to:
\begin{equation}
\dot S = \Phi^* - \left( rÊ+ \frac{a \Phi^*}{\kappa} \right) S ~.
\label{sys:eco2}
\end{equation}

Solving~(\ref{sys:eco2}) enables us to explicitly describe the market at each point of time. For $t_0 = 0$ and $r \ne - \frac{a \Phi^*}{\kappa}$ we have:
\begin{equation}
S(t)= S(t_0)e^{ -t (r+ \frac{a \Phi^*}{\kappa})}  + \frac{\Phi^*}{ r+ \frac{a \Phi^*}{\kappa} } [1-e^{ -t (r+ \frac{a \Phi^*}{\kappa})  } ] ~.
\label{sys:eco3}
\end{equation}

We notice that (\ref{sys:eco3}) converges if $r > - \frac{a \Phi^*}{\kappa}$ and we find:
\begin{equation}
S_{eq}=  \frac{\Phi^*}{ r+ \frac{a \Phi^*}{\kappa} } ~.
\label{sys:eco4}
\end{equation}

From equation (\ref{sys:eco3}) with $S(t_0)=0$, we deduce the analytical expression of $\Phi(t)$ where:
\begin{equation}
\Phi(t)= \Phi^* \frac{\Phi^*}{ \Phi^* + \frac{r \kappa}{a}  } [1-e^{ -t (rÊ+ \frac{a \Phi^*}{\kappa})  } ] ~.
\label{eqS:analytic_flows}
\end{equation}

If $r > - \frac{a \Phi^*}{\kappa}$,  $\Phi(t)$ converges to:
\begin{equation}
\Phi_{eq} = \Phi^* \frac{\Phi^*}{ \Phi^* + \frac{r \kappa}{a}} ~.
\label{eqS:analytic_eqflows}
\end{equation}

As expected from the analysis of the general case, we reach reference flow ($\Phi_{eq} = \Phi^*$) when stock loss is neglected ($r = 0$) and sub-optimal flow ($\Phi_{eq} < \Phi^*$) when stock loss is positive ($r > 0$). If $r \in ] - \frac{a \Phi^*}{\kappa}$, 0 [ flow converges and is over-optimal ($\Phi_{eq} > \Phi^*$). Finally if  $r \le - \frac{a \Phi^*}{\kappa}$, flow diverges to infinity ($\Phi \to \infty$).

For the realistic case where stock loss is strictly positive ($r > 0$), $\kappa$ and $r$ have a negative impact on equilibrium flow and we recover the key results of equations (\ref{eqS:suboptimalflows4})-(\ref{eqS:suboptimalflows5}). 

\subsubsection{Trade with external flows}

For sake of concision, we neglect stock loss ($r_S = r_D = 0$) and only allow for positive and symmetric external flow ($E_S = E_D = E > 0$). This corresponds to the case explored in the main text to show the impact of imports on epidemics (Figure 3C). We hence have: 
\begin{equation}
\begin{gathered}
\frac{dS}{dt}  =  \overbrace {N_S~\sigma_0~p^{\varepsilon_S}}^{\Sigma_\oplus} + E - \overbrace{ \frac{ \min \{ \Sigma_\oplus ; \Delta_\oplus \} }{\kappa} }^\Theta \overbrace{ \min \left\{ \frac{S}{N_S} ; \frac{D}{N_D} \right\}}^q ~, \hfill \\
\frac{dD}{dt}  = \overbrace {N_D~\delta_0~p^{- \varepsilon_D}}^{\Delta_\oplus} + E - \overbrace{ \Theta q }^\Phi ~, \hfill \\
\frac{dp}{dt} =  \mu (\Delta_\oplus - \Sigma_\oplus) p ~. \hfill
\label{sys:ref_market_and_E}
\end{gathered}
\end{equation}

As a side remark, the case where any values of $E_S$ and $E_D$ are allowed is not difficult to solve since the dynamics of $\frac{dp}{dt}$ do not depend on $S$ and $D$. As we do not use it in the main text, we do not develop it in the SI.

\paragraph{Equilibria.}~

Since we assume symmetrical external flow, the equilibrium in price is the same as the reference case. We hence have $\Theta_{eq} = \Theta^*$. $\Phi_{eq} = \Phi^* + E$ is the only value of $\Phi$ that satisfies $\frac{dS}{dt} = \frac{dD}{dt} = 0$. To determine the equilibrium values of supply and demand, we hence solve $q(S_{eq},D_{eq}) \equiv \min \{ \frac{S_{eq}}{N_S}; \frac{D_{eq}}{N_D} \} = \frac{\Phi_{eq}}{\Theta_{eq}} = \kappa \frac{\Phi^* + E}{\Phi^*}$. Due to the $\min$ function, the system is characterized by two sets of infinite equilibria: either $\{S_{eq} = \kappa N_S \frac{\Phi^* + E}{\Phi^*}$ and $D_{eq} \ge  \kappa N_D \frac{\Phi^* + E}{\Phi^*} \}$ or $\{ S_{eq} \ge \kappa N_S \frac{\Phi^* + E}{\Phi^*}$ and $D_{eq} =  \kappa N_D \frac{\Phi^* + E}{\Phi^*} \}$. It follows $q_{eq} = \kappa \frac{\Phi^* + E}{\Phi^*} > \kappa $

This simple example shows that external flows only act on the stocks of supply and demand and hence on the average stock exchanged per transaction $q$. The transaction rate $\Theta$ is unchanged, since it only depends on limited search and delivery budgets that depend on the numbers of suppliers, demanders and price. This finding has implications for epidemics since both $\Theta$ and $q$ determine the actual force of infection: if the probability of infection per good is already very large, increasing external flow will increase transmission but only moderately (see Figure 3C in the main text). 

\paragraph{Stability.}~

The proof is the same as the reference market. 

\subsection{Global sensitivity analysis (GSA) of the FTM model}

The FTM model includes many parameters, including initial conditions of state variables. To assess the robustness of our key analytical findings to uncertainty and variability in parameter values, we carry out a GSA on key economic outputs. 

We rank the relative importance of all parameters of potential importance with an improved version of the Morris method, a GSA technique used to screen the importance of parameters in high-dimensional models \citep{CampolongoEtal2007MorrisOptimalDesignEMS}. In a nutshell, the improved Morris method can discriminate the sign and overall influence of factors at a low computational cost and minor risk of error (see below for further details).

\subsubsection{Inputs and outputs explored}

Table \ref{tab:inputs_SA_Eco} describes the input parameters of the FTM model and associated ranges explored in the GSA. Each parameter $d$ with set in a range $[a ; b]$ is assumed to follow a probability distribution which is either uniform ($d \sim \mathcal{U}(a,b)$) or Ôlog-uniformÕ ($\log(d) \sim \mathcal{U}(\log_{10}(a), \log_{10}(b)$). We assume a constant number of suppliers ($N_S = 193,354$) and demanders ($N_D = 118,503$), but allow for a variable proportion of wholesalers in the system ($p_{N_{S \cap D}} \equiv \frac{N_{S \cap D}}{ \min \{ N_S, N_D \} } $). Baseline rates at which supply and demand are generated ($\sigma_0 = 39 $ per year and $\delta_0 = 64$ per year respectively) are also kept constant. Values of $N_S$, $N_D$, $\sigma_0$ and $\delta_0$ are derived from the French cattle market (Table~\ref{stab:parms}).

\begin{table}[H]
\caption{\bf{Input parameters and associated ranges explored in the GSA of the FTM model.} \hfill \label{tab:inputs_SA_Eco}}
\begin{center}
\begin{tabular}{llllll}
  \hline
 Parameter & Formal notation & Meaning & Range explored & Distribution \\ 
  \hline
   k & $\kappa$ & coefficient of friction & [$10^{-2}$ ; $10^2$] & log-uniform \\
   rs & $r_S$ & supply loss rate & [-0.05 ; 33] per year & uniform \\
   rd & $r_D$ & demand loss rate & [-0.05 ; 33] per year & uniform \\
   is & $E_S$ & supply import rate & [0 ; $\Phi^*$] per year & uniform \\
   id & $E_D$ & demand import rate & [0 ; $\Phi^*$] per year & uniform \\
   pNsd  & $p_{N_{S \cap D}}$ & proportion of wholesalers & [0.5 ; 1] & uniform \\ 
   St0 & $S(t_0)$ & initial supply stock & [$0.8 \kappa N_S$ ; $1.2 \kappa N_S$] & uniform \\ 
   Dt0 & $D(t_0)$ & initial demand stock & [$0.8 \kappa N_D$ ; $1.2 \kappa N_D$] & uniform \\ 
   pt0 & $p(t_0)$ & initial price & [0.8 ; 1.2] & uniform \\ 
   es & $\varepsilon_{S}$ & price elasticity of supply & [0 ; 3] & uniform \\ 
   ed & $\varepsilon_{D}$ & price elasticity of demand & [0 ; 3] & uniform \\ 
   mu & $\mu$ & pricing scale parameter & [0 ; $10^{-6} / \kappa$] & uniform \\
    \hline
\end{tabular}
\end{center}
\end{table}

The following economic outputs are analysed with the GSA: 
\begin{itemize}
\item the normalized equilibrium flow, $\tfrac{\Phi_{eq}}{\Phi^*} \in \mathbb{R}^+$,
\item the time to reach 95 \% of equilibrium flow, $t_{95}(\Phi_{eq}) \in [0 ; 200]$~year, 
\item the normalized flow at final time $t_f = 200$~year, $\tfrac{\Phi(t_f)}{\Phi^*} \in \mathbb{R}^+$,
\item the excess demand per agent at equilibrium (\ref{generalexcessdemand}), $\tfrac{E_{eq}(D_0 - (D_{min}^* - S_{min}^*),D_0)}{N} = \tfrac{D_{eq} - S_{eq} - \kappa (N_D - N_S)}{N} \in \mathbb{R}$,
\item the equilibrium price $p_{eq} \in \mathbb{R}^+$, 
\item and the time to reach 95 \% of equilibrium price, $t_{95}(p_{eq}) \in [0 ; 200]$~year. 
\end{itemize}

\subsubsection{Analysis with the improved Morris method}

\paragraph{Details on the method} ~\\

The Morris method is a very efficient GSA technique to explore numerically models with a large parametric space at a low computational cost and minor risk of error. The method is referred to as a global SA technique since it analyses distributions of local elementary effects based on various numerical evaluations of inputs sampled from the parametric space. The improved Morris method ensures that the coverage of the parametric space is improved compared with the standard Morris method by selecting a subset of evaluations that satisfy a maximin-type distance criterion in the parametric space. 

More precisely, the Morris method relies on calculating random trajectories with incremental variations along the $d$ parameters $1,...,i,...,d$  \citep[see][for details]{CampolongoEtal2007MorrisOptimalDesignEMS}. For a given trajectory, each parameter is selected randomly across $l$ levels, so that a set of $d+1$ values are explored sequentially by changing each one of the parameter at a time (OAT). A total of $r_{\max}$ trajectories are sampled randomly to constitute a basic OAT design. To ensure an optimal coverage of the parameter space, $r$ out $r_{\max}$ trajectories are then chosen to maximise the minimal Haussdorf distance between the $r_{\max}$ selected trajectories (maximin-type optimised design). Incremental ratios called elementary effects $EE$ are then computed. An $EE$ is formally defined as the difference in a given output induced by the difference in a jump of one level for a given parameter. Finally, a distribution of $r$ elementary effects is obtained for each parameter $i$, denoted $EE(i)$. Three simple metrics are derived from $EE(i)$: 
\begin{itemize}
\item $m^*$, the mean of the absolute value of $EE(i)$, which represents the overall influence of the parameter $i$ on the output,
\item $m$, the mean of $EE(i)$, which gives the overall sign of the influence of $i$ (and overall influence of $i$ provided the effect in monotonous),
\item $s$, the standard deviation of $EE(i)$, which jointly quantities the non-linear behaviour of $i$ and interactions of $i$ with other parameters $j$. 
\end{itemize}
Notice that as various trajectories are explored in the parameter space, the Morris method is a global SA tool. The improvement on the original Morris method consists in the maximin space-filling design and the introduction of $m^*$. The improved Morris method is implemented in the \textit{morris} function of the $R$ package \textit{sensitivity}. In our case we take $r_{\max}=1000$, $r=200$ (the value recommended in the literature is $50$), $l=5$. Hence, the impact of $d=12$ parameters are explored at the cost of $(d+1)*r = 2600$ simulations. In contrast, a standard SA based on an analysis of variance would require $l^d = 244140625$ simulations. 

\paragraph{Results} ~\\

The GSA confirms that the coefficient of friction $\kappa$ is a key parameter governing trade dynamics.  

\begin{table}[H]
\begin{minipage}{.5\textwidth}
\begin{table}[H]
\caption*{Sensitivity of $\tfrac{\Phi_{eq}}{\Phi^*}$}
\begin{center}
{\small
\begin{tabular}{rrrr}
  \hline
 & m* & m & s \\ 
   \hline
  k & 1.34 & -1.27 & 0.88 \\ 
  is & 0.37 & 0.36 & 0.33 \\ 
  id & 0.35 & 0.34 & 0.33 \\ 
  rs & 0.28 & -0.27 & 0.82 \\ 
  rd & 0.25 & -0.25 & 1.13 \\ 
  mu & 0.25 & 0.07 & 0.83 \\ 
  pt0 & 0.15 & 0.02 & 0.28 \\ 
  es & 0.15 & -0.05 & 0.32 \\ 
  ed & 0.13 & 0.01 & 0.23 \\ 
  St0 & 0.01 & 0.00 & 0.02 \\ 
  Dt0 & 0.01 & -0.00 & 0.01 \\ 
  pNsd & 0.00 & 0.00 & 0.00 \\ 
   \hline
\end{tabular}
}
\end{center}
\end{table}
\end{minipage}
\begin{minipage}{.5\textwidth}
\begin{table}[H]
\caption*{Sensitivity of $t_{95}(\Phi_{eq})$}
\begin{center}
{\small
\begin{tabular}{rrrr}
  \hline
 & m* & m & s \\ 
  \hline
k & 3.00 & 3.00 & 14.30 \\ 
  rs & 1.60 & -1.01 & 12.13 \\ 
  rd & 1.29 & -1.17 & 10.76 \\ 
  mu & 0.48 & -0.41 & 3.02 \\ 
  es & 0.28 & -0.25 & 2.45 \\ 
  id & 0.08 & 0.04 & 0.35 \\ 
  is & 0.06 & -0.04 & 0.16 \\ 
  ed & 0.05 & -0.04 & 0.39 \\ 
  pt0 & 0.05 & -0.01 & 0.13 \\ 
  St0 & 0.01 & -0.00 & 0.03 \\ 
  Dt0 & 0.01 & 0.00 & 0.03 \\ 
  pNsd & 0.00 & 0.00 & 0.00 \\ 
   \hline
\end{tabular}
}
\end{center}
\end{table}
\end{minipage}
\end{table}

\begin{table}[H]
\begin{minipage}{0.5\textwidth}
\begin{table}[H]
\caption*{Sensitivity of $\tfrac{\Phi(t_f)}{\Phi^*}$}
\begin{center}
{\small
\begin{tabular}{rrrr}
  \hline
 & m* & m & s \\ 
 \hline
ed & 5.26 & -5.12 & 62.64 \\ 
  mu & 5.08 & 0.21 & 41.98 \\ 
  rs & 3.08 & -3.06 & 30.63 \\ 
  k & 1.66 & -0.95 & 4.13 \\ 
  id & 1.45 & 1.45 & 13.58 \\ 
  rd & 0.44 & -0.43 & 2.55 \\ 
  pt0 & 0.36 & -0.12 & 2.21 \\ 
  is & 0.36 & 0.36 & 0.32 \\ 
  es & 0.15 & -0.05 & 0.32 \\ 
  St0 & 0.06 & -0.05 & 0.66 \\ 
  Dt0 & 0.05 & 0.04 & 0.50 \\ 
  pNsd & 0.00 & 0.00 & 0.00 \\ 
   \hline
\end{tabular}
}
\end{center}
\end{table}
\end{minipage}
\begin{minipage}{.5\textwidth}
\begin{table}[H]
\caption*{Sensitivity of $\tfrac{D_{eq} - S_{eq} - \kappa (N_D - N_S)}{N}$}  
\begin{center}
{\small
\begin{tabular}{rrrr}
  \hline
 & m* & m & s \\ 
 \hline
rs & 107.45 & 107.42 & 923.38 \\ 
  mu & 103.58 & 42.85 & 748.16 \\ 
  k & 70.25 & 16.57 & 326.82 \\ 
  rd & 62.45 & -61.45 & 626.41 \\ 
  es & 16.68 & -16.26 & 162.17 \\ 
  pNsd & 6.46 & 0.97 & 32.02 \\ 
  pt0 & 3.59 & -3.58 & 19.70 \\ 
  id & 2.70 & 2.60 & 9.54 \\ 
  is & 2.24 & -2.24 & 7.68 \\ 
  ed & 1.23 & 0.73 & 8.75 \\ 
  St0 & 1.16 & -1.15 & 6.14 \\ 
  Dt0 & 0.90 & 0.84 & 4.23 \\ 
   \hline
\end{tabular}
}
\end{center}
\end{table}
\end{minipage}
\end{table}

\begin{table}[H]
\begin{minipage}{.5\textwidth}
\begin{table}[H]
\caption*{Sensitivity of $p_{eq}$}
\begin{center}
{\small
\begin{tabular}{rrrr}
  \hline
 & m* & m & s \\ 
\hline
es & 0.77 & -0.71 & 5.89 \\ 
  rd & 0.73 & -0.73 & 7.04 \\ 
  k & 0.67 & -0.24 & 2.82 \\ 
  mu & 0.66 & 0.14 & 3.63 \\ 
  id & 0.39 & 0.39 & 2.39 \\ 
  is & 0.36 & -0.36 & 2.38 \\ 
  pt0 & 0.22 & 0.22 & 0.17 \\ 
  rs & 0.21 & 0.21 & 0.61 \\ 
  ed & 0.09 & 0.03 & 0.24 \\ 
  St0 & 0.02 & 0.02 & 0.05 \\ 
  Dt0 & 0.01 & -0.01 & 0.01 \\ 
  pNsd & 0.00 & 0.00 & 0.00 \\ 
   \hline
 \end{tabular}
}
\end{center}
\end{table}
\end{minipage}
\begin{minipage}{.5\textwidth}
\begin{table}[H]
\caption*{Sensitivity of $t_{95}(p_{eq})$}
\begin{center}
{\small
\begin{tabular}{rrrr}
  \hline
 & m* & m & s \\ 
  \hline
k & 0.55 & 0.50 & 0.92 \\ 
  mu & 0.30 & -0.29 & 0.60 \\ 
  pt0 & 0.08 & 0.07 & 0.22 \\ 
  rs & 0.05 & -0.00 & 0.08 \\ 
  id & 0.03 & 0.03 & 0.06 \\ 
  is & 0.01 & -0.01 & 0.02 \\ 
  rd & 0.01 & 0.00 & 0.03 \\ 
  ed & 0.01 & -0.01 & 0.01 \\ 
  Dt0 & 0.01 & -0.01 & 0.02 \\ 
  es & 0.01 & -0.01 & 0.01 \\ 
  St0 & 0.01 & 0.01 & 0.01 \\ 
  pNsd & 0.00 & 0.00 & 0.00 \\ 
   \hline
\end{tabular}
}
\end{center}
\end{table}
\end{minipage}
\end{table}

\clearpage

\section{Supplementary results on the market-epidemiological (ME) model}

\subsection{Analytical insights on the ME model}

\subsubsection{Presentation of the complete ME model}

Keeping the notations of the main text, the complete ME dynamics are given by the following set of ordinary differential equations: 
\begin{equation}
\begin{gathered}
\dot S = \overbrace { \underbrace{N_S^{XY}~\sigma_0~p^{\varepsilon_S}}_{\Sigma_\oplus} - \gamma \tfrac{N_S^Y}{N_S^{XY}}S + E_S}^\Sigma  - \overbrace {  \underbrace{\tfrac{\min \{ \Sigma_\oplus ;\Delta_\oplus \}}{\kappa}}_{\Theta} \underbrace{\min \{ \tfrac{S}{N_S^{XY}};\tfrac{D}{N_D^{XY}} \} }_q}^\Phi ~, \hfill \\    
\dot D = \overbrace { \underbrace{N_D^{XY}~\delta_0~p^{- \varepsilon_D}}_{\Delta_\oplus} - \gamma \tfrac{N_D^Y}{N_D^{XY}}D + E_D}^\Delta  - \overbrace { \Theta q}^\Phi ~, \hfill \\        
\dot p = \mu (\Delta - \Sigma) p ~, \hfill \\    
\dot N_{S\backslash{}D}^X = \nu N_{S\backslash{}D}^Z - \Lambda_{\overline{tr}}P_{RA} N_{S\backslash{}D}^X ~, \hfill \\
\dot N_{S\backslash{}D}^Y = \Lambda_{\overline{tr}}P_{RA} N_{S\backslash{}D}^X - \gamma N_{S\backslash{}D}^Y ~, \hfill \\
\dot N_{S\backslash{}D}^Z =   \gamma N_{S\backslash{}D}^Y - \nu N_{S\backslash{}D}^Z ~, \hfill \\ 
\dot N_{S \cap D}^X = \nu N_{S \cap D}^Z -  (\Lambda_{tr}+\Lambda_{\overline{tr}})P_{RA} N_{S \cap D}^X  ~, \hfill \\   
\dot N_{S \cap D}^Y = (\Lambda_{tr}+\Lambda_{\overline{tr}})P_{RA} N_{S \cap D}^X - \gamma N_{S \cap D}^Y  ~, \hfill \\   
\dot N_{S \cap D}^Z = \gamma N_{S \cap D}^Y - \nu N_{S \cap D}^Z ~, \hfill \\   
\dot N_{D\backslash{}S}^X = \nu N_{D\backslash{}S}^Z - (\Lambda_{tr}+\Lambda_{\overline{tr}}) P_{RA} N_{D\backslash{}S}^X  ~, \hfill \\
\dot N_{D\backslash{}S}^Y = (\Lambda_{tr}+\Lambda_{\overline{tr}}) P_{RA} N_{D\backslash{}S}^X - \gamma N_{D\backslash{}S}^Y ~, \hfill \\
\dot N_{D\backslash{}S}^Z =   \gamma N_{D\backslash{}S}^Y - \nu N_{D\backslash{}S}^Z ~. \hfill \\ 
\label{eqS:ecoepi}
\end{gathered}
\end{equation}

Here we restrict ourselves to the case of symmetric imports, i.e. $E_S = E_D = E \ge 0$.  

The forces of infection via trade routes ($\Lambda_{tr}$) and non-trade routes ($\Lambda_{\overline{tr}}$) are given by:
\begin{equation}
\begin{gathered}
\Lambda_{tr} =
\overbrace {\underbrace{ [ 1 - (1 - \phi)^q ] }_{P_{tr}(q)} \frac{\Theta }{N_D^{XY}} }^{\beta_{tr}}  \frac{N_S^Y}{N_S^{XY}} ~, \hfill \\
\Lambda_{\overline{tr}} = \beta_{\overline{tr}}  \frac{N^Y}{N^{XY}} ~. \hfill
\label{eqS:forces_infection}
\end{gathered}
\end{equation}

The risk aversion factor $P_{RA} \in [0, 1]$ is given by:
\begin{equation}
P_{RA} = \left(1 - \frac{N^Z}{N} \right)^\alpha ~.
\label{eqS:PRA}
\end{equation}

\subsubsection{Predicting the global behaviour of the ME model with the basic reproduction number $R_{0}$}

$R_0$, the basic reproduction number, is the average number of susceptible that will be infected along the course of an epidemic by a single infectious agent propagated in an initially disease-free population \citep{AndersonMay1991Book}. The global behavior of the coupled model depends on the position of $R_0$ with respect to 1. If $R_0 \le 1$, the epidemic will eventually die out. If $R_0 > 1$ the epidemic will invade the population. 

We derive $R_0$ using the next-generation matrix approach from \citet{DiekmannEtal2010Interface}. 

We take $S = \kappa N_S \tfrac{\Phi^* + E}{\Phi^*} $ and $D = \kappa N_D \tfrac{\Phi^* + E}{\Phi^*}$ as initial values for supply and demand, i.e. the equilibrium values of $S$ and $D$ in the FTM model when imports are not negligible (system~(\ref{sys:ref_market_and_E})). It follows the market is equilibrated before the epidemic onset, which implies, $\Theta = \Theta^* = \frac{\Phi^*}{\kappa}$, $q = \kappa \tfrac{\Phi^* + E}{\Phi^*}$ and $\Phi = \Phi^* + E$. 

To calculate $R_0$, we focus on the Ôdisease-free equilibriumÕ (DFE) where a small number of agents are initially infected. Since the market is initially equilibrated, we assume: $\Sigma \approx \Phi^* + E$ and $\Delta \approx \Phi^* + E$. Solving $\gamma \tfrac{N_S^Y}{N_S}S << \Phi^* + E$ and $\gamma \tfrac{N_D^Y}{N_D}D << \Phi^* + E$ with $S = \kappa N_S \tfrac{\Phi^* + E}{\Phi^*} $ and $D = \kappa N_D \tfrac{\Phi^* + E}{\Phi^*} $, we see that this approximation stands when $\max \{ N_S^Y; N_D^Y \} << \frac{\Phi^*}{\gamma \kappa}$. We now check if this assumption is in agreement with our livestock data. In practice, we have $\gamma_\text{max} = 33$ per year and $\kappa_\text{max} = 100$, so $\gamma_\text{max} \kappa_\text{max} \approx 10^3$ per year. $\Phi^* \approx 10^7$ per year (see Table~\ref{stab:parms} for reference values on $\kappa$ and $\Phi^*$). It follows $\frac{\Phi^*}{\gamma_\text{max} \kappa_\text{max}} \approx 10^4 \le \frac{\Phi^*}{\gamma \kappa}$. Our assumption on the values of $\Sigma$ and $\Delta$ at the DFE hence seems reasonable. 

At the DFE, we notice from system (\ref{eqS:ecoepi}) and equations (\ref{eqS:forces_infection})-(\ref{eqS:PRA}) that: 
\begin{equation}
\begin{gathered}
\Lambda_{tr} \approx  \overbrace { P_{tr}(\kappa \tfrac{\Phi^* + E}{\Phi^*}) \frac{\Phi^*}{\kappa N_D} }^{\beta_{tr}} \frac{{N_S^Y}}{{N_S}} ~, \hfill \\
\Lambda_{\overline{tr}} \approx \beta_{\overline{tr}}  \frac{N^Y}{N} ~, \hfill \\
P_{RA} \approx 1 ~. \hfill
\end{gathered}
\end{equation}

We can hence derive the infectious subsystem at the DFE: 
\begin{equation}
\begin{gathered}
\dot N_{S\backslash{}D}^Y = \hspace{2.5 cm} \beta_{\overline{tr}}  \frac{N_{S\backslash{}D}}{N} N^Y - \gamma N_{S\backslash{}D}^Y ~, \hfill \\
\dot N_{S \cap D}^Y = \beta_{tr} \frac{N_{S \cap D}}{N_S} N_S^Y +  \beta_{\overline{tr}} \frac{N_{S \cap D}}{N} N^Y - \gamma N_{S \cap D}^Y ~, \hfill \\   
\dot N_{D\backslash{}S}^Y = \beta_{tr} \frac{N_{D\backslash{}S}}{N_S} N_S^Y +  \beta_{\overline{tr}} \frac{N_{D\backslash{}S}}{N} N^Y - \gamma N_{D\backslash{}S}^Y . \hfill \\
\label{seqS:InfectiousSubsystem}
\end{gathered}
\end{equation}

We notice the right-hand side of system (\ref{seqS:InfectiousSubsystem}) is only a function of $N_S^Y$ and $N^Y$. We can simplify system (\ref{seqS:InfectiousSubsystem}) by direct summation over suppliers ($N_S^Y = N_{S\backslash{}D}^Y + N_{S \cap D}^Y$) and agents ($N^Y = N_S^Y + N_{D\backslash{}S}^Y$): 
\begin{equation}
\begin{gathered}
\dot N_S^Y = \beta_{tr} \frac{N_{S \cap D}}{N_S} N_S^Y +  \beta_{\overline{tr}} \frac{N_S}{N} N^Y - \gamma N_S^Y ~, \hfill \\  
\dot N^Y = \beta_{tr} \frac{N_D}{N_S} N_S^Y +  \beta_{\overline{tr}} N^Y - \gamma N^Y ~. \hfill \\   
\label{seqS:InfectiousSubsystem2}
\end{gathered}
\end{equation}

We decompose the Jacobian matrix $J$ from subsystem (\ref{seqS:InfectiousSubsystem2}) into the sum $T+F$, where $T = \left( {\begin{array}{*{20}{c}}
{{\beta _{tr}}\frac{{{N_{S \cap D}}}}{{{N_S}}}}&{{\beta _{\overline {tr} }}\frac{{{N_S}}}{N}}\\
{{\beta _{tr}}\frac{{{N_D}}}{{{N_S}}}}&{{\beta _{\overline {tr} }}}
\end{array}} \right)$ and $F = \left( {\begin{array}{*{20}{c}}{ - \gamma }&0\\ 0&{ - \gamma } \end{array}} \right)$ are the matrices accounting respectively for transmission and transitions from an epidemiological point-of-view. 

We finally derive the next-generation matrix with large domain given by:
\begin{equation}
K_L =  - T F^{-1} = \left( {\begin{array}{*{20}{c}}
{\frac{{{\beta _{tr}}}}{\gamma }\frac{{{N_{S \cap D}}}}{{{N_S}}}}&{\frac{{{\beta _{\overline {tr} }}}}{\gamma }\frac{{{N_S}}}{N}}\\
{\frac{{{\beta _{tr}}}}{\gamma }\frac{{{N_D}}}{{{N_S}}}}&{\frac{{{\beta _{\overline {tr} }}}}{\gamma }}
\end{array}} \right)
\label{seqS:KL}
\end{equation}
 
$R_0$ is the leading eigenvalue of $K_L$. The two eigenvalues of $K_L$, denoted $a$, are obtained by solving the equation $\det (K_L - a I_2) = 0$, i.e. by solving the equation:
\begin{equation}
(\frac{{{\beta _{tr}}}}{\gamma }\frac{{{N_{S \cap D}}}}{{{N_S}}} - a)(\frac{{{\beta _{\overline {tr} }}}}{\gamma } - a) - (\frac{{{\beta _{tr}}}}{\gamma }\frac{{{N_D}}}{{{N_S}}})(\frac{{{\beta _{\overline {tr} }}}}{\gamma }\frac{{{N_S}}}{N}) = 0 ~.
\label{seqS:eigsKL}
\end{equation}

Equation (\ref{seqS:eigsKL}) is quadratic in $a$ with a positive discriminant. It follows equation (\ref{seqS:eigsKL}) has two real roots $a_1 \le a_2$ with $R_0 = a_2$ given by: 
\begin{equation}
\begin{gathered}
R_0 = \frac {R_0^{tr} + R_0^{\overline{tr}} + \sqrt{ (R_0^{tr} - R_0^{\overline{tr}} )^2 + 4 R_0^{tr} R_0^{\overline{tr}} \frac{N_S N_D}{N N_{S \cap D}} }}{2}\text{, with $R_0^{tr}$ and $R_0^{\overline{tr}}$ given by:} \hfill \\
R_0^{tr} = \frac{\beta_{tr}}{\gamma} \frac{N_{S \cap D}}{N_S}~, \hfill \\
R_0^{\overline{tr}} = \frac{\beta_{\overline{tr}}}{\gamma}~.  \hfill \\
\label{eqS:R0}
\end{gathered}
\end{equation}

If we assume that all agents are wholesalers ($N_S = N_D = N_{S \cap D} = N$), $R_0$ simplifies to $R_0 = R_0^{tr} + R_0^{\overline{tr}}$. 

\subsubsection{Insights on the impacts of epidemics on market dynamics}

Epidemics impact the market by decreasing the number of active suppliers and demanders and depleting the stocks of supply and demand. Assuming trade flow is equilibrated at the reference market level, it follows epidemics act by decreasing the net creations rates of supply and demand: $\Sigma \le \Phi^*$ and $\Delta \le \Phi^*$. The latter implies $\Phi \le \Phi^*$, i.e. trade flow is decreased by epidemics.  

\begin{figure}[h]
\begin{center}
\includegraphics[width=\textwidth]{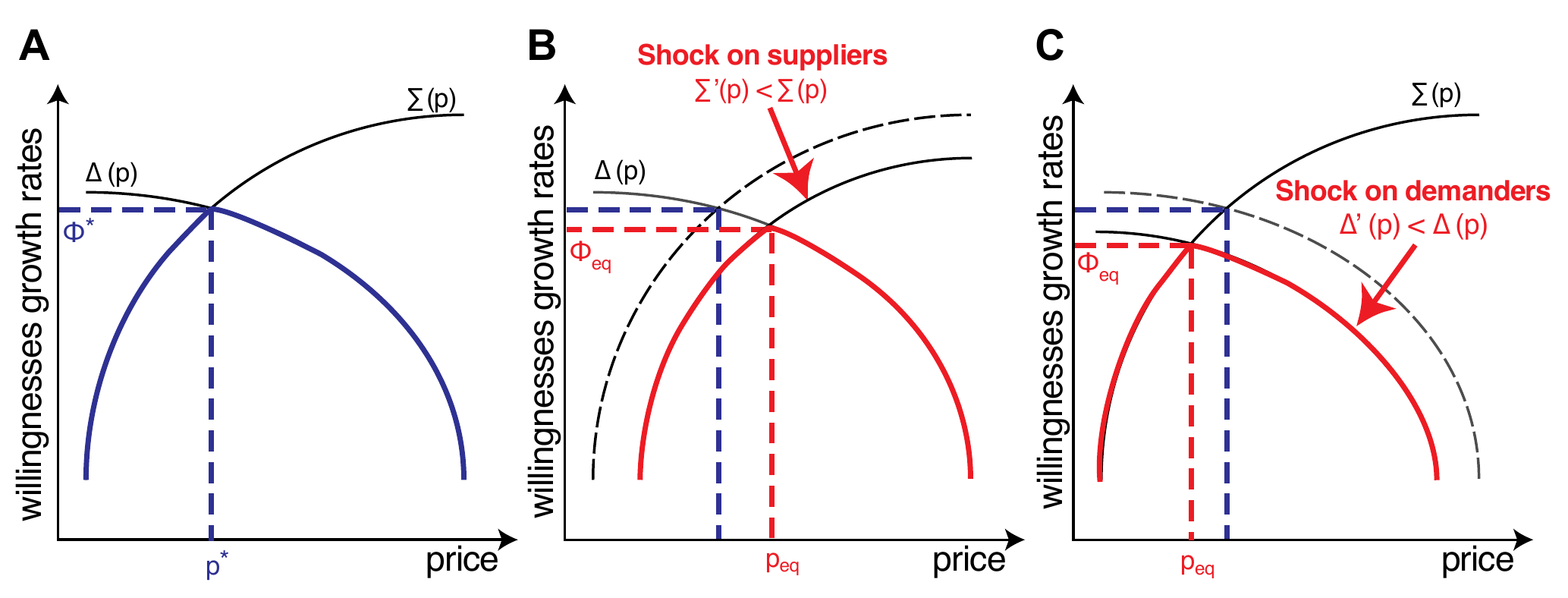}
\end{center}
\caption[Impact of shocks such as epidemics on trade flow and prices]{{\bf Impact of shocks such as epidemics on trade flow and prices}
\small 
\label{sfig:FlowsPriceShocks}}
\end{figure}

The evolution of price is less intuitive. Generally speaking, an equilibrium price will satisfy $\Sigma(S,N_S^{XY},p_{eq}) = \Delta(D,N_D^{XY},p_{eq})$. We assume trade is the only path of transmission. Since trade-related transmission is directed from suppliers to demanders, demanders will be more impacted than suppliers. From the constraint $\Sigma = \Delta$, it directly follows that equilibrium prices will tend to decrease when submitted to epidemiological shocks (Fig. \ref{sfig:FlowsPriceShocks}). This is not necessarily true any more if other paths of transmission are included. 

\clearpage

\subsection{Numerical exploration of the ME model}

\subsubsection{Impacts of frictional-trade dynamics with risk aversion on disease dynamics (additional results)}

\begin{figure}[h]
\begin{center}
\includegraphics[width=\textwidth]{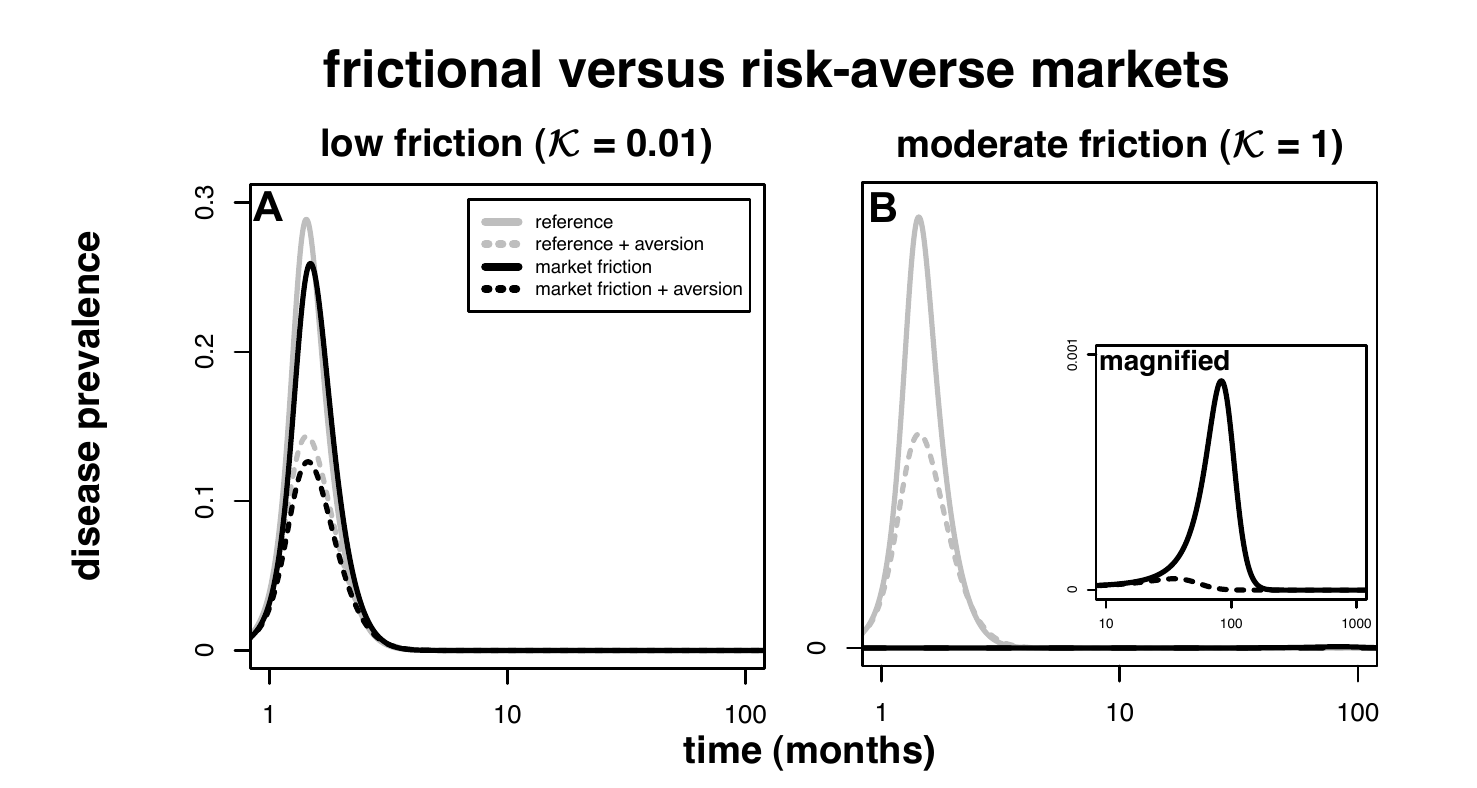}
\end{center}
\caption[Impacts of frictional-trade dynamics with risk aversion on disease dynamics (additional results)]
{{\bf Impacts of frictional-trade dynamics with risk aversion on disease dynamics (additional results).}
\small Evolution of disease prevalence ($Y(t)/N$) as function of time with low (\textbf{A}) and moderate friction (\textbf{B}). Meaning of colours/lines and values of parameters are the same as in Figure 3A-B of the main text. 
\label{sfig:prevalence}}
\end{figure}

\clearpage

\subsubsection{Impacts of epidemics on trade dynamics}

The main text focuses on the impacts of frictional-trade dynamics on epidemics. Here we show an alternative point-of-view: the impact of epidemics on overall trade dynamics captured by the evolution of trade flow and price.

\begin{figure}[h]
\begin{center}
\includegraphics[width=\textwidth]{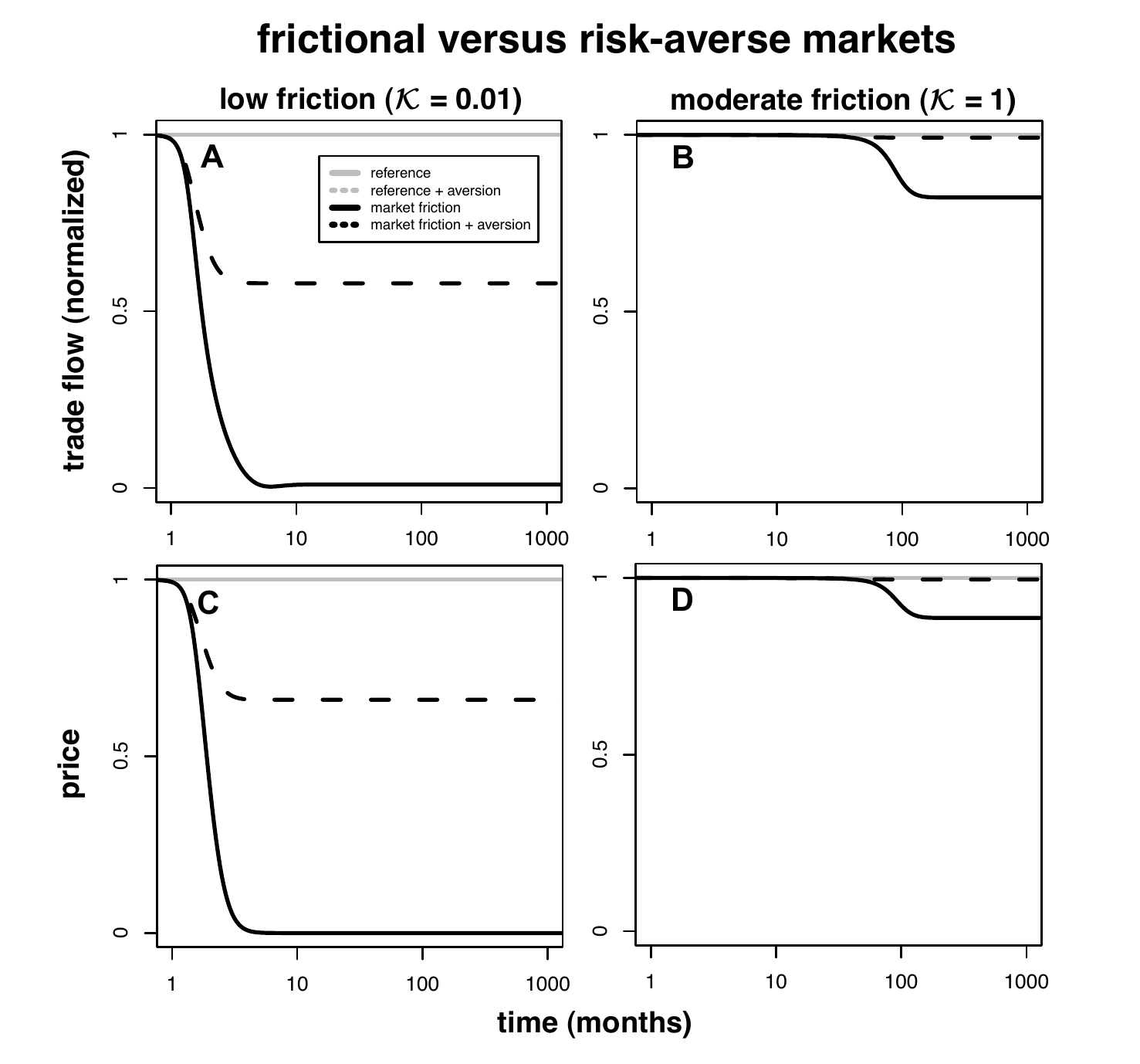}
\end{center}
\caption[ Impacts of epidemics on trade dynamics.]
{{\bf Impacts of epidemics on trade dynamics}
\small Evolution of normalized trade flow ($\frac{\Phi(t)}{\Phi^*}$, \textbf{A-B}) and price ($p(t)$, \textbf{C-D}) as function of time with low (\textbf{A-C}) and moderate friction (\textbf{B-D}). Meaning of colours/lines and values of parameters are the same as in Figure 3A-B of the main text. 
\label{sfig:Epi_on_Market}}
\end{figure}

\clearpage

\subsubsection{GSA of the ME model}

As for the FTM model, we assess the robustness of our conclusions on key economic and epidemiological outputs with the Morris method.

\paragraph{Inputs and outputs explored} ~\\

Table \ref{tab:inputs_SA_Coupled} describes the input parameters and associated ranges explored in the sensitivity analyses. Except when stated otherwise, notations and parameters are the same as described in the GSA of the FTM model. 

\begin{table}[H]
\caption{\bf{Input parameters and associated ranges explored in the GSA of the ME model.} \hfill \label{tab:inputs_SA_Coupled}}
\begin{center}
{\footnotesize
\begin{tabular}{llllll}
  \hline
 Parameter & Formal notation & Meaning & Range explored & Distribution \\ 
  \hline
   k & $\kappa$ & coefficient of friction & [$10^{-2}$ ; $10^2$] & log-uniform \\
   i & $E_S = E_D = E$ & supply and demand import rates (assumed equal) & [0 ; $\Phi^*$] per year & uniform \\
   pNsd  & $p_{N_{S \cap D}}$ & proportion of wholesalers & [0.5 ; 1] & uniform \\ 
   St0 & $S(t_0)$ & initial supply stock & [$0.8 \kappa N_S \tfrac{\Phi^* + E}{\Phi^*}$ ; $1.2 \kappa N_S  \tfrac{\Phi^* + E}{\Phi^*}$] & uniform \\ 
   Dt0 & $D(t_0)$ & initial demand stock & [$0.8 \kappa N_D  \tfrac{\Phi^* + E}{\Phi^*}$ ; $1.2 \kappa N_D  \tfrac{\Phi^* + E}{\Phi^*}$] & uniform \\ 
   pt0 & $p(t_0)$ & initial price & [0.8 ; 1.2] & uniform \\ 
   es & $\varepsilon_{S}$ & price elasticity of supply & [0 ; 3] & uniform \\ 
   ed & $\varepsilon_{D}$ & price elasticity of demand & [0 ; 3] & uniform \\ 
   mu & $\mu$ & pricing scale parameter & [0 ; $10^{-6} / \kappa$] & uniform \\
   alpha & $\alpha$ & level of aversion to risk & [0 ; 8] & uniform \\ 
   gamma  & $\gamma$ & rate of detection and removal & [0.5 ; 33] per year & uniform \\ 
   nu  & $\nu$ & rate of market reentry & [0 ; 33] per year & uniform \\ 
   phi & $\phi$ & probability of infection per good exchanged & [0 ; $\phi^{\max}$] & uniform \\
   betabar & $\beta_{\overline{tr}}$ & non-trade-related rate of transmission & [0 ; $\beta_{\overline{tr}}^{\max}$] per year & uniform\\
    \hline
    \multicolumn{5}{p{\textwidth}}{\small To avoid unrealistic epidemics, we first set upper bounds on the trade-related ($R_{0}^{tr}$) on non-trade-related ($R_{0}^{\overline{tr}}$) components of $R_0$: $R_{0}^{{tr},\max} = 10$ and $R_{0}^{\overline{tr},\max} = 10$ respectively. Then we calculate the corresponding upper bounds for the probability of infection per good exchanged ($\phi^{\max}$) and the non-trade-related rate of transmission ($\beta_{\overline{tr}}^{\max}$) where ${\phi^{\max}~=~1-(1-P_{tr}^{\max})^{\tfrac{\Phi^*}{\kappa (\Phi^*+E)}}}$ with $P_{tr}^{\max} = \min \{1 ; \frac{ \gamma R_{0}^{{tr},\max} \kappa N_S N_D }{\Phi^* N_{S \cap D}} \}$ and $\beta_{\overline{tr}}^{\max} = \gamma \min \{ R_{0}^{\overline{tr},\max}, 2R_{0}^{tr} \}$.}
\end{tabular}
}
\end{center}
\end{table}

The following economic and epidemiological outputs are analysed with the GSA: 
\begin{itemize}
\item the endemic proportion of susceptible, $x_{eq} \in  [0 ; 1]$,
\item the time to reach a real coupled equilibrium (in supply, demand, price and epidemiological categories) or an apparent coupled equilibrium (in trade flow, price and epidemiological categories), $t_{eq} \in [0 ; 200]$~year,  
\item the ratio of equilibrium flow to disease-free equilibrium flow, $\tfrac{\Phi_{eq}}{\Phi^*} \in \mathbb{R}^+$,
\item the equilibrium price, $p_{eq} \in \mathbb{R}^+$, 
\item the basic reproduction number, $R_0 \in \mathbb{R}^+$, 
\item the endemic proportion of infected, $y_{eq} \in [0 ; 1]$, 
\item the proportion of infected at the peak, $y_{\max} \in [0 ; 1]$, 
\item and the total number of cases at equilibrium normalized by total agents, $[\text{total cases}]_{eq} / N  \in \mathbb{R}^+$. 
\end{itemize}

\clearpage

\paragraph{Results} ~\\

The GSA confirms that the coefficient of friction $\kappa$ is a key parameter governing trade and epidemiological dynamics. For all outputs explored, the effect of $\kappa$ systematically outweighs the effect of adaptive risk aversion (RA) controlled by parameter $\alpha$.  

\begin{table}[H] 
\begin{minipage}{.5\textwidth}
\begin{table}[H]
\caption*{Sensitivity of $x_{eq}$}
\begin{center}
{\small
\begin{tabular}{rrrr}
  \hline
 & m* & m & s \\ 
  \hline
k & 0.45 & 0.45 & 0.56 \\ 
  phi & 0.42 & -0.42 & 0.53 \\ 
  gamma & 0.34 & 0.34 & 0.45 \\ 
  betabar & 0.23 & -0.23 & 0.31 \\ 
  alpha & 0.15 & 0.15 & 0.25 \\ 
  nu & 0.07 & 0.03 & 0.19 \\ 
  pNsd & 0.06 & -0.06 & 0.10 \\ 
  i & 0.04 & -0.04 & 0.10 \\ 
  pt0 & 0.01 & -0.00 & 0.03 \\ 
  mu & 0.01 & -0.01 & 0.03 \\ 
  ed & 0.00 & -0.00 & 0.01 \\ 
  es & 0.00 & 0.00 & 0.00 \\ 
  St0 & 0.00 & 0.00 & 0.00 \\ 
  Dt0 & 0.00 & -0.00 & 0.00 \\ 
   \hline
\end{tabular}
}
\end{center}
\end{table}
\end{minipage}
\begin{minipage}{.5\textwidth}
\begin{table}[H]
\caption*{Sensitivity of $t_{eq}$}
\begin{center}
{\small
\begin{tabular}{rrrr}
  \hline
 & m* & m & s \\ 
 \hline
k & 65.44 & 57.95 & 97.86 \\ 
  mu & 34.44 & -0.54 & 79.03 \\ 
  gamma & 24.55 & -15.57 & 55.10 \\ 
  es & 21.11 & -2.34 & 55.29 \\ 
  ed & 20.68 & 2.25 & 57.11 \\ 
  pt0 & 17.89 & 1.59 & 47.08 \\ 
  nu & 14.40 & -5.87 & 41.73 \\ 
  i & 11.85 & 2.57 & 37.64 \\ 
  St0 & 11.56 & 0.99 & 38.74 \\ 
  pNsd & 11.25 & -2.12 & 35.31 \\ 
  betabar & 10.70 & 1.75 & 30.25 \\ 
  phi & 7.59 & 0.51 & 19.83 \\ 
  Dt0 & 7.47 & -1.55 & 28.68 \\ 
  alpha & 3.93 & -2.13 & 20.82 \\ 
   \hline
\end{tabular}
}
\end{center}
\end{table}
\end{minipage}
\end{table}

\begin{table}[H]
\begin{minipage}{0.5\textwidth}
\begin{table}[H]
\caption*{Sensitivity of $\tfrac{\Phi_{eq}}{\Phi^*}$}
\begin{center}
{\small
\begin{tabular}{rrrr}
  \hline
 & m* & m & s \\ 
 \hline
i & 0.94 & 0.94 & 0.13 \\ 
  k & 0.35 & 0.20 & 0.56 \\ 
  phi & 0.29 & -0.29 & 0.42 \\ 
  gamma & 0.23 & 0.02 & 0.42 \\ 
  nu & 0.16 & 0.14 & 0.37 \\ 
  betabar & 0.13 & -0.13 & 0.24 \\ 
  alpha & 0.12 & 0.12 & 0.22 \\ 
  mu & 0.10 & 0.10 & 0.19 \\ 
  pt0 & 0.09 & 0.01 & 0.22 \\ 
  pNsd & 0.05 & -0.05 & 0.12 \\ 
  ed & 0.05 & -0.02 & 0.11 \\ 
  es & 0.04 & -0.03 & 0.11 \\ 
  St0 & 0.00 & 0.00 & 0.01 \\ 
  Dt0 & 0.00 & 0.00 & 0.01 \\ 
   \hline
\end{tabular}
}
\end{center}
\end{table}
\end{minipage}
\begin{minipage}{.5\textwidth}
\begin{table}[H]
\caption*{Sensitivity of $p_{eq}$}
\begin{center}
{\small
\begin{tabular}{rrrr}
  \hline
 & m* & m & s \\ 
  \hline
pt0 & 0.13 & 0.13 & 0.17 \\ 
  k & 0.11 & 0.09 & 0.33 \\ 
  mu & 0.11 & -0.04 & 0.23 \\ 
  phi & 0.10 & -0.08 & 0.25 \\ 
  nu & 0.06 & 0.06 & 0.22 \\ 
  gamma & 0.06 & -0.01 & 0.18 \\ 
  ed & 0.05 & 0.04 & 0.19 \\ 
  alpha & 0.04 & 0.02 & 0.16 \\ 
  es & 0.04 & 0.02 & 0.13 \\ 
  i & 0.02 & 0.00 & 0.14 \\ 
  betabar & 0.02 & 0.01 & 0.06 \\ 
  pNsd & 0.01 & 0.01 & 0.05 \\ 
  St0 & 0.01 & 0.01 & 0.02 \\ 
  Dt0 & 0.00 & -0.00 & 0.01 \\ 
   \hline 
\end{tabular}
}
\end{center}
\end{table}
\end{minipage}
\end{table}

\clearpage

\begin{table}[H]
\begin{minipage}{.5\textwidth}
\begin{table}[H]
\caption*{Sensitivity of $R_0$}
\begin{center}
{\small
\begin{tabular}{rrrr}
  \hline
 & m* & m & s \\ 
  \hline
k & 7.13 & -6.84 & 9.88 \\ 
  phi & 6.92 & 6.92 & 8.37 \\ 
  gamma & 5.35 & -5.35 & 7.48 \\ 
  betabar & 3.31 & 3.31 & 4.14 \\ 
  pNsd & 0.90 & 0.51 & 1.38 \\ 
  i & 0.40 & 0.40 & 0.76 \\ 
  St0 & 0.00 & 0.00 & 0.00 \\ 
  Dt0 & 0.00 & 0.00 & 0.00 \\ 
  pt0 & 0.00 & 0.00 & 0.00 \\ 
  es & 0.00 & 0.00 & 0.00 \\ 
  ed & 0.00 & 0.00 & 0.00 \\ 
  mu & 0.00 & 0.00 & 0.00 \\ 
  alpha & 0.00 & 0.00 & 0.00 \\ 
  nu & 0.00 & 0.00 & 0.00 \\ 
   \hline 
 \end{tabular}
}
\end{center}
\end{table}
\end{minipage}
\begin{minipage}{.5\textwidth}
\begin{table}[H]
\caption*{Sensitivity of $y_{eq}$}
\begin{center}
{\small
\begin{tabular}{rrrr}
  \hline
 & m* & m & s \\ 
 \hline
gamma & 0.35 & -0.35 & 0.51 \\ 
  k & 0.22 & -0.22 & 0.38 \\ 
  phi & 0.21 & 0.21 & 0.39 \\ 
  nu & 0.14 & 0.14 & 0.30 \\ 
  betabar & 0.12 & 0.12 & 0.21 \\ 
  alpha & 0.06 & -0.06 & 0.10 \\ 
  pNsd & 0.03 & 0.03 & 0.06 \\ 
  i & 0.02 & 0.02 & 0.04 \\ 
  pt0 & 0.00 & -0.00 & 0.02 \\ 
  mu & 0.00 & 0.00 & 0.02 \\ 
  ed & 0.00 & -0.00 & 0.00 \\ 
  es & 0.00 & -0.00 & 0.00 \\ 
  Dt0 & 0.00 & 0.00 & 0.00 \\ 
  St0 & 0.00 & -0.00 & 0.00 \\ 
   \hline
\end{tabular}
}
\end{center}
\end{table}
\end{minipage}
\end{table}

\begin{table}[H]
\begin{minipage}{0.5\textwidth}
\begin{table}[H]
\caption*{Sensitivity of $y_{max}$}
\begin{center}
{\small
\begin{tabular}{rrrr}
  \hline
 & m* & m & s \\ 
 \hline
gamma & 0.35 & -0.35 & 0.46 \\ 
  k & 0.34 & -0.34 & 0.44 \\ 
  phi & 0.31 & 0.31 & 0.44 \\ 
  betabar & 0.18 & 0.18 & 0.23 \\ 
  alpha & 0.07 & -0.07 & 0.11 \\ 
  nu & 0.06 & 0.06 & 0.11 \\ 
  pNsd & 0.05 & 0.05 & 0.08 \\ 
  i & 0.03 & 0.03 & 0.06 \\ 
  pt0 & 0.01 & 0.00 & 0.02 \\ 
  mu & 0.00 & 0.00 & 0.02 \\ 
  ed & 0.00 & -0.00 & 0.00 \\ 
  es & 0.00 & -0.00 & 0.00 \\ 
  Dt0 & 0.00 & 0.00 & 0.00 \\ 
  St0 & 0.00 & -0.00 & 0.00 \\ 
   \hline
\end{tabular}
}
\end{center}
\end{table}
\end{minipage}
\begin{minipage}{.5\textwidth}
\begin{table}[H]
\caption*{Sensitivity of $[\text{total cases}]_{eq} / N$}
\begin{center}
{\small
\begin{tabular}{rrrr}
  \hline
 & m* & m & s \\ 
\hline
k & 4.50 & -2.15 & 8.26 \\ 
  phi & 3.42 & 2.97 & 6.68 \\ 
  mu & 3.26 & -3.19 & 8.27 \\ 
  gamma & 2.31 & -1.13 & 5.60 \\ 
  betabar & 2.13 & 1.01 & 5.70 \\ 
  nu & 1.98 & 1.36 & 5.28 \\ 
  pt0 & 1.26 & 0.31 & 4.93 \\ 
  alpha & 1.21 & -0.63 & 3.21 \\ 
  i & 0.90 & 0.60 & 3.70 \\ 
  es & 0.85 & -0.45 & 3.36 \\ 
  pNsd & 0.83 & 0.33 & 2.62 \\ 
  St0 & 0.60 & 0.33 & 3.64 \\ 
  ed & 0.55 & -0.42 & 2.58 \\ 
  Dt0 & 0.37 & 0.32 & 1.91 \\ 
   \hline
\end{tabular}
}
\end{center}
\end{table}
\end{minipage}
\end{table}

\end{document}